\documentclass[aps,prb,10pt,twocolumn,amsmath,amssymb]{revtex4-2}

\usepackage{graphicx}
\usepackage{amsmath}
\usepackage{dcolumn}
\usepackage{bm}
\usepackage{xcolor}
\usepackage{graphicx}
\usepackage{physics}
\usepackage{mathptmx}
\usepackage{lipsum}
\usepackage{fixltx2e}
\usepackage{gensymb} % Provide \degree
\usepackage[caption=false]{subfig}
\usepackage[version=4]{mhchem} % Chemical formulas using \ce{BaTiO3}

\begin{document}

\title{Type II multiferroic order in two-dimensional transition metal halides from first principles spin-spiral calculations}

\author{Joachim Sødequist}
\affiliation{Computational Atomic-Scale Materials Design (CAMD), Department of Physics, Technical University of Denmark, 2800 Kgs. Lyngby, Denmark}%Lines break automatically or can be
\author{Thomas Olsen}
\email{tolsen@fysik.dtu.dk}
\affiliation{Computational Atomic-Scale Materials Design (CAMD), Department of Physics, Technical University of Denmark, 2800 Kgs. Lyngby, Denmark}%Lines break automatically or can be

\date{\today}

\begin{abstract}
We present a computational search for spin spiral ground states in two-dimensional transition metal halides that are experimentally known as van der Waals bonded bulk materials. Such spin spirals break the rotational symmetry of the lattice and lead to polar ground states where the axis of polarization is strongly coupled to the magnetic order (type II multiferroics). We apply the generalized Bloch theorem in conjunction with non-collinear density functional theory calculations to find the spiralling vector that minimizes the energy and then include spin-orbit coupling to calculate the preferred orientation of the spin plane with respect to the spiral vector. We find a wide variety of magnetic orders ranging from ferromagnetic, stripy anti-ferromagnetic, 120$^\circ$ non-collinear structures and incommensurate spin spirals. The latter two introduce polar axes and are found in the majority of materials considered here. The spontaneous polarization is calculated for the incommensurate spin spirals by performing full supercell relaxation including spinorbit coupling and the induced polarization is shown to be strongly dependent on the orientation of the spiral planes. We also test the effect of Hubbard corrections on the results and find that for most materials LDA+U results agree qualitatively with LDA. An exception is the Mn halides, which are found to exhibit incommensurate spin spiral ground states if Hubbard corrections are included whereas bare LDA yields a 120$^\circ$ non-collinear ground state.% All results are in qualitative agreement with experimental predictions from the bulk phases.
\end{abstract}

\maketitle

\section{Introduction}

The recent discovery of ferromagnetic order in two-dimensional (2D) CrI$_3$ \cite{Huang2017} has initiated a vast interest in 2D magnetism \cite{Gibertini2019, Gong2019, Jiang2021}. Several other materials have subsequently been demonstrated to preserve magnetic order in the monolayer limit when exfoliated from magnetic van der Waals bonded compounds and the family of 2D magnets is steadily growing. A crucial requirement for magnetic order to persist in the 2D limit is the presence of magnetic anisotropy that breaks the spin rotational symmetry that would otherwise render magnetic order at finite temperatures impossible by the Mermin-Wagner theorem \cite{Mermin1966}. This is exemplified by the cases of 2D CrBr$_3$ \cite{Zhang2019,Sun2021} and CrCl$_3$ \cite{Cai2019, Bedoya-Pinto2021}, which are isostructural to CrI$_3$ and while the former remains ferromagnetic in the atomic limit due to easy-axis anisotropy (like CrI$_3$) the latter has a weak easy plane that forbids proper long range order. Other materials with persisting ferromagnetic order in the 2D limit include the metallic compounds Fe$_{3/4/5}$GeTe$_2$ \cite{Fei2018,Seo2020,Li2020} and the anisotropic insulator CrSBr \cite{Lee2021}, which has an easy-axis aligned with the atomic plane. Finally, FePS$_3$ \cite{Lee2016} and MnPS$_3$ \cite{Long2020} constitute examples of in-plane anti-ferromagnets that preserve magnetic order in the monolayer limit due to easy-axis anisotropy, whereas the magnetic order is lost in monolayers of the isostructural easy-plane compound NiPS$_3$ \cite{Kim2019}. The 2D materials mentioned above all constitute examples of rather simple collinear magnets. However, the ground state of three-dimensional magnetic materials often exhibit complicated non-collinear order that gives rise to a range of interesting properties \cite{Qin2020}. Such materials, are so far largely lacking from the field of 2D magnetism and the discovery of new non-collinear 2D magnets would greatly enhance the possibilities of constructing versatile magnetic materials using 2D magnets as building blocks \cite{Sierra2021}. 

The ground state of the classical isotropic Heisenberg model can be shown to be a planar spin spiral characterised by a propagation vector $\mathbf{Q}$ \cite{Kaplan1959} and such spin configurations thus comprise a broad class of states that generalise the concept of ferromagnetism and anti-ferromagnetism. In fact, spin spiral order is rather common in layered van der Waals bonded materials \cite{mcguire2017crystal} and it is thus natural to investigate the ground state order of the corresponding monolayers for spin spiral order. Moreover, for non-bipartite magnetic lattices the concept of anti-ferromagnetism is not unique. This is exemplified by the abundant example of the triangular lattice where one may consider the cases of anti-aligned ferromagnetic stripes or 120$^\circ$ non-collinear order, which can be represented as spin spirals of $\mathbf{Q}=(1/2,0)$ and $\mathbf{Q}=(1/3,1/3)$ respectively \cite{Chernyshev2009,Maksimov2019}. The concept of spin spirals thus constitute a general framework for specifying the magnetic order, which may or may not be commensurate with the crystal lattice.

Finite spin spiral vectors typically break symmetries inherent to the crystal lattice and may thus induce physical properties that are predicted to be absent if one only considers the crystal symmetries. In particular, the spin spiral may yield a polar axis that lead to ferroelectric order \cite{Kimura2003}. Such materials are referred to as type II multiferroics and examples include MnWO$_4$ \cite{PhysRevLett.97.097203}, CoCr$_2$O$_4$ \cite{PhysRevLett.96.207204}, LiCu$_2$O$_2$ \cite{PhysRevLett.98.057601} and LiCuVO$_4$ \cite{doi:10.1143/JPSJ.76.023708} as well as the triangular magnets CuFeO$_2$ \cite{PhysRevB.73.220401}, CuCrO$_2$ \cite{PhysRevB.73.220401}, AgCrO$_2$ \cite{PhysRevLett.101.067204} and MnI$_2$ \cite{PhysRevLett.106.167206}. In addition to these materials, 2D NiI$_2$ has recently been shown to host a spin spiral ground state that induces a spontaneous polarization \cite{Song2022} and 2D NiI$_2$ thus comprises the first example of a 2D type II multiferroic.% Since spiral order is common in layered van der Waals bonded magnets \cite{mcguire2017crystal} it is likely that exfoliated monolayers from these compounds may exhibit type II multiferroic order as well.

The prediction of new materials with certain desired properties can be vastly accelerated by first principles simulations. In general, the search for materials with spin spiral ground states is complicated by the fact that the magnetic order requires large super cells in the simulations. However, if one neglects spinorbit coupling, spin spirals of arbitrary wavevectors can be represented in the chemical unit cell by utilising the generalized Bloch theorem that encodes the spiral in the boundary conditions \cite{Sandratskii_1986, Sandratskii1991}. This method has been applied in conjunction with density functional theory (DFT) to a wide range of materials and typically produces results that are in good agreement with experiments \cite{Bylander1998,Kurz2001,Sandratskii2007,Zimmermann2019,Gutzeit2022}.

In the present work we use DFT simulations in the framework of the generalized Bloch theorem to investigate the magnetic ground state of monolayers derived from layered van der Waals magnets. We then calculate the preferred orientation of the spiral plane by adding a single component of the spinorbit coupling in the normal direction of various trial spiral planes. This yields a complete classification of the magnetic ground state for these materials under the assumption that higher order spin interactions can be neglected. On the other hand, the effect of higher order spin interactions can be quantified by deviations between spin spiral energies in the primitive unit cell and a minimal super cell. The results for all compounds are discussed and compared with existing knowledge from experiments on the parent bulk materials. Finally, we analyse the spontaneous polarization in all cases where an incommensurate ordering vector is predicted.

The paper is organised as follows. In Sec. \ref{sec:theory} we summarise the theory used to obtain spin spiral ground states based on the generalized Bloch theorem and briefly outline the implementation. In Sec. \ref{sec:results} we present the results and summarise the magnetic ground states of all the investigated materials. Sec. \ref{sec:conclusion} provides a conclusion and outlook.

%\begin{itemize}
%    \item The \cite{mcguire2017crystal} problem \cite{sandratskii2017insight} is predicting plausible magnetic ground state is a difficult problem, since the periodicity can vary arbitrarily far away from that of the crystal. In the the collinear approximation, one can use magnetic space groups to simplify and enumerate plausible ground states based on symmetry \cite{horton2019high}.
%    \item In non-collinear calculations such symmetry analysis cannot be done because each atom has its own quantisation axis. This is particular important on anti-ferromagnetic materials such as the kagome and triangular lattices where a collinear antiferromagnetic state leads to magnetic frustration.
%    \item In this paper, we approach this problem by implementing and benchmarking the Generalized Bloch's Theorem (GBT) on magnetic 2D materials. We build upon the density functional theory code GPAW, using plane-wave basis functions in the projector augmented-wave (PAW) formalism.
%\end{itemize}
\begin{figure*}[tb]
    \centering
    \subfloat[]{\parbox[b][3cm][t]{.8\textwidth}{\includegraphics[page=37, width=0.8\textwidth]{Tikz_figures.pdf}}}\hfill
    \subfloat[]{\parbox[b][3cm][t]{.15\textwidth}{\includegraphics[width=0.15\textwidth]{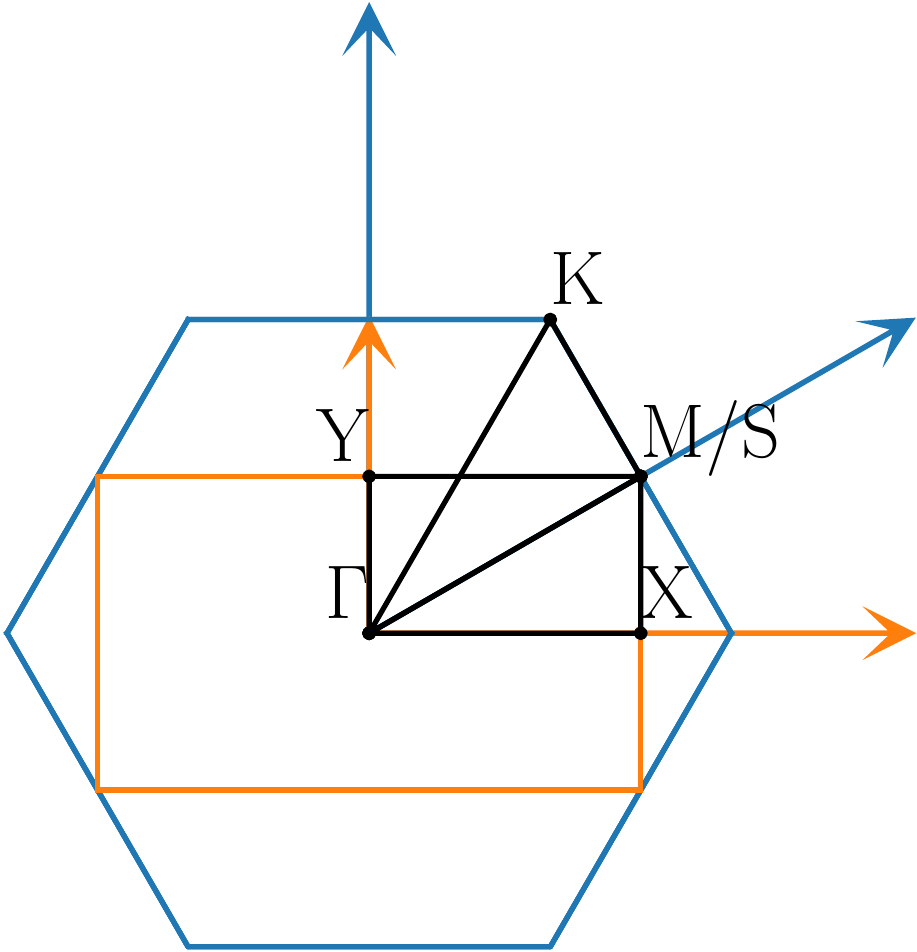}}}
    \caption{(a) Examples of magnetic structures in the triangular lattice. The $\mathbf{Q}=(1/3,1/3)$ (corresponding to the high symmetry point K) is the classical ground state in the isotropic Heisenberg model with nearest neighbour antiferromagnetic exchange and is degenerate with $\mathbf{Q}=(-1/3,-1/3)$. The stripy antiferromagnetic $\mathbf{Q}=(1/2,0)$ (corresponding to the high symmetry point M) is only found for CoI$_2$ in the present study and is degenerate with $\mathbf{Q}=(0,1/2)$ and $\mathbf{Q}=(1/2,1/2)$. The incommensurate spiral with $\mathbf{Q}=(0.14, 0.14)$ corresponds to the prediction of NiI$_2$ in the present work. The rectangular cell with $\mathbf{Q}=(0,1/2)$ is a bicollinear antiferromagnet that corresponds to superpositions of (0, $\pm$1/4) states in the primitive cell. (b) Brillouin zone of the hexagonal (blue) and rectangular (orange) unit cell. The high symmetry band paths used to sample the spiral ordering vectors are shown in black.}
    \label{fig:spirals}
\end{figure*}
\section{Theory}\label{sec:theory}
\subsection{Generalized Bloch's Theorem}
The Heisenberg model plays a prominent role in the theory of magnetism and typically gives an accurate account of the fundamental magnetic excitations as well as the thermodynamic properties of a given material. In the isotropic case it can be written as
\begin{align}\label{eq:heisenberg}
    H=-\frac{1}{2}\sum_{ij}J_{ij}\mathbf{S}_i\cdot\mathbf{S}_j,
\end{align}
where $\mathbf{S}_i$ is the spin operator for site $i$ and $J_{ij}$ is the exchange coupling between sites $i$ and $j$. In a classical treatment, the spin operators are replaced by vectors of fixed magnitude and it can be shown that the classical energy is minimised by a planar spin spiral \cite{Kaplan1959}. Such a spin configuration is characterised by a wave vector $\mathbf{Q}$, which is determined by the set of exchange parameters $J_{ij}$. The spin at site $i$ is rotated by an angle $\mathbf{Q}\cdot\mathbf{R}_i$ with respect to the origin and the wave vector may or may not be commensurate with the lattice.

In a first principles framework it is thus natural to search for planar spin spiral ground states that give rise to periodically modulated magnetisation densities satisfying
\begin{align}\label{eq:mag}
    \mathbf{m}_\mathbf{q}(\mathbf{r}+\mathbf{R}_i)=\mathrm{U}_{\mathbf{q},\mathbf{R}_i}\mathbf{m}_\mathbf{q}(\mathbf{r}).
\end{align}
Here $\mathbf{R}_i$ is a lattice vector (of the chemical unit cell) and $\mathrm{U}_{\mathbf{q},\mathbf{R_i}}$ is a rotation matrix that rotates the magnetisation by an angle $\mathbf{q}\cdot\mathbf{R}_i$ around the normal of the spiral plane. In the absence of spinorbit coupling we are free to perform a global rotation of the magnetisation density and we will fix the spiral plane to the $xy$-plane from hereon. In the framework of DFT, the magnetisation density \eqref{eq:mag} gives rise to an exchange-correlation magnetic field satisfying the same symmetry under translation. If spinorbit coupling is neglected the Kohn-Sham Hamiltonian thus commutes with the combined action of translation (by a lattice vector) and a rotation of spinors by the angle $\mathbf{q}\cdot\mathbf{R}_i$. % The existence of a generalized Bloch's theorem can be seen from the fact that when neglecting spin-orbit coupling, the magnetic Schrödinger equation is invariant with respect to arbitrary spin rotations.
%This symmetry allows us to construct a spin rotation operator which commutes with the Hamiltonian.
%Then in a periodic crystal, the independent commutation of translations and spin rotations imply that we can construct a combined product symmetry operation which also commutes with the Hamiltonian.
%Now in order to create a non-trivial spin state, one has to create a \textbf{r}-dependent spin rotation operator.
%In constructing this operator, the diagonal of the density matrix remains unchanged by translation invariance of the density, we thus see the $m_z$ component is also constant under the operator.
%It is therefore natural that the representation of the spin rotation operator should be $R_{\mathbf{q}}(\mathbf{r})=\exp(-i\sigma_z\varphi/2)$ where $\varphi = \mathbf{q}\cdot\mathbf{r}$.
%Thus, we construct a flat single fourier component spin spiral in the x-y plane with wavelength $\lambda = 1/|\mathbf{q}|$, propagating in the direction of $\mathbf{q}$.
%The generalized Bloch's theorem can then be states as
This implies that the Kohn-Sham eigenstates can be written as
\begin{equation}\label{eq:GBT}
\psi_{\mathbf{q},\mathbf{k}}(\mathbf{r})=e^{i\mathbf{k}\cdot\mathbf{r}} U_\mathbf{q}^\dag(\mathbf{r})
\begin{pmatrix}
u^{\uparrow}_{\mathbf{q},\mathbf{k}}(\mathbf{r})\\
u^{\downarrow}_{\mathbf{q},\mathbf{k}}(\mathbf{r})
\end{pmatrix}
\end{equation}
where $u^{\uparrow}_{\mathbf{q},\mathbf{k}}(\mathbf{r})$ and $u^{\downarrow}_{\mathbf{q},\mathbf{k}}(\mathbf{r})$ are periodic in the chemical unit cell and the spin rotation matrix is given by
\begin{align}\label{eq:U}
    U_\mathbf{q}(\mathbf{r})=
\begin{pmatrix}
e^{i\mathbf{q}\cdot\mathbf{r}/2} & 0\\
0 & e^{-i\mathbf{q}\cdot\mathbf{r}/2}
\end{pmatrix}
\end{align}
This is known as the generalized Bloch Theorem (GBT) and the Kohn-Sham equations can then be written as
\begin{align}
    H_\mathbf{q,k}^\mathrm{KS}u_\mathbf{q,k}=\varepsilon_\mathbf{q,k}u_\mathbf{q,k}
\end{align}
where the generalized Bloch Hamiltonian:
\begin{align}
    H_\mathbf{q,k}^\mathrm{KS}&=e^{-i\mathbf{k}\cdot\mathbf{r}}
U_\mathbf{q}(\mathbf{r})H^\mathrm{KS}U_\mathbf{q}^\dag(\mathbf{r})e^{i\mathbf{k}\cdot\mathbf{r}}
\end{align}
is periodic in the unit cell. Here $\mathbf{k}$ is the crystal momentum, $\mathbf{q}$ is the spiral wave vector and $H^\mathrm{KS}$ is the Kohn-Sham Hamiltonian, which couples to the spin degrees of freedom through the exchange-correlation magnetic field.% $\mathbf{B}_\mathrm{xc}[\mathbf{m}_\mathbf{q}]$.

In the present work, we will not consider constraints besides the boundary conditions defined by Eq. \eqref{eq:mag}. For a given $\mathbf{q}$ we can thus obtain a unique total energy $E_\mathbf{q}$ and the magnetic ordering vector is determined as the point where $E_\mathbf{q}$ has a minimum (denoted by $\mathbf{Q}$) when evaluated over the entire Brillouin zone. However, if the chemical unit cell contains more than one magnetic atom there may be different local extrema corresponding to different intracell alignments of magnetic moments. In order ensure that the correct ground state is obtained it is thus pertinent to perform a comparison between calculations that are initialised with different relative magnetic moments. As a simple example of this, one may consider a honeycomb lattice of magnetic atoms where the ferromagnetic and anti-ferromagnetic configurations both correspond to $\mathbf{q=0}$, but are distinguished by different intracell orderings of the local magnetic moments. We will discuss this in the context of CrI$_3$ in section \ref{sec:AB3}.%In general, the energy of arbitrary spin spirals are described by $E(\mathbf{q}, \theta^\alpha, \xi^\alpha)$, where $\theta^\alpha$ and $\xi^\alpha$ describe the atomic dependent out-of-plane rotation and intra-cell relative phase, respectively. 
%These parameters are not explicit in the GBT, and so are not constrained parameters in the self-consistent search for the energy minimum.
%They are still important to consider as implicit parameters of the magnetic moments initial condition.
%Consider if the magnetisation is initialised with $m_z = 0$ thereby choosing $\theta^\alpha = \frac{\pi}{2}$ then by construction in local spin density approximation (LSDA) $B_z = 0$, thus only flat spin spiral may be found.
%In a similar manner, choosing $\theta^\alpha = 0$ the GBT boundary conditions do nothing and only collinear states may be found. 
%However, if the initial magnetic state is in between these scenarios then it may converge into either flat or conic spin spirals with appropriate $\mathbf{q}=\mathbf{q}_c$ or a $\mathbf{q}=0$ collinear state.
%Furthermore, we consider only collinear intra-cell phases $\xi^\alpha \in \{0, \pi\}$ if the chemical unit cell of the particular material contains .
%XXX Question: if I create a supercell with 3 magnetic moments inside, and initiallize FM with spinspiral K, does density converge to K, answer no in VCl2, but it does become non-collinear

%These functions retain rotational freedom to converge to any non-collinear structure inside the chemical cell.
%If a material is sensitive to the intial conditions of the calculation, it might not be possible to complete full Brillouin zone scans in the spiral parameter q.

We also note that the true magnetic ground state is not necessarily representable by the ansatz \eqref{eq:mag} and one is therefore not guaranteed to find the ground state by searching for spin spirals based on the minimal unit cell. In figure \ref{fig:spirals} we show four examples of possible magnetic ground states of the triangular lattice. Three of these correspond to spin spirals of the minimal unit cell while the fourth - a bicollinear antiferromagnet - requires a larger unit cell. The bicollinear state may arise as a consequence of higher order exchange interactions, which tend to stabilize linear combinations of degenerate single-$q$ states.% The magnetic structures which we can calculate is clearly limited by the single fourier component we can specify by $\varphi = \mathbf{q}\cdot\mathbf{r}$. It is therefore worthwhile to consider the space of q-representable magnetic structures. Take for example a cubic unit cell with one magnetic atom, we can represent the antiferromagnetic types A, C and G with this unit cell and $q_A=(0, 0, 0.5)$, $q_C=(0.5, 0.5, 0)$ and $q_G=(0.5, 0.5, 0.5)$ respectively. However in order to create bicollinear type E antiferromagnetic order, we need to construct a 2x1 supercell and choose relative phases $\xi^0 = 0, \xi^1 = \pi$. If the primitive unit cell had contained two in-equivalent magnetic moments, such an arbitrary enlargement would not be necessary, but still the complexity in $\xi^\alpha$ must be dealt with.

\subsection{Spinorbit coupling}\label{sec:soc}
In the presence of spinorbit coupling, the spin spiral plane will have a preferred orientation and the magnetic ground state is thus characterised by a normal vector $\mathbf{\hat n}_0$ of the spiral plane as well as the spiral vector $\mathbf{Q}$. Spinorbit coupling is, however, incompatible with application of the GBT and has to be approximated in a post processing step when working with the spin spiral representation in the chemical unit cell. It can be shown that first order perturbation theory only involves contributions from the spinorbit components orthogonal to the plane \cite{Heide2009}
\begin{align}
    \langle\psi_{\mathbf{q},\mathbf{\hat n}}|\mathbf{L}\cdot\mathbf{S}|\psi_{\mathbf{q},\mathbf{\hat n}}\rangle=\langle\psi_{\mathbf{q},\mathbf{\hat n}}|(\mathbf{L}\cdot \mathbf{\hat n})(\mathbf{S}\cdot\mathbf{\hat n})|\psi_{\mathbf{q},\mathbf{\hat n}}\rangle,
\end{align}
and this term is thus expected to yield the most important contribution to the spinorbit coupling. Since $(\mathbf{L}\cdot \mathbf{\hat n})(\mathbf{S}\cdot\mathbf{\hat n})$ commutes with a spin rotation around the axis $\mathbf{\hat n}$, the spin spiral wavefunctions remain eigenstates when such a term is included in $H^\mathrm{KS}$. This approach was proposed by Sandratskii \cite{sandratskii2017insight} and we will refer to it as the projected spinorbit coupling (PSO). For the spin spiral calculations in the present work we include spinorbit coupling non-selfconsistently by performing a full diagonalization of the $H^\mathrm{KS}_{\mathbf{q},\mathbf{k}}$ including the PSO. The magnetic ground state is then found by evaluating the total energy at all normal vectors $\mathbf{\hat n}$, which will yield $\mathbf{\hat n}_0$ as the normal vector that minimizes the energy.% For the case of $\mathbf{q}=\mathbf{0}$ this procedure simply corresponds to finding the magnetic easy axis.

\subsection{Computational Details}
The GBT has been implemented in the electronic structure software package GPAW \cite{Enkovaara2010}, which is based on the projector augmented wave method (PAW) and plane waves. The implementation uses a fully non-collinear treatment within the local spin density approximation where both the interstitial and atom-centered PAW regions are handled non-collinearly. Spinorbit coupling is included non-selfconsistently \cite{Olsen2016a} as described in Section \ref{sec:soc}. The implementation is described in detail in Appendix \ref{sec:implementation} and benchmarked for fcc Fe in Appendix \ref{sec:benchmark}. We find good agreement with previous results from the literature and we also assert that results from spin spiral calculations within the GBT agree exactly with supercell calculations without spinorbit in the case of bilayer CoPt. Finally, we compare the results of the PSO approximations with full inclusion of spinorbit coupling for both supercells and GBT spin spirals of the CoPt bilayer. We find exact agreement between the PSO in the supercell and GBT spin spiral and the approximation only deviates slightly compared to full spinorbit coupling for the supercell calculations.

All calculations have been carried out with a plane wave cutoff of 800 eV, a $k$-point density of 14 {\AA} and a Fermi smearing of 0.1 eV. The structures and initial magnetic moments are taken from the Computational Materials Database (C2DB) \cite{Haastrup2018,Gjerding2021}.% and for the cases where two magnetic atoms are present in the chemical unit cell we have carried out the calculation for both ferromagnetic and antiferromagnetic alignment of the intracell moments for all wave vectors $\mathbf{q}$. 
 In order to find the value of $\mathbf{Q}$, which describes the ground state magnetic order, we calculate $E_\mathbf{q}$ along a representative path connecting high symmetry points in the Brillouin zone. While the true value of $\mathbf{Q}$ could be situated away from such high symmetry lines we deem this approach sufficient for the present study. %The workflow is able to consistently converge the magnetic structure for various different GBT boundaries.
\begin{table*}[tb]
\begin{tabular}{l|l|l|l|l|l|l|l|l}
      & $\mathbf{Q}$ & $\mathrm{E}_{min}$ [meV] & $(\theta, \varphi)$ & Exp. IP order      & BW [meV] & PSO BW [meV] & $m_\Gamma$ $[\mu_\mathrm{B}]$ & $\Delta\varepsilon_\mathbf{Q}$ [eV] \\ \hline
TiBr$_2$ & (1/3, 1/3)   & -78.12                   & (90,90)           & -                  & 78.1     & 0.6          & 1.5        & 0.0                                 \\
TiI$_2$  & (1/3, 1/3)   & -44.33                   & (90,90)           & -                  & 44.3     & 1.0          & 1.9        & 0.0                                 \\
NiCl$_2$ & (0.06, 0.06) & -0.81                    & (90,31)           & FM $\parallel$     & 45.2     & 0.0          & 2.0        & 0.81                                \\
NiBr$_2$ & (0.11, 0.11) & -8.62                    & (44,0)           & FM $\parallel$, HM & 50.7     & 0.3          & 2.0        & 0.62                                \\
NiI$_2$  & (0.14, 0.14) & -28.48                   & (64,0)           & HM                 & 68.3     & 4.1          & 1.8        & 0.28                                \\
VCl$_2$  & (1/3, 1/3)   & -60.07                   & (90,0)            & $120\degree$       & 60.1     & 0.1          & 3.0        & 0.96                                \\
VBr$_2$  & (1/3, 1/3)   & -36.21                   & (90,18)           & $120\degree$       & 36.2     & 0.1          & 3.0        & 0.9                                 \\
VI$_2$   & (0.14, 0.14) & -4.43                    & (6,0)           & stripe                  & 9.8      & 0.7          & 3.0        & 0.96                             \\  
MnCl$_2$ & (1/3, 1/3)   & -20.48                   & (90,15)           & stripe or HM & 20.5     & 0.0          & 5.0        & 1.92                                \\
MnBr$_2$ & (1/3, 1/3)   & -20.13                   & (90,15)           & stripe $\parallel$       & 20.1     & 0.1          & 5.0        & 1.76                                \\
MnI$_2$  & (1/3, 1/3)   & -21.32                   & (0,0)           & HM                 & 21.3     & 1.1          & 5.0        & 1.41                                \\
FeCl$_2$ & (0, 0)       & 0.0                      & (0, 0)$^*$            & FM $\perp$         & 115.2    & 0.5$^*$            & 4.0        & 0.0                                 \\
FeBr$_2$ & (0, 0)       & 0.0                      & (0, 0)$^*$            & FM $\perp$         & 81.3     & 0.8$^*$            & 4.0        & 0.0                                 \\
FeI$_2$  & (0, 0)       & 0.0                      & (0, 0)$^*$            & stripe $\perp$     & 36.5     & 1.9$^*$            & 4.0        & 0.0                                 \\
CoCl$_2$ & (0, 0)       & 0.0                      & (90,90)$^*$            & FM $\parallel$     & 46.0     & 1.2$^*$            & 3.0        & 0.0                                 \\
CoBr$_2$ & (0.03, 0.03) & -0.04                    & (0,0)           & FM $\parallel$     & 21.2     & 0.1          & 3.0        & 0.0                                 \\
CoI$_2$  & (1/2, 0)     & -20.95                   & (90,90)           & HM                 & 41.7     & 5.6          & 1.2        & 0.0                            
\end{tabular}
\caption{Summary of magnetic properties of the AB$_2$ compounds. The ground state ordering vector is denoted by $\mathbf{Q}$ and $E_\mathrm{min}$ is the ground state energy relative to the ferromagnetic state. The normal vector of the spiral plane is defined by the angles $\theta$ and $\varphi$ (see text). We also display the experimental in-plane order of the parent layered compound (Exp. IP order). In addition we state the spin spiral band width BW, the magnetic moment per unit cell in the ferromagnetic state $m_\Gamma$ and the band gap at the ordering vector $\Delta\varepsilon_\mathbf{Q}$. For the case of NiI$_2$, $m_\Gamma$ deviates from an integer value because the ferromagnetic state is metallic in LDA (whereas the spin spiral ground state has a gap). The cases of FeX$_2$, CoCl$_2$ and CoBr$_2$ are half metals, which enforces integer magnetic moment despite the metallic ground state. The asterisks indicate ferromagnets where full spinorbit coupling was included and the angles then refer to the direction of the spins rather that the spiral plane normal vector.}
\label{tab:AB2Result}
\end{table*}
\section{results}\label{sec:results}
A comprehensive review on the magnetic properties of layered transition metal halides was provided in Ref. \cite{mcguire2017crystal}. Here we present spin spiral calculations and extract the magnetic properties of the corresponding monolayers. In addition to the magnetic moments, the properties are mainly characterised by a spiral ordering vector $\mathbf{Q}$ and the normal vector to the spin spiral plane $\mathbf{\hat n}_0$. The materials either have AB$_2$ or AB$_3$ stoichiometries and we will discuss these cases separately below. 

We have performed LDA and LDA+U calculations for all materials. In most cases, the Hubbard corrections does not make any qualitative difference although the spiral ordering vector does change slightly and we will not discuss these calculations further here. The Mn halides comprise an exception to this where LDA+U calculations differ significantly from those of bare LDA and the LDA+U calculations will be discussed separately for these materials below.

For the AB$_2$ materials, we find 12 that exhibit a spiral order that breaks the crystal symmetry and yields a ferroelectric ground state. For six of these compounds we have calculated the spontaneous polarization by performing full relaxation (including self-consistent spinorbit coupling) in supercells hosting the spiral order.

\subsection{Magnetic ground state of AB$_2$ materials}
The AB$_2$ materials all have space group $P\bar3m1$ corresponding to monolayers of the CdI$_2$ (or CdCl$_2$) prototype. The magnetic lattice is triangular and a few representative possibilities for the magnetic order is illustrated in figure \ref{fig:spirals}. The magnetic properties of all the considered compounds are summarized in table \ref{tab:AB2Result}. In addition to the ordering vector $\mathbf{Q}$ we provide the angles $\theta$ and $\phi$, which are the polar and azimuthal angles of $\mathbf{\hat n}_0$ with respect to the out-of-plane direction and the ordering vector respectively. It will be convenient to consider three limiting cases of the orientation of spin spiral planes: the proper screw ($\theta=90, \varphi=0$), the out-of-plane cycloid ($\theta=90, \varphi=90$) and the in-plane cycloid ($\theta=0, \varphi=0$). We also provide the ground state energy relative to  the ferromagnetic configuration ($\mathbf{Q}=(0,0)$), the band gap, the spin spiral band width, which reflects the strength of the magnetic interactions and the PSO band width, which is the energy difference between the easy and hard orientations of the spiral plane. The magnetic moments are calculated as the total moment in the unit cell using the ferromagnetic configurations without spinorbit interaction and thus yields an integer number of Bohr magnetons for insulators. The magnitude of the local magnetic moments (obtained by integrating the magnetization density over the PAW spheres) in the ground state are generally found to be very close to the moments in the ferromagnetic configuration, unless explicitly mentioned. The spin spiral energy dispersions are provided for all AB$_2$ materials in the supporting information. The different classes of materials are described in detail below.
%The spin spiral calculations has been included into a high throughput workflow which has been tested on 2D monolayers of experimentally relevant vdW transition metal halides \cite{mcguire2017crystal}.
%The resulting structures can be found in tables \ref{tab:AB2Result} and XXX(AB3).
%The \ce{AB2} monolayer has triangular symmetry, and is often found to be anti-ferromagnetic. It is therefore expected that the groundstate tend to non-collinear order due to geometric frustration. In general, we find good agreement with the experimental in-plane magnetic orders. We find all three commensurate orders $q_\Gamma$, $q_M$ and $q_K$ present. Several incommensurate helimagnets with long and moderate wavelengths are distinct in the sense that small-q helimagnets have FM NN exchange coupling, while the latter are NN AFM.

\begin{figure*}[tb]
\subfloat[]{\includegraphics[page=39, width=0.35\textwidth]{Tikz_figures.pdf}}
\subfloat[]{\includegraphics[width=0.333\textwidth]{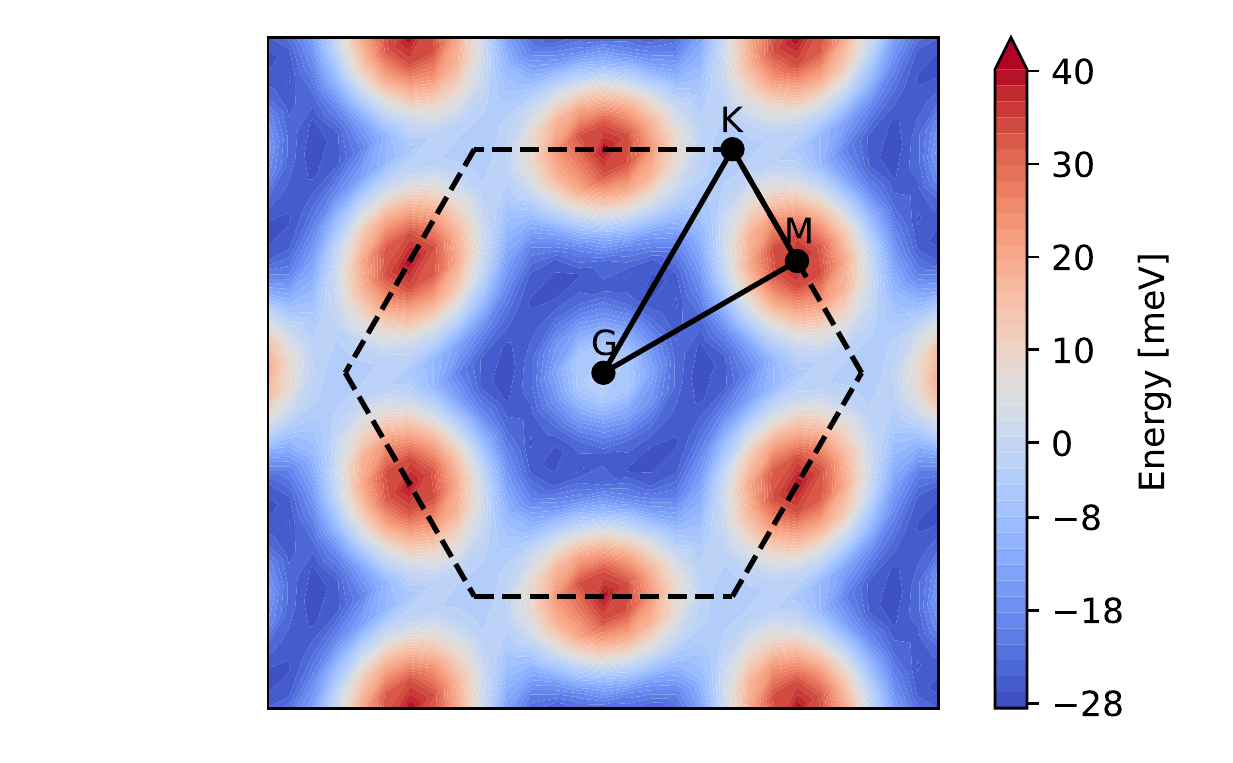}}
\subfloat[]{\includegraphics[page=38, width=0.35\textwidth]{Tikz_figures.pdf}}
% note black line required for vertical stack
\caption{Spin spiral energy of NiI$_2$. Left: the spin spiral energy as a function of $\mathbf{q}$ without spinorbit coupling. Center: Spin spiral energy in evaluated in entire Brillouin zone. Right: spiral energy as a function of spiral plane orientation evaluated at the minimum $\mathbf{Q}=(0.14, 0.14)$. The spiral plane orientation is parameterized in terms of the polar angle $\theta$ and azimuthal angle $\varphi$ (measured from $\mathbf{Q}$) of the spiral plane normal vector.}
\label{fig:NiI2}
\end{figure*}

\subsubsection*{NiX$_2$}
The nickel halides all have ground states with incommensurate spiral vectors between $\Gamma$ and K. Experimentally, both NiI$_2$ and NiBr$_2$ in bulk form have been determined to have incommensurate spiral vectors \cite{ADAM19801, day1980incommensurate, KUINDERSMA1981231} in qualitative agreement with the LDA results. The case of NiCl$_2$, however, have been found to have ferromagnetic intra-layer order whereas we find a rather small spiral vector of $\mathbf{Q}=(0.06, 0.06)$. 

In bulk NiI$_2$ the experimental ordering vector $\mathbf{Q}_\mathrm{exp} = (0.1384, 0, 1.457)$ has an in-plane component in the $\Gamma\mathrm{M}$-direction with a magnitude of roughly $1/7$ of a reciprocal lattice vector, while for the monolayer we find $\mathbf{Q} = (0.14, 0.14, 0)$, which is in the $\Gamma\mathrm{K}$-direction. Evaluating the spin spiral energy in the entire Brillouin zone, however, reveals a nearly degenerate ring encircling the $\Gamma$-point with a radius of roughly 1/5 of a reciprocal lattice vector. The point $\mathbf{q}_\mathrm{M} = (0.21, 0)$ thus comprises a very shallow saddle point with an energy that exceeds the minimum by merely 2 meV. This is illustrated in figure \ref{fig:NiI2}. We also show a scan of the spin spiral energy (within the PSO approximation) as a function of orientation of the spin spiral plane on a path that connects the limiting cases of in-plane cycloid, out-of-plane cycloid and proper screw. An unconstrained spin spiral calculation using the rectangular unit cell of figure \ref{fig:spirals} does not reveal any new minima in the energy, which implies that the ground state is well represented by a single-$q$ spiral and that higher order exchange interactions are neglectable in NiI$_2$. The normal vector of the spiral makes an angle of 64{\degree} with the out-of-plane direction. This orientation is in good agreement with the experimental assignment of a proper screw (along $\mathbf{Q}_\mathrm{exp} = (0.1384, 0, 1.457)$), which corresponds to a tilt of 55\degree$\pm$10{\degree} with respect to the $c$-axis \cite{KUINDERSMA1981231}, but disagrees with the model proposed in Ref. \cite{Song2022} where the spiral was found to be a proper screw.% with respect to a propagation vector in the plane.% at $\mathbf{q}_M = (0.21, 0, 0)$. %Thus we see in LDA tends to favour the perpendicular spin structure here, being close to the maximally perpendicular order $\mathbf{q}=M/2$ and $\mathbf{q}=K/2$. In Fig. \ref{fig:NiI2} we see in the full brilluoin zone of spin spirals that these two minima are connected by a one dimensional path forming a nearly degenerate subspace. Additionally, the energy landscape has the expected six-fold symmetry without spin orbit interactions, which when included lowers the symmetry to threefold. However, applying the PSO approximation to the non-orthogonal space of spin spirals is ambiguous because 

%In \ce{NiI2} the incommensurate spin spiral order couples to polarization of the lattice through higher order exchange interactions \cite{katsura2005spin}. The resulting polarization is given by $\mathbf{P}\propto \mathbf{e}_{ij}\cross(\mathbf{S}_i\cross\mathbf{S}_j)$. It is often assumed that the spiral order of the bulk phase extends naturally to the monolayer order by setting $q_z = 0$, however this is important to verify because a proper screw ground state the $\mathbf{q}$ is parallel to the spin plane normal axis which yield $\mathbf{P}=0$. The different spin plane rotations is evaluated non-self consistently using the PSO for the minimum energy spin spiral. The is exemplified for NiI$_2$ in Fig. \ref{fig:NiI2}, where the ground state orientation is found at $\theta, \phi = (64\degree, 0 \degree)$, which agrees with experiments within experimental uncertainty.

At low temperatures NiBr$_2$ has been reported to exhibit $\mathbf{Q}_\mathrm{exp}=(x, x, 3/2)$ where x changes continuously from 0.027 at 4.2 K to 0.09 at 22.8 K and then undergoes first order transition at 24 K to intra-layer ferromagnetic order \cite{adam1980neutron}. The structure predicted here is close to the one observed in bulk at 22.8 K. The discrepancy could be due to the magnetoelastic deformation \cite{tokunaga2011multiferroicity} that has been associated with the modulation of the spiral vector. This effect could in principle be captured by relaxing the structure in supercell calculations, but the small wavelength spirals require prohibitively large supercells and are not easily captured by first principles methods. It is also highly likely that LDA is simply not accurate enough to describe the intricate exchange interactions that define the true ground state in this material.

Bulk NiCl$_2$ is known to be an inter-layer antiferromagnet with ferromagnetically ordered layers \cite{pollard1982electronic}. We find the ground state to be a long wavelength incommensurate spin spiral with $\mathbf{Q}=(0.06, 0.06)$, which is in rather close proximity to ferromagnetic order. The ground state energy is less than 1 meV lower than the ferromagnetic state, but we cannot say at present whether this is due to inaccuracies of LDA or if the true ground state indeed exhibits spiral magnetic order in the monolayer limit.%   while in the monolayer limit all interlayer interactions disappear and so we find it to no longer exhibit clear intralayer ferromagnetism but rather a long wavelength spin spiral, albeit with energies less than 1 meV compared to the ferromagnetic state. 
%The spin plane is in the \textit{ab} plane

%Several factors are important when evaluating spin spiral in a high-throughput schemes. Due to the intracell non-collinear freedom we have to deal with badly- or non-converged results for every requested $q$. In valid converged results we assert that the converged intracell magnetic moments are sufficiently close to the initiated magnetic moments. The magnitude is allowed to vary as long as they remain finite on magnetic sites. The projected magnetic moments should remain its sign, so that we can distinguish between intracell FM and AFM order, which thus define the higher optical branches. 
%\begin{figure*}[tb]
%\subfloat[]{\includegraphics[page=39, width=\columnwidth]{Tikz_figures.pdf}}
%\subfloat[]{\includegraphics[page=38, width=\columnwidth]{Tikz_figures.pdf}}
% note black line required for vertical stack

%\subfloat[]{\includegraphics[width=\columnwidth]{NiI2_fullbz_nosoc.png}}
%\caption{NiI$_2$ spin spiral figures}
%\label{fig:NiI2}
%\end{figure*}

\subsubsection*{VX$_2$}
The three vanadium halides are insulators and whereas VCl$_2$ and VBr$_2$ are found to form $\mathbf{Q} = (1/3,1/3)$ spiral structures, VI$_2$ has an incommensurate ground state with $\mathbf{Q} = (0.14,0.14)$. The magnetic ground state of VCl$_2$ and VBr$_2$ is in good agreement with experiments on bulk materials where both have been found to exhibit out-of-plane 120{\degree} order \cite{kadowaki1985neutron}. This structure is expected to arise from strong nearest neighbour anti-ferromagnetic interactions between the V atoms. The case of VI$_2$ has a significantly smaller spiral band width, signalling weaker exchange interactions compared to VCl$_2$ and VBr$_2$. A collinear energy mapping based on the Perdew-Burke-Ernzerhof (PBE) exchange-correlation functional \cite{Gjerding2021} yields a weakly ferromagnetic nearest neighbour interaction for VI$_2$ and strong anti-ferromagnetic interactions for VCl$_2$ and VBr$_2$. This is in agreement with the present result, which indicate that the magnetic order of VI$_2$ is not dominated by nearest neighbour interactions. 

\begin{figure}[tb]
    \centering
    \includegraphics[width=0.5\textwidth]{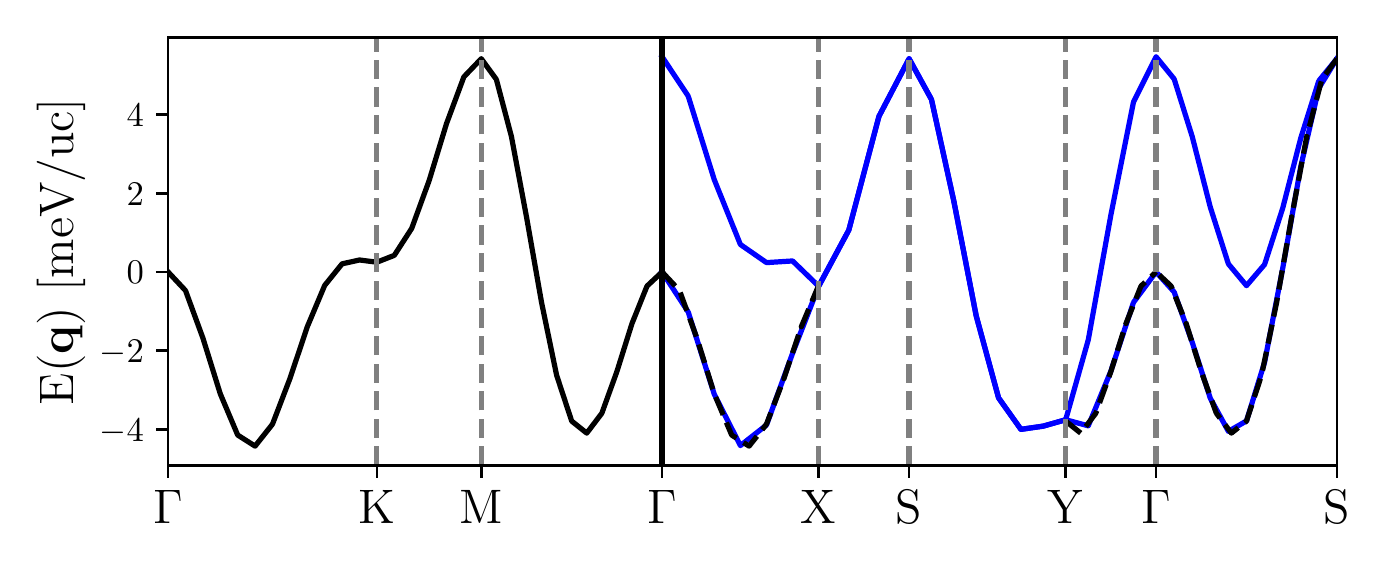}
    \caption{Spin spiral energies of VI$_2$ obtained from the primitive cell (black) and the rectangular super cell (blue). The dashed lines repeat the primitive cell results on the corresponding super cell path.}
    \label{fig:VI2}
\end{figure}

Experimentally \cite{kuindersma1979magnetic}, the bulk VI$_2$ magnetic order has been found to undergo a phase transition at 14.4 K from a 120{\degree} state to a bicollinear state with $\mathbf{Q} = (1/2, 0)$, where the spins are perpendicular to $\mathbf{Q}$ and tilted by $29\degree$ from the z-axis. Such a bicollinear state implies that the true ground state is a double-$q$ state stabilized by higher order spin interactions and cannot be represented as a spin spiral in the primitive unit cell. To check whether LDA predicts the experimental ground state we have therefore performed spiral calculations in the rectangular cell shown in figure \ref{fig:spirals}. The result is shown in figure \ref{fig:VI2} along with the spiral calculation in the primitive cell and we do not find any new minima in the super cell calculation. We have initalized angles in the super cell caluculation such that they corresponds to bicollinear order and the angles are observed to relax to the single-$q$ spin spiral of the primitive cell. It is likely that LDA is insufficient to capture the subtle higher order exchange interactions in this material, but it is possible that the monolayer simply has a magnetic order that differs from the individual layers in the bulk material. 
%The Vanadium halides have been found to contain strong antiferromagnetic interactions, which leads to geometric frustration already for nearest neighbour interactions. As a result, bulk VBr$_2$ and VCl$_2$ have been experimentally determined to exhibit a non-collinear 120\degree magnetic structure. This corresponds to a spin spiral with $\mathbf{Q} = (1/3,/1/3)$ as shown in figure \ref{fig:spirals}. Our calculations predict that this magnetic structure is retained in the monolayer limit for all the halides.
%In the PSO approximation we find that the three vanadium halides prefer out-of-plane spiral planes. The energy, however, is rather insensitive to the orientation of the spiral and the subspace of in-plane normal vectors form a nearly degenerate subspace of ground states with a slight preference of proper screws for VCl$_2$ and VBr$_2$. The ground state of VI$_2$ is found to be close to the out-of-plane cycloid with a normal vector to the spiral plane forming a 6\degree angle with $\mathbf{Q}$.

In the PSO approximation we find that VCl$_2$ and VBr$_2$ prefer out-of-plane spiral planes. The energy is rather insensitive to $\varphi$ forming a nearly degenerate subspace of ground states with a slight preference of the proper screw. The ground state of VI$_2$ is found to be close to the in-plane cycloid with a normal vector to the spiral plane forming a 6{\degree} angle with $\mathbf{Q}$.
The spinorbit corrections in VI$_2$ are also found to be the smallest compared to other iodine based transition metal halides studied here and the ground state energy only deviates by $0.7$ meV per unit cell from the out-of-plane cycloid, which constitutes the orientation of the spin plane with highest energy.

\subsubsection*{MnX$_2$}
The manganese halides are all found to form 120{\degree} ground states, which is in agreement with previous theoretical studies \cite{li2020high} using PBE. In contrast to the other insulators studied in the present work, however, we find that the results are qualitatively sensitive to the inclusion of Hubbard corrections. This was also found in Ref. \cite{torelli2019high}, where the sign of the nearest neighbour exchange coupling was shown to change sign when a Hubbard U parameter was included in the calculations. With U = 3.8 eV we find that all three compounds has spiral ground states with incommensurate spiral vector $\mathbf{Q} = (0.11, 0.11, 0)$. Moreover, spin spiral band width in the LDA+U calculations decrease by more than an order of magnitude compared to the bare LDA calculations. 

The experimental magnetic structure of the manganese halides are rather complicated, exhibiting several magnetic phase transitions in a range of 0.1 K below the initial ordering temperature. In particular MnI$_2$ (MnBr$_2$) has been found to have three (two) complex non-collinear phases \cite{SATO1995224}, and MnCl$_2$ has two complex phases that are possibly collinear \cite{wilkinson1958neutron}. % In particular the energies of the ferromagnetic becomes lower than the nearest neighbour antiferromagnets, this has been attributed to changes in the nearest neighbour heisenberg exchange interaction \cite{torelli2019high}. 
%In pristine bulk MnCl$_2$ and MnBr$_2$ has been experimentally found in the bicollinear \textit{uudd} order, which corresponds to $\mathbf{q}=Y$ shown in Fig \ref{fig:spirals}. We find for both MnCl$_2$ in Fig \ref{fig:MnCl2} and MnBr$_2$ the predicted monolayer ground state to be helical with a hubbard U, while the bicollinear order is only a local minimum.

%The experimental ground state in MnCl$_2$ intercalated monolayers has been reported to be a helimagnetic ground state with $\mathbf{Q}_{exp} = (0.153, 0.153, 0)$ which is in decent agreement with the present result.
%The Manganese TMHs have likewise been found to have lowest energy at $\mathbf{q} = K$. 
The experimental ground state of bulk MnCl$_2$ not unambiguously known, but under the assumption of collinearity a possible ground state contains 15 Mn atoms in an extended stripy pattern \cite{wilkinson1958neutron}. Due to the weak and subtle nature of magnetic interactions in the manganese compounds, however, it is not unlikely that the ground state in the monolayers can differ from that of bulk. This is corroborated by an experimental study of MnCl$_2$ intercalated by graphite where a helimagnetic ground state with $\mathbf{Q}_\mathrm{exp} = (0.153, 0.153)$ was found \cite{wiesler1995determination}. This is rather close to our predicted ordering vector obtained from LDA+U.

Experimentally, bulk MnBr$_2$ is found to exhibit a stripy bicollinear \textit{uudd} order at low temperatures \cite{wollan1958neutron}. The order cannot be represented by a spiral in the minimal cell, but requires calculations in rectangular unit cells with spiral order $\mathbf{Q} = (0, 1/2)$ similar to VI$_2$ discussed above. We have calculated the high symmetry band path required to show this order and do not find any new minima. It is likely that the situation resembles MnCl$_2$ where a single-$q$ spiral has been observed for decoupled monolayers in agreement with our calculations. %However, since the band width of the LDA+U calculations is only 1.5 meV we cannot exclude that low energy minima arises in the super cell calculations. The energy gain could be below the required accuracy of the calculations.´%TODO: In figure \ref{fig:NiI2}, we compare the spiral calculations with LDA+U for both the minimal unit cell and the rectangular unit cell. The results predict an incommensurate spin spiral ground state at $\mathbf{Q}=(0.11,0.11)$ in the minimal unit cell. The experimentally reported ground state corresponds to the high symmetry point Y in the rectangular unit cell, which is seen to comprise a local minimum that is close to being degenerate with the global minimum.  

%However, even including this search still yields $\mathbf{q} = K$ in the minimal unit cell as the minimum, instead we find that the minimum in MnX$_2$ is sensitive to hubbard U, in which case we find a helimagnetic ground state for all three Mn based structures. This is encouraging because in graphene intercalated monolayer MnCl$_2$ has been found to have a helimagnetic ground state with $\mathbf{q} = (0.153, 0.153, 0)$ which is in decent agreement with the extended search seen in figure \ref{fig:MnCl2}. MnI$_2$ is found to exhibit helimagnetic order as well in decent agreement with our calculations. The bicollinear calculations with hubbard U for bromide and chloride shows new local minima at the stripe \textit{uudd} order, which is the one stabilized in bulk samples, but the global minimum remain the finite $q$ helimagnet. We therefore predict that MnBr$_2$ would behave much like MnCl$_2$ and likewise manifest a helimagnet in the monolayer limit.

%The ground state of bulk MnI$_2$ has been shown to be rather sensitive to temperature, where the ground state is helimagnetic with $\mathbf{Q}_1=(0.18,0,0.44)$, undergoing two phase transitions changing to $\mathbf{Q}=(0.10,0.10,0.5)$ at 4 K.%The situation appears to even more complicated for bulk MnI$_2$,  but undergoes a phase transition . %[MnI2 experimental: (0.181, 0, 0.439) @ 3.4K $\rightarrow$ (0.1025, 0.1025, 0.5) @ 4K]

\subsubsection*{FeX$_2$}
We find all the iron halides to have ferromagnetic ground states. For FeCl$_2$ and FeBr$_2$ this is in agreement with the experimentally determined magnetic order for the bulk compounds \cite{wilkinson1959neutron}. In contrast, FeI$_2$ has been reported to exhibit a bicollinear antiferromagnetic ground state \cite{gelard1974magnetic} similar to the case of MnBr$_2$ discussed above. It is again possible that the ground state of the monolayer (calculated here) could differ from the magnetic ground state of the bulk compound as has been found for MnCl$_2$. 

LDA predict the three compounds to be half metals, meaning that the majority spin bands are fully occupied and only the minority bands have states at the Fermi level. This enforces an integer number of Bohr magnetons (four) per unit cell at any $\mathbf{q}$-vector in the spin spiral calculations. Thus longitudinal fluctuations are expected to be strongly suppressed in iron halides and it is likely that these materials can be accurately modelled by Heisenberg Hamiltonians despite the itinerant nature of the electronic structure.

%metallic but still the magnetic moments $\mu_q = 3.5\mu_B$ of all different spin spirals vary very little, which indicates perturbative approaches in the transversal direction is a good description.

%The projected spin orbit corrections to the spin spirals of FeCl$_2$ and FeBr$_2$ is a constant shift at all $q$ of $2.3$ meV and $7$ meV respectively. Additionally these have the largest bandwidth of the TMHs suggesting strong intralayer exchange interactions. FeI$_2$ has comparatively smaller spiral energies, and also a larger spin orbit correction on the order of the spiral energies and the in-plane spin plane orientation has lowest energy. Seeing as these energies are on the same order of magnitude, treating spin orbit perturbatively is perhaps not well justified.  
The projected spin orbit coupling is not applicable to collinear structures and we therefore include full spin orbit coupling, which is compatible with the $\mathbf{Q}=(0,0)$ ground state. We find that all the iron compounds have an out-of-plane easy axis, which is in agreement with experiments. The bandwidth provided in table \ref{tab:AB2Result} then simply corresponds to the magnetic anisotropy energy which is smallest for FeCl$_2$ and increases for the heavier Br and I compounds as expected.

\subsubsection*{CoX$_2$}
%The cobalt halides have a $3d^7$ electronic configuration for the Co atoms, and so both high- and low-spin states are possible. We find a high spin state for chloride and bromide consistent with experiments, but the iodine variant finds the low spin configuration. 
We predict CoCl$_2$ to have an in-plane ferromagnetic ground state in agreement with the experimentally determined magnetic order of the bulk compound \cite{wilkinson1959neutron}. CoBr$_2$ is found to have a long wavelength spin spiral with $\mathbf{Q}=(0.03, 0.03)$. The spiral energy in the vicinity of the $\Gamma$-point is, however, extremely flat with almost vanishing curvature and the ground state energy is merely 0.04 meV lower than the ferromagnetic state. We regard this as being in agreement with the experimental report of intra-layer ferromagnetic order in the bulk compound \cite{wilkinson1959neutron}. 

The case of CoI$_2$ deviates substantially from the other two halides. CoCl$_2$ and CoBr$_2$ are half-metals with $m=3\;\mu_\mathrm{B}$ per unit cell, whereas CoI$_2$ is an ordinary metal with $m\approx1.2\;\mu_\mathrm{B}$ per unit cell. We find the magnetic ground state of CoI$_2$ to be stripy anti-ferromagnetic with $\mathbf{Q}=(1/2,0)$, whereas experiments on the bulk compound have reported helimagnetic in-plane order with $\mathbf{Q}_\mathrm{exp}=(1/6, 1/8, 1/2)$ in the rectangular cell \cite{MEKATA1992859}. We note, however, that the calculated local magnetic moments vary strongly with $\mathbf{q}$ (up to 0.5 $\mu\mathrm{B}$) in the spin spiral calculations, which signals strong longitudinal fluctuations. This could imply that the material comprises a rather challenging case for DFT and LDA may be insufficient to treat this material properly. %  the iodine TMH is found to have lowest energy in the collinear stripe antiferromagnetic phase with $q = M$ as seen in figure \ref{fig:spirals}, while the experimental structure a commensurate spiral with $q = (1/8, 0, 0)$ with in-plane spin orientation. Computationally CoI2 is quite complex since it is metallic and the magnetic moment changes by $0.4 \mu_B$ for the antiferromagnetic configurations compared to the ferromagnetic ones. 
%CoI$_2$ should first order phase transition to $\mathbf{q} = (1/6, 1/8, 0)$ in the $a\cross\sqrt{3}a$ unit cell at 9.4K \cite{mekata1992first} 

\subsection{Spontaneous polarization of AB$_2$ materials}
The materials in table \ref{tab:AB2Result} that exhibit spin spiral ground states are expected to introduce a polar axis due to spinorbit coupling and thus allow for spontaneous electric polarization. The stripy antiferromagnet with $\mathbf{Q}=(1/2, 0)$ preserves a site-centered inversion center and remains non-polar. In addition, the case of $\mathbf{Q}=(1/3, 1/3)$ with in-plane orientation of the spiral plane breaks inversion symmetry, but retains the three-fold rotational symmetry (up to translation of a lattice vector) and therefore cannot acquire components of in-plane polarization.

To investigate the effect of symmetry breaking we have constructed $7\times1$ supercells of VI$_2$ and the Ni halides and performed a full relaxation of the $\mathbf{q}=(1/7,0)$ spin spiral commensurate with the supercell. This is not exactly the spin spirals found as the ground state from LDA, but we will use these to get a rough estimate of the spontaneous polarization. We note that this is very close to the in-plane component of $\mathbf{Q}_\mathrm{exp}$ for bulk NiI$_2$, which is found to be nearly degenerate with the predicted ground state (see figure \ref{fig:NiI2}). The other materials exhibit similar near-degeneracies, but the calculated polarization could be sensitive to which spiral ordering vector is used. We have chosen to focus on the incommensurate spirals, but note that all the $\mathbf{Q}=(1/3,1/3)$ materials of table \ref{tab:pol} are expected to introduce a spontaneous polarization as well. Besides the incommensurate spirals we thus only include the cases of MnBr$_2$ and MnI$_2$ where the $\mathbf{Q}=(1/3, 1/3)$ spirals may be represented in $\sqrt{3}\times\sqrt{3}$ supercells. The former case represents an example of a proper screw while the latter is an in-plane cycloid. The experimental order in the Mn halides materials is complicated, and our LDA+U calculations yield an ordering vector that differs from that of LDA. However, here we mostly consider these examples for comparison and to check the symmetry constraints on the polarization in the $\mathbf{Q}=(1/3, 1/3)$ spirals.% In the former case we obtain a polarization of 194 fC/m directed along the ordering vector. The orientation of the polarization is fixed by the fact that a $\mathbf{Q}=(1/3, 1/3)$ proper screw has a two-fold rotational symmetry axis along $\mathbf{Q}$ combined with time-reversal symmetry. In the case of MnI$_2$ we obtain a pure out-of plane polarization of 70 pC/m as expected for a $\mathbf{Q}=(1/3, 1/3)$ in-plane cycloid where in-plane polarization is forbidden by symmetry.

In order to calculate the spontaneous polarization we relax the atomic positions in the super cells both with and without spinorbit coupling (included self-consistently) and calculate the 2D polarization from the Berry phase formula \cite{Gjerding2021}. The results are summarized in Tab. \ref{tab:pol}. We can separate the effect of relaxation from the pure electronic contribution by calculating the polarization (including spin-orbit) of the structures that were relaxed without spinorbit coupling. These numbers are stated in brackets in table \ref{tab:pol} as well as the total polarization (including relaxation) and the angles that define the orientation of the spiral plane with respect to $\mathbf{Q}$. The self-consistent calculations yield the optimal orientations of the spiral planes without the PSO approximations and it is reassuring that the orientation roughly coincides with the results of the GBT and the PSO approximation.% (note that $\varphi$ is measured with respect to $\mathbf{Q}$ and is thus expected to deviate by 30{\degree} from the values in table \ref{tab:AB2Result} where $\mathbf{Q}$ is in the $\Gamma$K direction).  

The magnitude of polarization largely scales with the atomic number of ligands (as expected from the strength of spinorbit coupling) and the iodide compounds thus produce the largest polarization. The in-plane cycloid in MnI$_2$ only give rise to out-of-plane polarization as expected from symmetry and the $\mathbf{Q}=(1/3,1/3)$ proper screw in MnBr$_2$ has polarization that is strictly aligned with $\mathbf{Q}$. The latter results is expected for any proper screw in the $\Gamma$K-direction because $\mathbf{Q}$ then coincides with a two-fold rotational axis and the ground state remains invariant under the combined action of this rotation and time-reversal symmetry. Since the polarization is not affected by time-reversal it must be aligned with the two-fold axis. The polarization vectors of the remaining materials (except for NiCl$_2$) are roughly aligned with the intersection between the spiral plane and the atomic plane. 

It is interesting to note that the calculated magnitudes of total polarization are 5-10 times larger than the prediction from the pure electronic contribution where the atoms were not relaxed with spinorbit coupling. We also tried to calculate the polarization by using the Born effective charge tensors (without spin-orbit) and the atomic deviations from the centrosymmetric positions. However, this approximation severely underestimates the polarization and even produces the wrong sign of the polarization in the case of NiBr$_2$ and NiI$_2$. To obtain reliable values for the polarization it is thus crucial to include the relaxation effects and take the electronic contribution properly into account (going beyond the Born effective charge approximation).
In Ref. \cite{Song2022} a value of 141 fC/m was predicted in 2D NiI$_2$ from the gKNB model \cite{Xiang2011} and this is comparable to the values found in table \ref{tab:pol} without relaxation effects. When relaxation is included we find a magnitude of 1.9 pC/m for NiI$_2$, which is an order of magnitude larger compared to the previous prediction. The results are, however, not directly comparable since Ref. \cite{Song2022} considered a spiral along the $\Gamma$K direction whereas the present result is for a spiral along $\Gamma$M. We note that Ref. \cite{Song2022} finds the polarization to be aligned with $\mathbf{Q}$ in agreement with the symmetry considerations above. %.Similarly, bulk MnI$_2$ has been predicted to have polarization of 71.4 $\mu C/m^2$, which corresponds to a 2D polarization of 49 $fC/m$ (by multiplication of the bulk layer separation). 
Finally, the values for the spontaneous polarization in table \ref{tab:pol} may be compared with those of ordinary 2D ferroelectrics, which are typically on the order of a few hundred pC/m for in-plane ferroelectrics and a few pC/m for out-of-plane ferroelectrics \cite{Kruse2022} . %The spontaneous polarization of the type II multiferroics studied here are thus orders of magnitude smaller compared to those of ordinary ferroelectrics. and TODO?
%The cases of NiBr$_2$ and NiI$_2$ have similar orientation of spin spiral planes and we observe that the direction of spontaneous polarization is similar although NiI$_2$ exhibits somewhat larger magnitude due to the stronger spinorbit coupling induced by the I atoms. The case of NiCl$_2$ has a spin plane orientation (proper screw) which only deviate slightly from NiBr$_2$, but the direction of polarization is reversed compared to NiBr$_2$. The origin of this qualitative difference is not completely clear at present, but it could be related to the nature of the spinorbit mediated magnetic interactions, which are likely to originate from the Ni atoms in NiCl$_2$, whereas the halide atoms are expected to deliver the dominating spinorbit effects in NiBr$_2$ and NiI$_2$. 

In all of these type II multiferroics, the orientation of the induced polarization depends on the direction of the ordering vector, which may thus be switched by application of an external electric field. We have checked explicitly that the sign of polarization is changed if we relax a right-handed instead of a left-handed spiral (corresponding to a reversed ordering vector). The small values of spontaneous polarization in these materials implies that rather modest electric fields are required for switching the ordering vector and thus comprise an interesting alternative to standard multiferroics such as BiFeO$_3$ and YMnO$_3$, where the coercive electric fields are orders of magnitude larger.
%\begin{table}[tb]
%\begin{tabular}{c|r|r|r|r|r|r}
%      & $(\theta,\varphi)$ & $Z^M_\parallel$ & $Z^M_\perp$ & $P_x$ & $P_y$  & $P_z$
%     \\ \hline
%VI$_2$ & (11, 30) & 1.64 & 0.34 & -47 & 82  & 0.091 \\     
%NiCl$_2$ & (90, 0) & 1.86 & 0.39 & -53 & 31  & -9.7 \\  
%NiBr$_2$ & (69, 20) & 1.77 & 0.34 & 59 & -53  & 9.4 \\   
%NiI$_2$ & (70, 30) & 1.47 & 0.25 & 127 & -220  & 0.33
%\end{tabular}
%\caption{Orientation of spin planes, transition metal Born charges and 2D polarization (in fC/m) of VI$_2$ and the Ni halides obtained from $7\times1$ supercell calculations with left-handed $\mathbf{Q}$=(1/7, 0) spirals.}
%\label{tab:pol}
%\end{table}

\subsection{Magnetic ground state of AB$_3$ materials}\label{sec:AB3}
The AB$_3$ materials all have space group $P\bar3m1$ corresponding to monolayers of the BI$_3$ (or AlCl$_3$) prototype. The magnetic lattice is the honeycomb motif, thus hosting two magnetic ions in the primitive cell. Several materials of this prototype have been characterized experimentally, but here we only present results for the Cr compounds. This is due to the fact that experimental data of in-plane order is missing for all but CrX$_3$, FeCl$_3$ and RuCl$_3$. Moreover, all magnetic compounds were found to have a simple ferromagnetic ground state. RuCl$_3$ is a well known insulator with stripy antiferromagnetic in-plane order. However, bare LDA finds a metallic state and both Hubbard corrections and self-consistent spinorbit coupling are required to obtain the correct insulating state \cite{kim2015kitaev}. The latter is incompatible with the GBT approach and we have not pursued this further here.
Bulk FeCl$_3$ is known to be an insulating helimagnet with $\mathbf{Q} = (\frac{4}{15}, \frac{1}{15}, \frac{3}{2})$ \cite{cable1962neutron}, while we find the monolayer to be a metallic ferromagnet. 

For CrI$_3$ we compare the spin spiral dispersion to the spiral energy determined by a third nearest neighbour energy mapping procedure. The prototype thus serves as a testing ground for applying unconstrained GBT to materials with multiple magnetic atoms in the unit cell. We analyse the intracell angle between the Cr atoms of CrI$_3$ and provide an expression for generating good initial magnetic moments for GBT calculations. We finally discuss the observed deviations from the classical Heisenberg model and to what extend the flat spiral spectrum can be used to obtain the magnon excitation spectrum.
\begin{table}[tb]
\begin{tabular}{c|c|c|c|c}
      & $(\theta,\varphi)$ & $P_\parallel$ & $P_\perp$  & $P_z$
     \\ \hline
%VI$_2$ & (11, 30) & -145 (-75) & 250 (67)  & 0.05 (0.11) \\     
VI$_2$ & (11, 0) & -0.6 (-31) & 290 (96)  & 0.05 (0.11) \\     
%NiCl$_2$ & (90, 0) & -70 (-8.6) & 47 (12.1)  & 3.5(-5.1) \\  
NiCl$_2$ & (90, -30) & -37 (-1.4) & 76 (15)  & 3.5(-5.1) \\  
%NiBr$_2$ & (69, 20) & -159 (-21) & 300 (25) & 26 (37) \\   
NiBr$_2$ & (69, -10) & 12 (-6) & 340 (32) & 26 (37) \\   
%NiI$_2$ & (70, 30) & -950 (-240) & 1630 (320)  & -0.18 (12) \\
NiI$_2$ & (70, 0) & -8 (-48) & 1890 (400)  & -0.18 (12) \\
MnBr$_2$ & (90, 0) & 430 (38) & 0 (0.02)  & 0 (0) \\
MnI$_2$ & (0, 0) & 0 (0.6) & 0.3 (-7)  & -260 (-105)
\end{tabular}
\caption{Orientation of spin planes, and 2D polarization (in fC/m) of selected transition metal halides. $P_\parallel$ denotes the polarization along $\mathbf{Q}$, while $P_\perp$ denotes the polarization in the atomic plane orthogonal to $\mathbf{Q}$ and $P_z$ is the polarization orthogonal to the atomic plane. The numbers in brackets are the polarization values obtained prior to relaxation of atomic positions. We have used $7\times1$ supercells for the V and Ni halides and $\sqrt{3}\times\sqrt{3}$ supercells for the Mn halides. All calculations are set up with left-handed spirals. The numbers in brackets state the spontaneous polarization without relaxation effects.}
\label{tab:pol}
\end{table}
\subsubsection*{CrX3}
The chromium trihalides are of considerable interest due to the versatile properties that arise across the three different halides. Monolayer CrI$_3$ was the first 2D monolayer that were demonstrated to host ferromagnetic order below 45 K \cite{Huang2017} and has spurred intensive scrutiny in the physics of 2D magnetism. The magnetic order is governed by strong magnetic easy-axis anisotropy, which is accurately reproduced by first principles simulations \cite{Lado2017,Torelli2018}. In contrast, monolayers of CrCl$_3$ exhibit ferromagnetic interactions as well, but no proper long range order due easy-plane anisotropy. Instead, these monolayers exhibit Kosterlitz-Thouless physics, which give rise to quasi long range order below 13 K \cite{Bedoya-Pinto2021}.

The GBT is not really necessary to find the ground state of the monolayer chromium halides. They are all ferromagnetic and insulating and only involve short range exchange interactions that are readily obtained from collinear energy mapping methods \cite{Lado2017, Torelli2018, Olsen2019}. Nevertheless, the gap between the acoustic and optical magnons in bulk CrI$_3$ has been proposed to arise from either (second neighbor) Dzyalosinskii-Moriya interactions \cite{Chen2018a} or Kitaev interactions \cite{Xu2018a, Lee2020}. The former could in principle be extracted directly from planar spin-spiral calculations \cite{sandratskii2017insight}, while the latter requires conical spin spirals. The origin of this gap is, however, still subject to debate \cite{Do2022} and here we will mainly focus on the magnetic interactions that do not rely on spinorbit coupling. In the following we will focus on CrI$_3$ as a representative member of the family.% XXX biquadratic exchange \cite{Kartsev2020}.
\begin{figure*}[tb]
    \centering
    \includegraphics[width=1\textwidth]{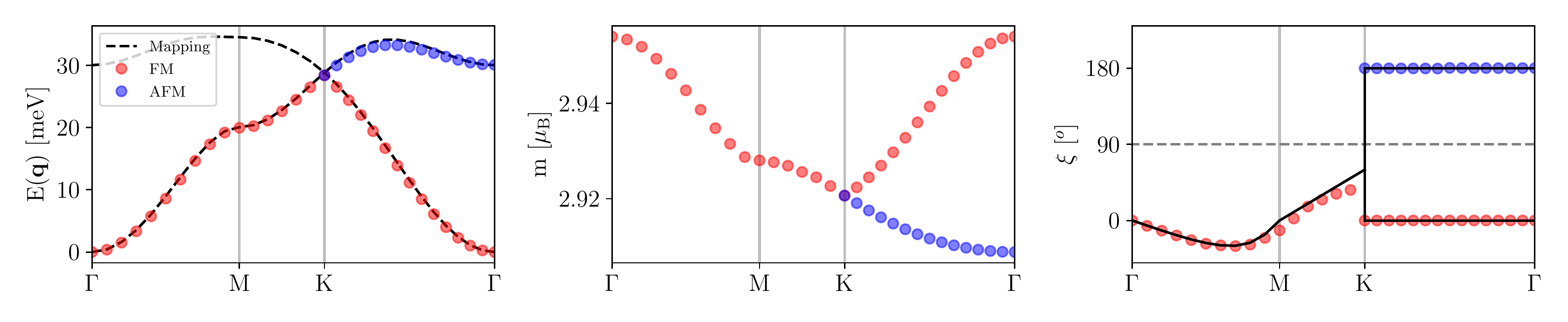}
    \caption{(Left: spin spiral energies of CrI$_3$ compared to  third nearest neighbour energy mapping. Right: angles beteen the two magnetic moments. The spin spirals are initialised with angles determined by Eq. \eqref{eq:nnminimize} which are shown in black. The moments are collinear on the $\Gamma K$ path and so the AFM solution is also quasi-stable in DFT. Center: the magnitude of local magnetic moments along the spiral path.}
    \label{fig:CrI3}
\end{figure*}

The honeycomb lattice contains two magnetic atoms per unit cell and the magnetic moments at the two sites will in general differ by an angle $\xi$. Since we do not impose any constraints except for the boundary conditions specified by $\mathbf{q}$, the angle will be relaxed to its optimal value when the Kohn-Sham equations are solved self-consistently. The convergence of $\xi$, may be a tedious process since the total energy has a rather weak dependence on $\xi$. For a given $\mathbf{q}$ the classical energy of the model \eqref{eq:heisenberg} is minimized by the angle $\xi^0$ given by
\begin{align}
    \tan{\xi^0}=-\frac{\mathrm{Im}J^{12}(\mathbf{q})}{\mathrm{Re}J
    ^{12}(\mathbf{q})},
    \label{eq:nnminimize}
\end{align}
where 
\begin{align}
    J^{12}(\mathbf{q})=\sum_iJ^{12}_{0i}e^{-i\mathbf{q}\cdot\mathbf{R}_i}
\end{align}
is the Fourier transform of the inter-sublattice exchange coupling. If one assumes nearest neighbour interactions only, $\xi^0$ becomes independent of exchange parameters and the resulting expression thus comprises a suitable initial guess for the inter-sublattice angle. We note that the classical spiral energy is independent of $\xi$ (in the absence of spinorbit coupling) when $J^{12}(\mathbf{q})=0$ and the angle may be discontinuous at such $\mathbf{q}$-points. %In such cases the initial guess is chosen to be the $\xi^0 = \mathbf{q}\cdot \mathbf{a}$, where $\mathbf{a}$ is the inter-sublattice vector. 
This occurs for example in the magnetic honeycomb lattice at the K-point ($\mathbf{q}=(1/3,1/3)$). In general, Eq. \eqref{eq:nnminimize} has two solutions that differ by $\pi$ and only one of these minimzes the energy while the other maximizes it. The maximum energy constitutes an "optical" spin spiral branch, which is if interest if one wishes to extract the exchange coupling constants. %Along the $\Gamma\mathrm{K}$ line, $\xi=0$ yields the minimum energy, but it is also possible to converge the local maxium at $\xi=\pi$, which yields an "optical" spin spiral state. 

The spiral energies of CrI$_3$ (with optimized intracell angles) are shown in figure \ref{fig:CrI3}, where we show both the ferromagnetic ($\xi=0$) and the antiferromagnetic ($\xi=\pi$) results on the $\Gamma\mathrm{K}$ path. We also show the spiral energy obtained from the model \eqref{eq:heisenberg} with exchange parameters calculated from a collinear energy mapping using four differnet spin configurations. We get $J_1=2.47$ meV,
$J_2=0.682$ meV and $J_3=-0.247$ meV for the first, second and nearest neighbour interactions respectively, which is in good agreement with previous LDA calculations \cite{Olsen2021}. The model spiral energy is seen to agree very well with that obtained from the GBT, which largely validates such a three-parameter model (when spinorbit is neglected). We do, however, find a small deviation in the regions between high-symmetry points. This is likely due to higher order exchange interaction, which will deviate in the two approaches. For example, a biquadratic exchange term \cite{Gutzeit2022}, will cancel out in any collinear mapping, but will influence the energies obtained from the GBT. Biquadratic exchange parameters could thus be extracted from the deviation between the two calculations. 

In figure \ref{fig:CrI3} we also show the calculated values of $\xi$ and the magnitude of the local magnetic moment at the Cr sites along the path. The self-consistent intracell angles are found to match very well with the initial guess, except for a slight deviation on the Brillouin zone boundary. This corroborates the fact that exchange couplings beyond second neighbours are insignificant (the second nearest neighbor coupling is an intra-sublattice interaction and does not influence the angle). 

It is also rather instructive to analyze the variation in the magnitude of local magnetic moments. In general, the mapping of electronic structure problems to Heisenberg types of models like \eqref{eq:heisenberg} rests on an adiabatic assumption where it is assumed that the magnitude of the moments are fixed. However, the present variation in the magnitude of moments does not imply a breakdown of the adiabatic assumption, but reflects that DFT should be mapped to a quantum mechanical Heisenberg model rather than a classical model. In particular, the ratio of spin expectation values between the ferromagnetic ground state and the (anti-ferromagnetic) state of highest energy is approximately $\langle S_i\rangle_\mathrm{AFM}/\langle S_i\rangle_\mathrm{FM}=0.83$ in the quantized model \cite{Torelli2020}. While this ratio is somewhat smaller than the difference between ferromagnetic and anti-ferromagnetic moments found here, the result does imply that the magnitude of moments should depend on $\mathbf{q}$. And the fact that the $\mathbf{q=0}$ anti-ferromagnetic moments are smaller than the ferromagnetic ones in a self-consistent treatments reflects that DFT captures part of the quantum fluctuations inherent to the model \eqref{eq:heisenberg}.

We note that the spin spiral energy $E_\mathbf{q}$ calculated from the isotropic Heisenberg model using the optimal angle given by Eq. \eqref{eq:nnminimize} is related to the dynamical excitations (magnon energies) by $\omega_\mathbf{q}^\pm=E_\mathbf{q}^\pm/S$ and the spiral energies thus comprise a simple method to get the magnetic excitation spectrum. However, even if a model like \eqref{eq:heisenberg} fully describes a magnetic material (no anisotropy or higher order terms) there will be a systematic error in the extracted exchange parameters (and resulting magnon spectrum) if the parameters are extracted by mapping to the classical model. The reason is, that the classical energies correspond to expectation values of spin configurations with fixed magnitude of the spin, which is not accommodated in a self-consistent approach. This error is directly reflected by the variation of the magnitude of moments in figure \ref{fig:CrI3}. The true exchange parameters can only be obtained either by mapping to eigenstates of the model \cite{Torelli2020} or by considering infinitesimal rotations of the spin, which may be handled non-selfconsistently using the magnetic force theorem \cite{Liechtenstein1987,Bruno2003,Halilov1998,Zimmermann2019,Durhuus2022}. Nevertheless, the deviations between exchange parameters obtained from classical and quantum mechanical energy mapping typically deviates by less than 5 {\%} \cite{Torelli2020} and for insulators it is a good approximation to extract the magnon energies from planar spiral calculations although the mapping is only strictly valid in the limit of small $\mathbf{q}$.
%whereas the same q-vector with smooth noncollinear order with $\xi = 2\pi/3$ is mapped to the $\mathbf{K}$ point in the primitive cell, and that is found to have higher energy. 

%\begin{figure}[tb]
%    \centering
%    \includegraphics[width=0.33\textwidth]{NiI2_bandpath.png}
%    \caption{High symmetry band path for spin spiral ground state calculations of \ce{NiI2}.}
%    \label{fig:NiI2}
%\end{figure}

\section{Conclusion and outlook}\label{sec:conclusion}
In conclusion, we have demonstrated the abundance of spiral magnetic order in 2D transition metal dichalcogenides from first principles calculations. The calculations imply that type II multiferroic order is rather common in these materials and we have calculated the spontaneous polarization in a selected subset of these using fully relaxed structures in super cells. While the super cell calculations does not correspond to the exact spirals found from the GBT, the calculations show that relaxation effects plays a crucial role for the induced polarization and should be taken into account in any quantitative analysis. The spontaneous polarization in type II multiferroics is in general rather small compared to what is found in ordinary 2D ferroelectrics and could imply that the chirality of spirals are switchable by small electric fields. It would be highly interesting to calculate the coercive field for switching in these materials, but due to the importance of relaxation effects and spin-orbit coupling this is a non-trivial computation that cannot simply be obtained from the Born effective charges and force constant matrix.

The GBT comprises a powerful framework for extracting the magnetic properties of materials from first principles. In addition to the single-$q$ states considered here, one may use super cells to extract the importance of higher order exchange interactions and unravel the possibility of having multi-$q$ ground states. In addition, for non-centrosymmetric materials, the PSO approach may be readily applied to obtain the Dzyaloshinskii-Moriya interactions, which may lead to Skyrmion lattice ground states or stabilize other multi-$q$ states.

\onecolumngrid
\section{Appendix}
\subsection{Implementation}\label{sec:implementation}
In the PAW formalism we expand the spiral spinors using the standard PAW transformation \cite{blochl}
\begin{align}
    \psi_{\mathbf{q},\mathbf{k}}(\mathbf{r}) &= \hat{\mathcal{T}} \tilde{\psi}_{\mathbf{q},\mathbf{k}}(\mathbf{r})= \tilde{\psi}_{\mathbf{q},\mathbf{k}}(\mathbf{r})+\sum_a \sum_i (\phi_i^a(\mathbf{r}) - \tilde{\phi}_i^a(\mathbf{r}))\int d\mathbf{r} [\tilde{p}_i^{a}(\mathbf{r})]^*\tilde{\psi}_{\mathbf{q},\mathbf{k}}(\mathbf{r}),
\end{align}
where $\tilde{\psi}_{\mathbf{q},\mathbf{k}}(\mathbf{r})$ is a smooth (spinor) pseudo-wavefunction that coincides with $\psi_{\mathbf{q},\mathbf{k}}(\mathbf{r})$ outside the augmentation spheres and deviates from $\psi_{\mathbf{q},\mathbf{k}}(\mathbf{r})$ by the second term inside the augmentation spheres. The all-electron wavefunction $\psi_{\mathbf{q},\mathbf{k}}(\mathbf{r})$ is thus expanded in terms of (spinor) atomic orbitals $\phi_i^a$ inside the PAW spheres and the expansion coefficients are given by the overlap between the pseudowavefunction and atom-centered spinor projector functions $\tilde p_i^a$. Using Eq. \eqref{eq:GBT} we may write this as
\begin{align}
    \psi_{\mathbf{q},\mathbf{k}}(\mathbf{r})
                 &= e^{i\mathbf{k}\cdot\mathbf{r}}U^\dag_\mathbf{q}(\mathbf{r})\tilde{u}_\mathbf{q,k}(\mathbf{r})+\sum_a \sum_i (\phi_i^a(\mathbf{r}) - \tilde{\phi}_i^a(\mathbf{r}))\int d\mathbf{r} [\tilde{p}_i^{a}(\mathbf{r})]^*e^{i\mathbf{k}\cdot\mathbf{r}}U^\dag_\mathbf{q}(\mathbf{r})\tilde{u}_\mathbf{q,k}(\mathbf{r})\notag\\               &=e^{i\mathbf{k}\cdot\mathbf{r}}U^\dag_\mathbf{q}(\mathbf{r})\tilde{u}_\mathbf{q,k}(\mathbf{r})+\sum_a \sum_i (\phi_i^a(\mathbf{r}) - \tilde{\phi}_i^a(\mathbf{r}))\int d\mathbf{r} [e^{-i\mathbf{k}\cdot\mathbf{r}}U_\mathbf{q}(\mathbf{r})\tilde{p}_i^{a}(\mathbf{r})]^*\tilde{u}_\mathbf{q,k}(\mathbf{r})\notag\\
 &=e^{i\mathbf{k}\cdot\mathbf{r}}U^\dag_\mathbf{q}(\mathbf{r})\tilde{u}_\mathbf{q,k}(\mathbf{r})+\sum_a \sum_i (\phi_i^a(\mathbf{r}) - \tilde{\phi}_i^a(\mathbf{r}))\int d\mathbf{r} [\tilde{p}_{i,\mathbf{q,k}}^a(\mathbf{r}) ]^*\tilde{u}_\mathbf{q,k}(\mathbf{r})\notag\\
 &\equiv\mathcal{T}_\mathbf{q,k}\tilde{u}_\mathbf{q,k}(\mathbf{r})           ,
\end{align}
where $U_\mathbf{q}(\mathbf{r})$ was given in Eq. \eqref{eq:U} and we defined
\begin{align}\label{eq:projector}
    \tilde{p}_{i,\mathbf{q,k}}^a(\mathbf{r}) 
    &= e^{-i\mathbf{k}\cdot\mathbf{r}}U_{\mathbf{q}} (\mathbf{r})\tilde{p}_i^{a}(\mathbf{r}).
\end{align}
The PAW transformed Kohn-Sham equations then read
\begin{align}
    \tilde{H}_{\mathbf{q},\mathbf{k}}\tilde{u}_{\mathbf{q},\mathbf{k}}(\mathbf{r}) = \epsilon_\mathbf{q,k}S_\mathbf{q,k}\tilde{u}_{\mathbf{q},\mathbf{k}}(\mathbf{r}),
\end{align}
with
\begin{align}
    \tilde{H}_\mathbf{q,k}=\mathcal{T}_\mathbf{q,k}^\dag H\mathcal{T}_\mathbf{q,k},\qquad S_\mathbf{q,k}=\mathcal{T}_\mathbf{q,k}^\dag \mathcal{T}_\mathbf{q,k}.
\end{align}
Calculations in the framework of the GBT thus requires two modifications compared to the approach for solving the ordinary Kohn-Sham equations in the PAW formalism. 1) The $k$-dependence of the standard Bloch Hamiltonian is replaced by $\mathbf{k}\rightarrow\mathbf{k}\mp\mathbf{q}/2$ for spin-up and spin down components respectively. 2) Different spin dependent projector functions has to be applied when calculating the projector overlaps with the spin-up and spin-down components of the psudowavefunctions (see Eq. \eqref{eq:projector}).
\vspace{5mm}

\twocolumngrid

\begin{figure}[tb]
    \centering
    \includegraphics[width=0.45\textwidth]{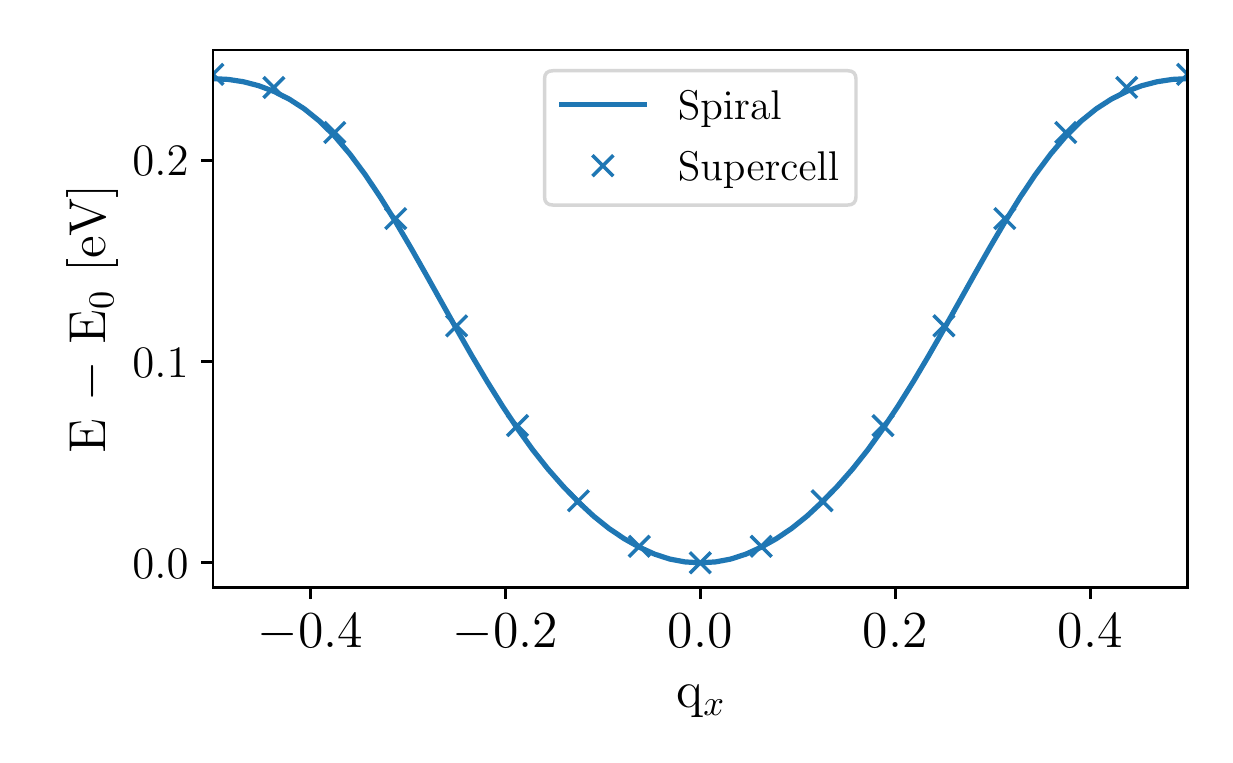}
    \caption{Comparison between GBT spin spiral calculations and supercell calculations without spinorbit coupling in monolayer CoPt.}
    \label{fig:CoPtsupercell}
\end{figure}
\begin{figure}[tb]
    \centering
    \includegraphics[width=0.45\textwidth]{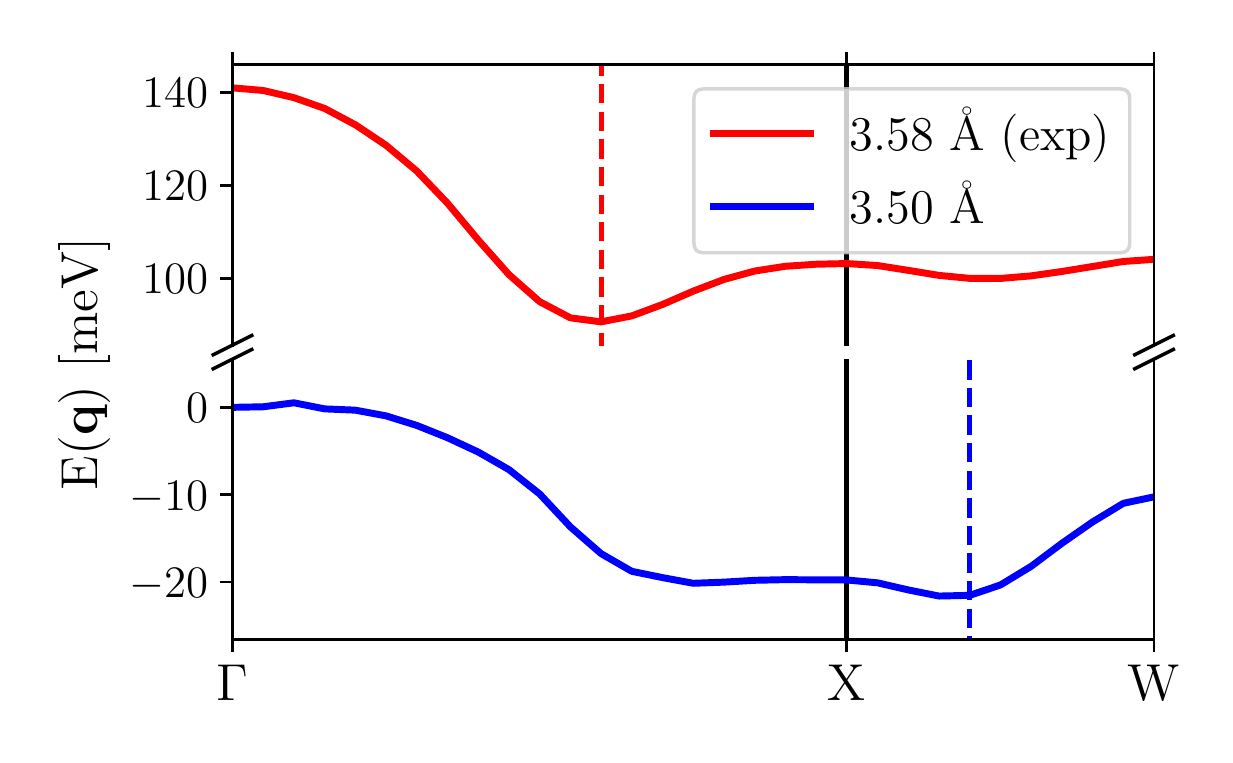}
    \caption{Spin spiral energies of fcc Fe for the experimental lattice constant (red) and a strained latice constant, which is known to reproduce the experimental spin spiral order in (blue). The dashed vertical lines indicate the minima found in Ref. \cite{marsman2002broken}.}
    \label{fig:fcciron}
\end{figure}
\begin{figure}[tb]
    \centering
    \includegraphics[width=0.45\textwidth]{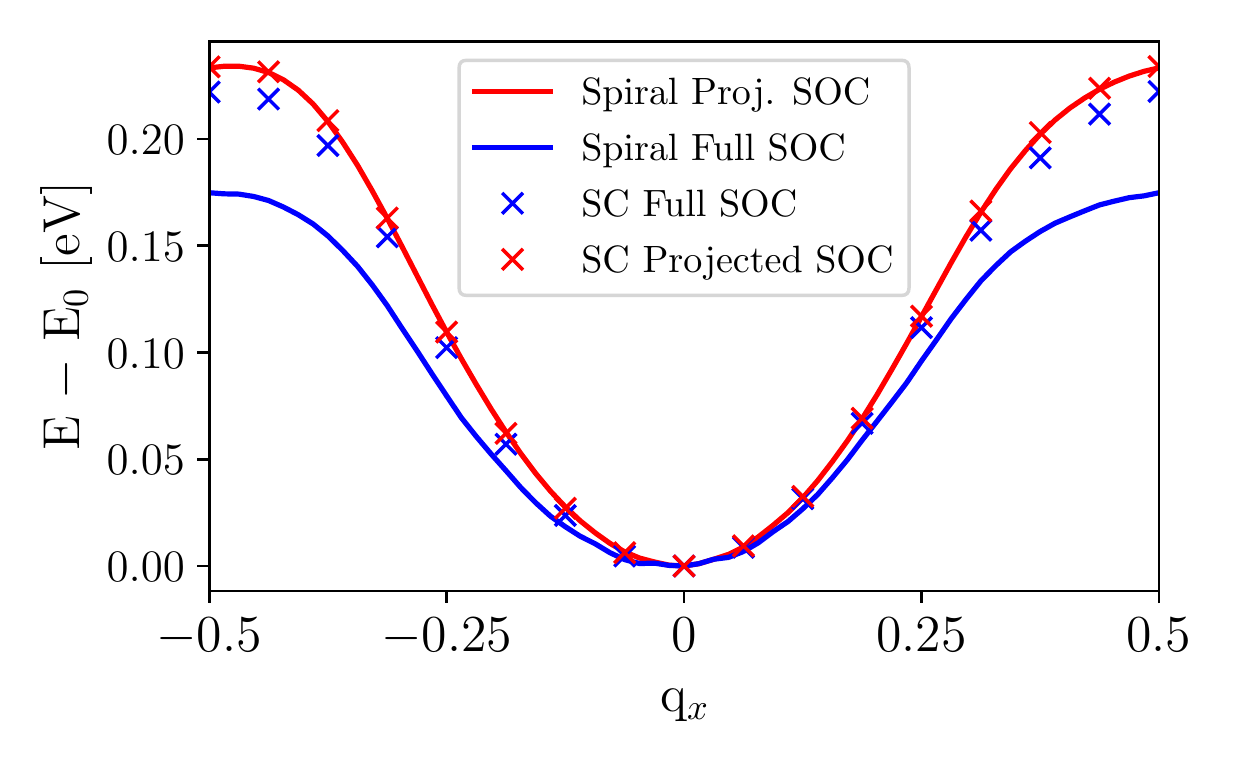}
    \caption{Comparison between GBT spin spiral calculations and supercell calculations with projected and full spinorbit coupling in monolayer CoPt.}
    \label{fig:CoPt_energy}
\end{figure}
\subsection{Benchmark}\label{sec:benchmark}
The LDA implementation of the GBT have been tested by checking that our results agree with similar calculations from the literature and by verifying internal consistency by comparing with super cell calculations. The case of fcc Fe has been found to have a spin spiral ground state \cite{Y_Tsunoda_1987} and the calculation of the ordering vector $\mathbf{Q}$ has been become a standard benchmark for spin spiral implementations \cite{kurz2004ab}. In previous simulations the ordering vector was found to be rather sensitive to the lattice constant and in figure \ref{fig:fcciron} we show the spin spiral energies along the $\Gamma$XW path using the experimental lattice constant as well as the lattice constant which has been found to reproduce the experimental ordering vector \cite{marsman2002broken}. The calculated value of $\mathbf{Q}$ is in good agreement with previous reports in both cases \cite{garcia2004first}. We also confirm a similar low energy barrier between the two local minima, as is expected from LDA \cite{knopfle2000spin}.

In order to check internal consistency we have investigated the case of monolayer CoPt \cite{sandratskii2017insight} where we compare spin spiral energies calculated using the GBT with energies calculated from super cells. We thus construct a $16x1$ super cell of the CoPt monolayer and consider spirals with $\mathbf{q}_\mathrm{c}=(\frac{n}{16})$ in units of reciprocal lattice vectors. This allows us to extract 16 different spiral energies in the supercell using standard non-collinear DFT. In order to compare the two methods we have used a $k$-point grid of $16\times16\times1$ for the GBT and $1\times16\times1$ for the supercell and a plane wave cutoff of 700 eV for both calculations. In Fig. \ref{fig:CoPtsupercell} we compare the results without spinorbit coupling and find excellent agreement between supercell and GBT calculations. We note that when spinorbit coupling is neglected one has $E_{\mathbf{q}}=E_{-\mathbf{q}}$. 
%The spin spiral and supercell total energies agree when spin orbit is neglected as is seen in figure \ref{fig:CoPtsupercell}. We find that spin spiral energies interpolate well to longer commensurate spirals. Looking into the converged magnetic state, we also find very good agreement with the spin spiral primitive cell and the first cell in the supercell.

Since spinorbit coupling is incompatible with the GBT one has to resort to approximate schemes to include it in the calculations. In the present work we have used the PSO method proposed by Sandratskii \cite{sandratskii2017insight}. In Fig. \ref{fig:CoPt_energy} we compare spin spiral calculations with supercell calculations where the spinorbit coupling has been included either fully or by the PSO method. The PSO method is fully compatible with the GBT and we find excellent agreement between the spin spiral energies calculated with the GBT and with supercells. The PSO approach is, however an approximation and the correct result can only be obtained from the supercell using the full spinorbit coupling. We see that the PSO calculations are in good agreement with those obtained from full spinorbit coupling but overestimates the energies at the Brillouin zone boundary by a few percent. In contrast, if one tries to include the full spinorbit operator in the GBT calculations (by diagonalizing $H^\mathrm{KS}$ including spinorbit coupling on a basis of GBT eigenstates without spinorbit coupling) the energies are severely underestimated with respect to the exact result (from the supercell calculation). We note that the spiral energies including spinorbit coupling shows a slight asymmetry between points at $q$ and $-q$, which can be related to the Dzyaloshinskii-Moriya interactions in the system \cite{sandratskii2017insight}.

\newpage
\bibliography{bibliography.bib}
\newpage
\onecolumngrid
\setcounter{section}{0}

\section*{Supplementary Information - Type II multiferroic order in two-dimensional transition metal halides from first principles spin-spiral calculations}

\subsection*{\normalfont Joachim Sødequist$^1$ and Thomas Olsen$^{1, *}$}
\vspace{-4mm}
\subsubsection*{$^1$CAMD, Department of Physics, Technical University of Denmark, 2820 Kgs. Lyngby Denmark}
\setcounter{section}{0}
\setcounter{figure}{0}

\section{Spin spiral dispersions}
\begin{figure*}[h!]
    \centering
    \includegraphics[width=0.32\textwidth]{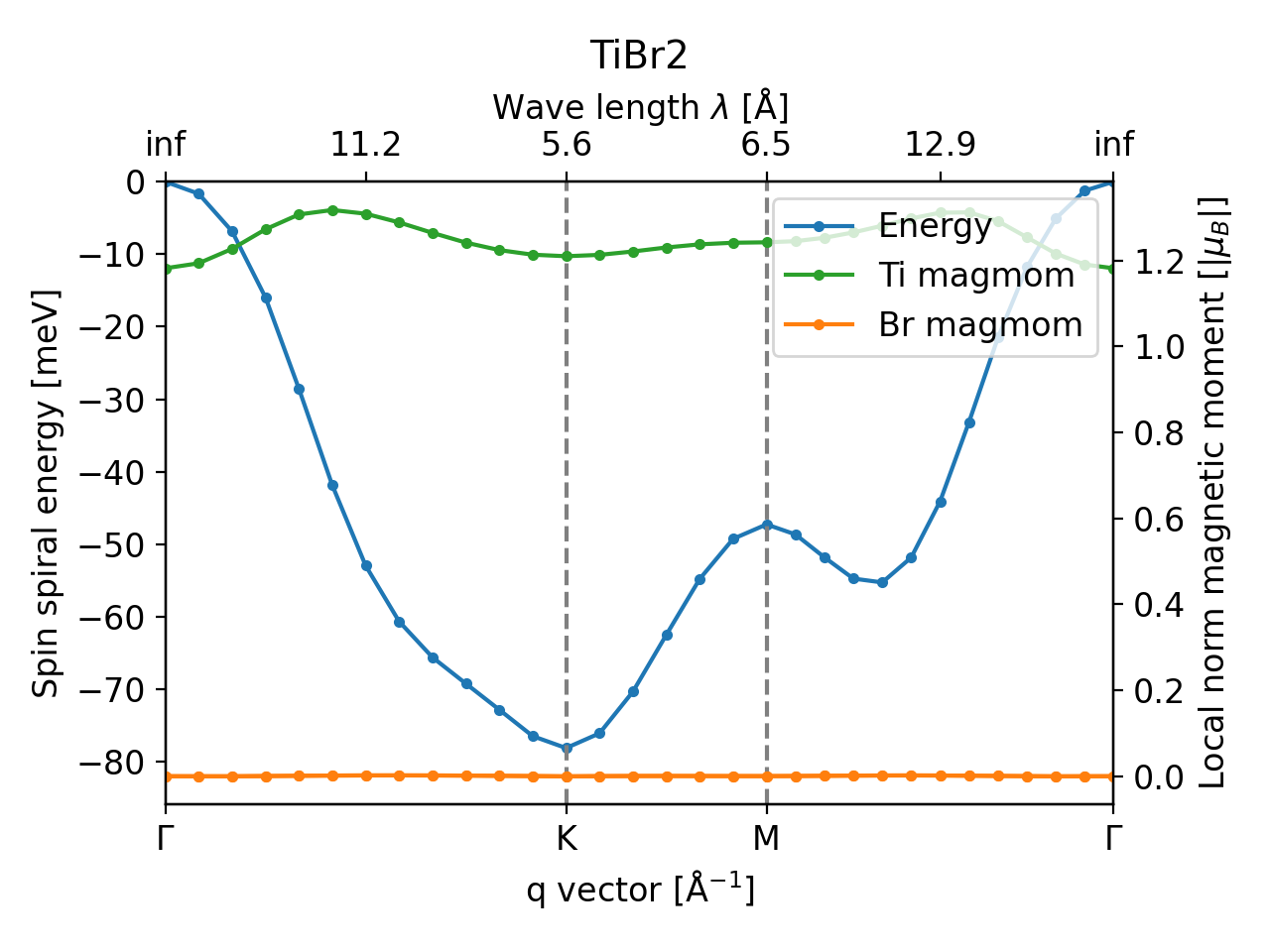}
    \includegraphics[width=0.32\textwidth]{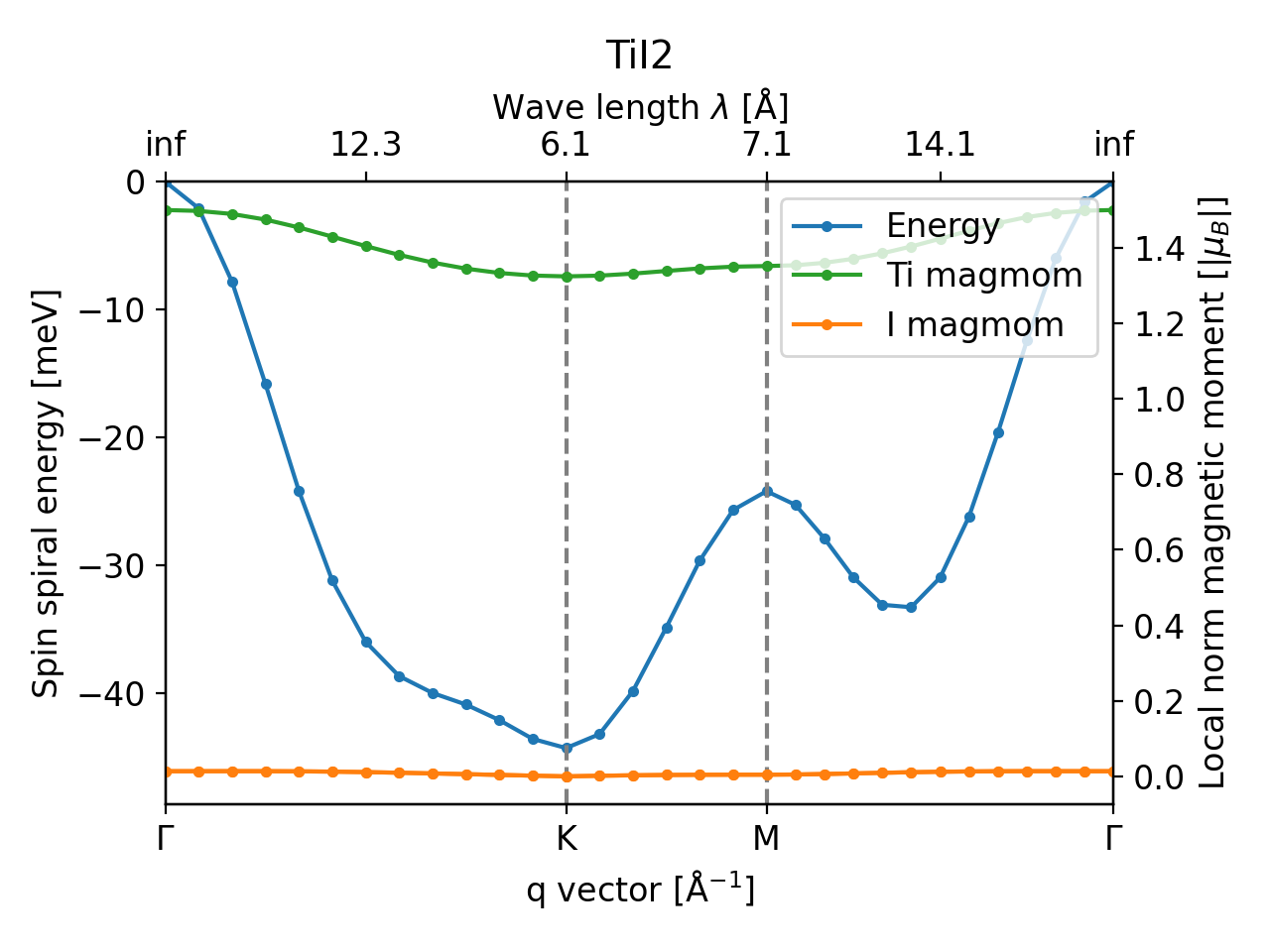}
    \includegraphics[width=0.32\textwidth]{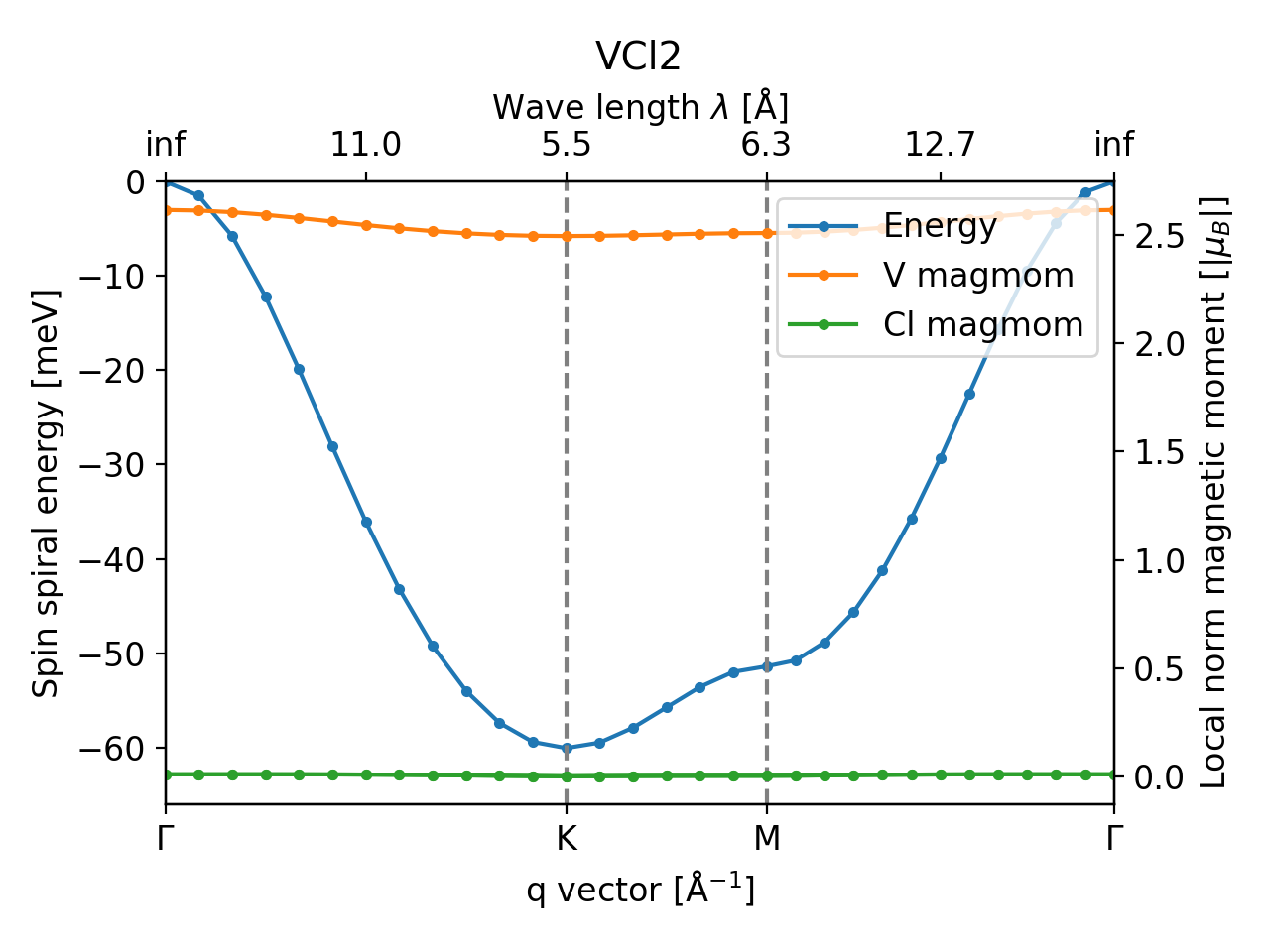}
    \includegraphics[width=0.32\textwidth]{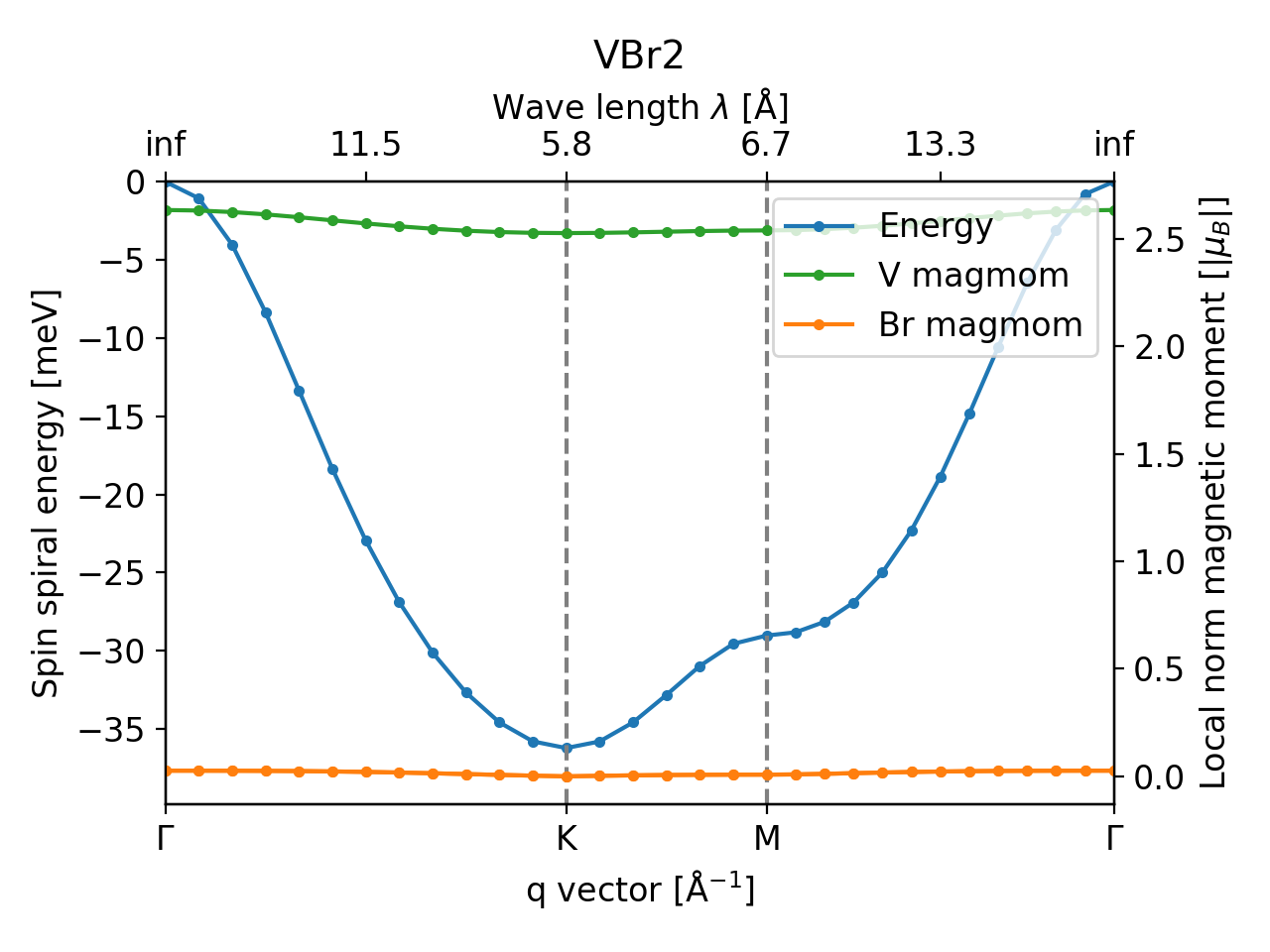}
    \includegraphics[width=0.32\textwidth]{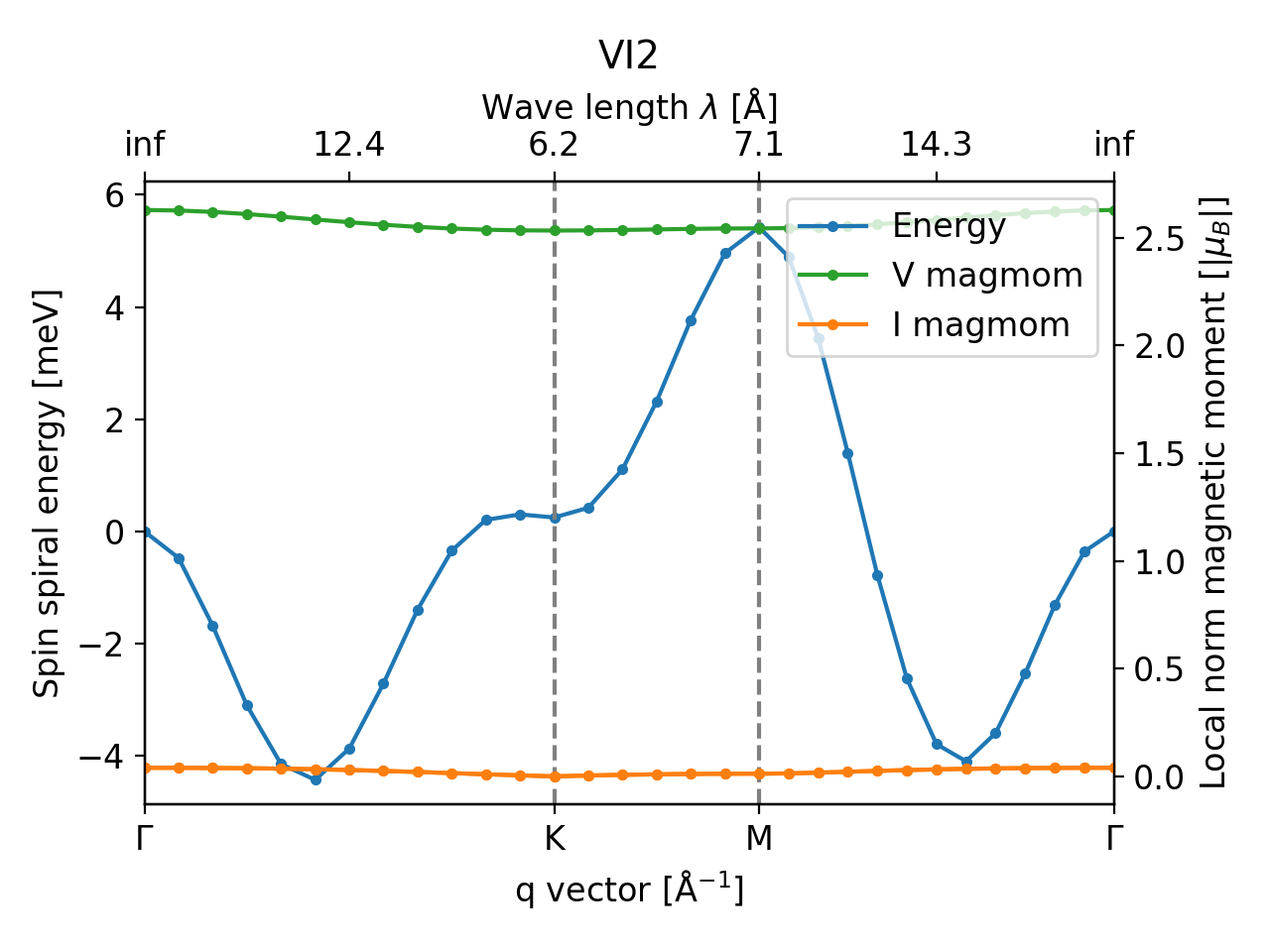}
    \includegraphics[width=0.32\textwidth]{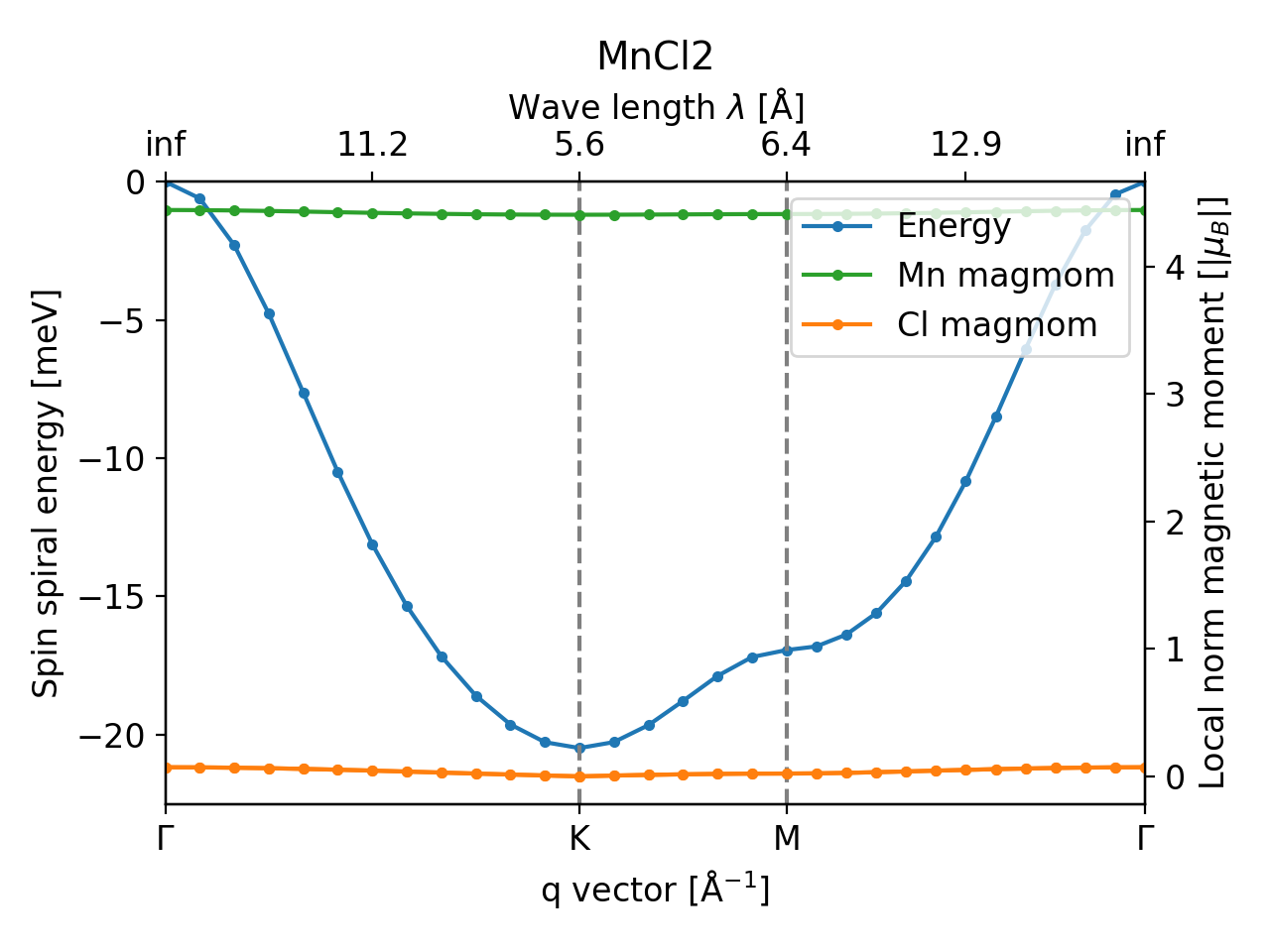}
    \includegraphics[width=0.32\textwidth]{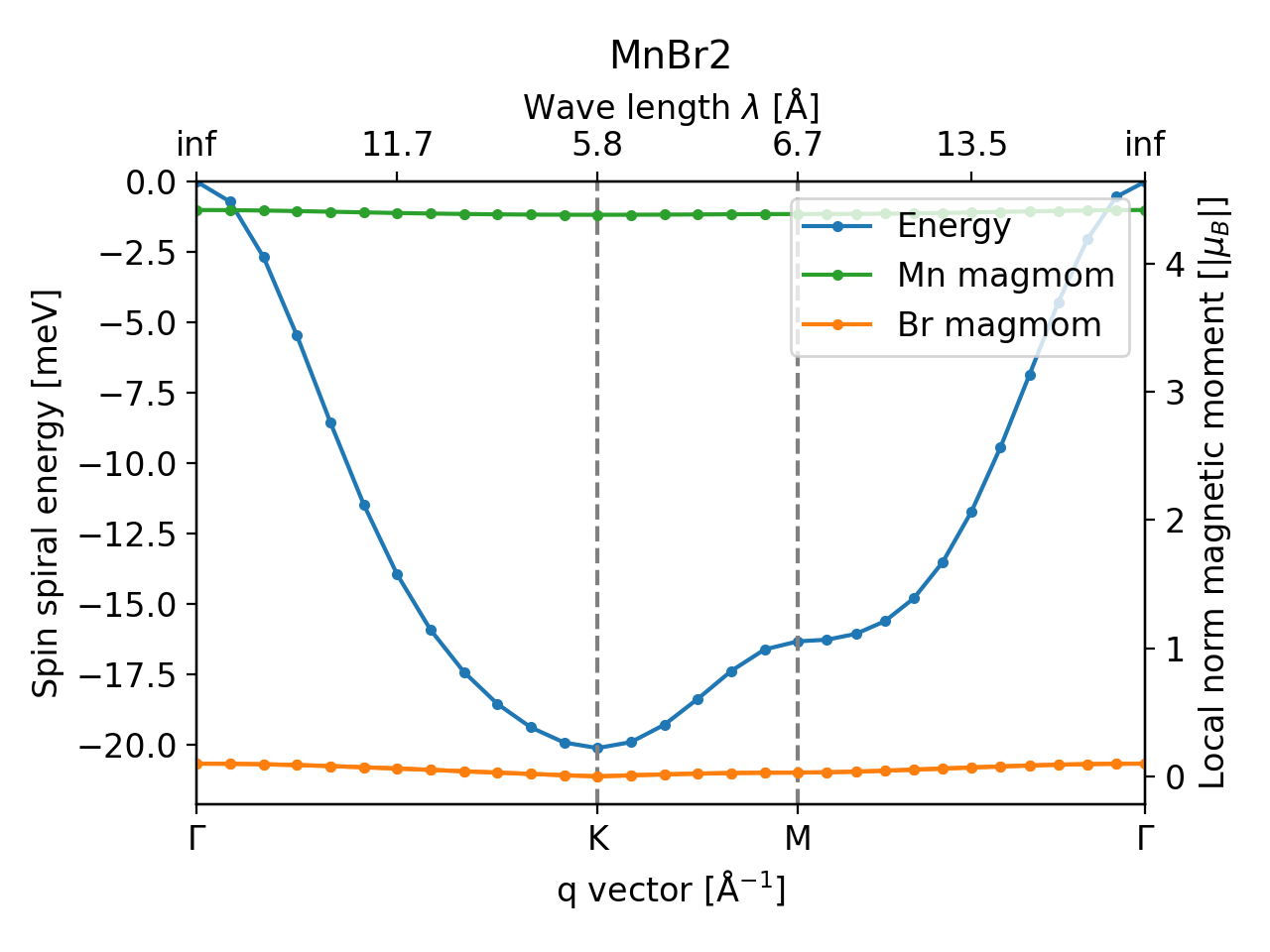}
    \includegraphics[width=0.32\textwidth]{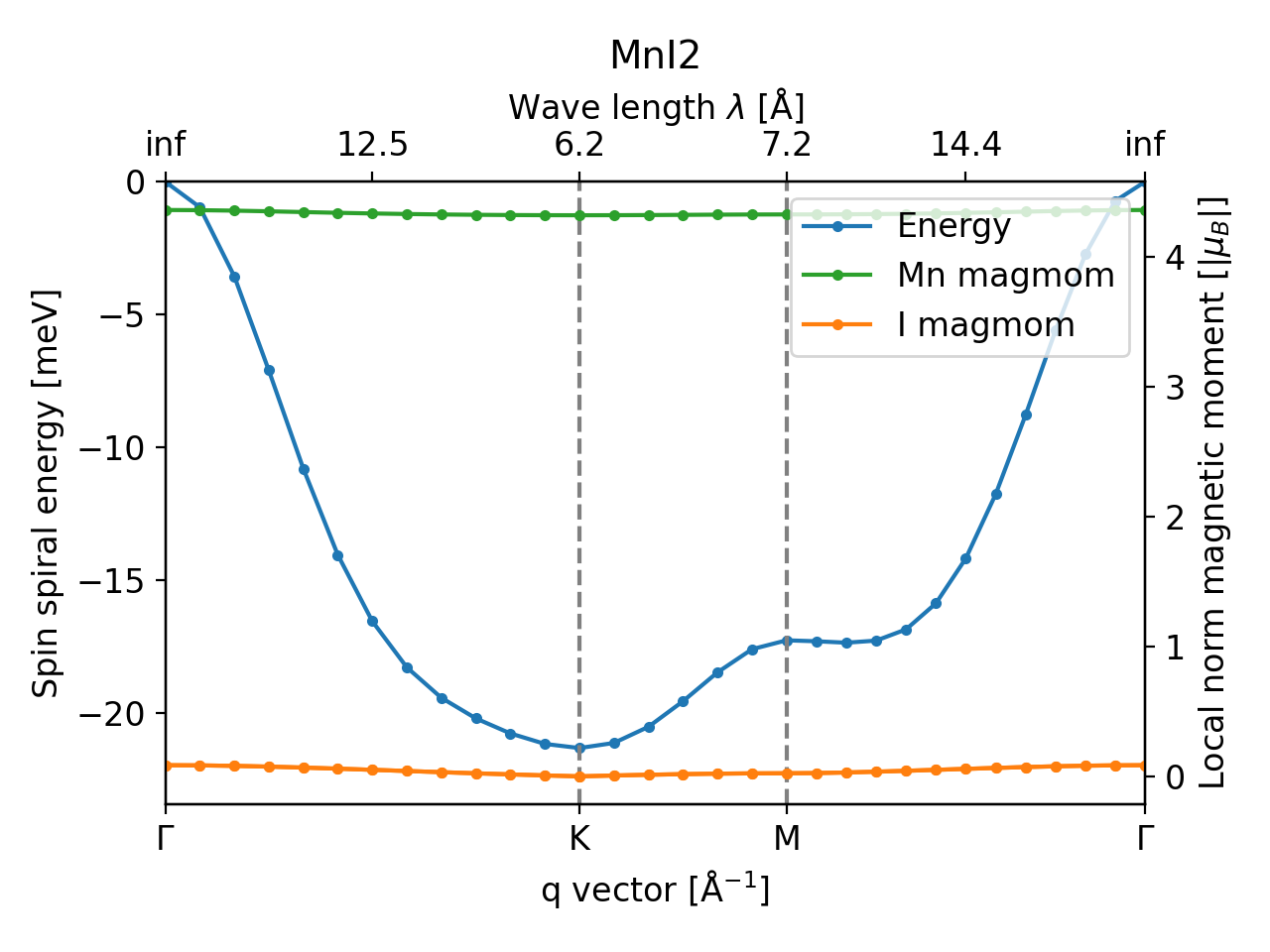}
    \includegraphics[width=0.32\textwidth]{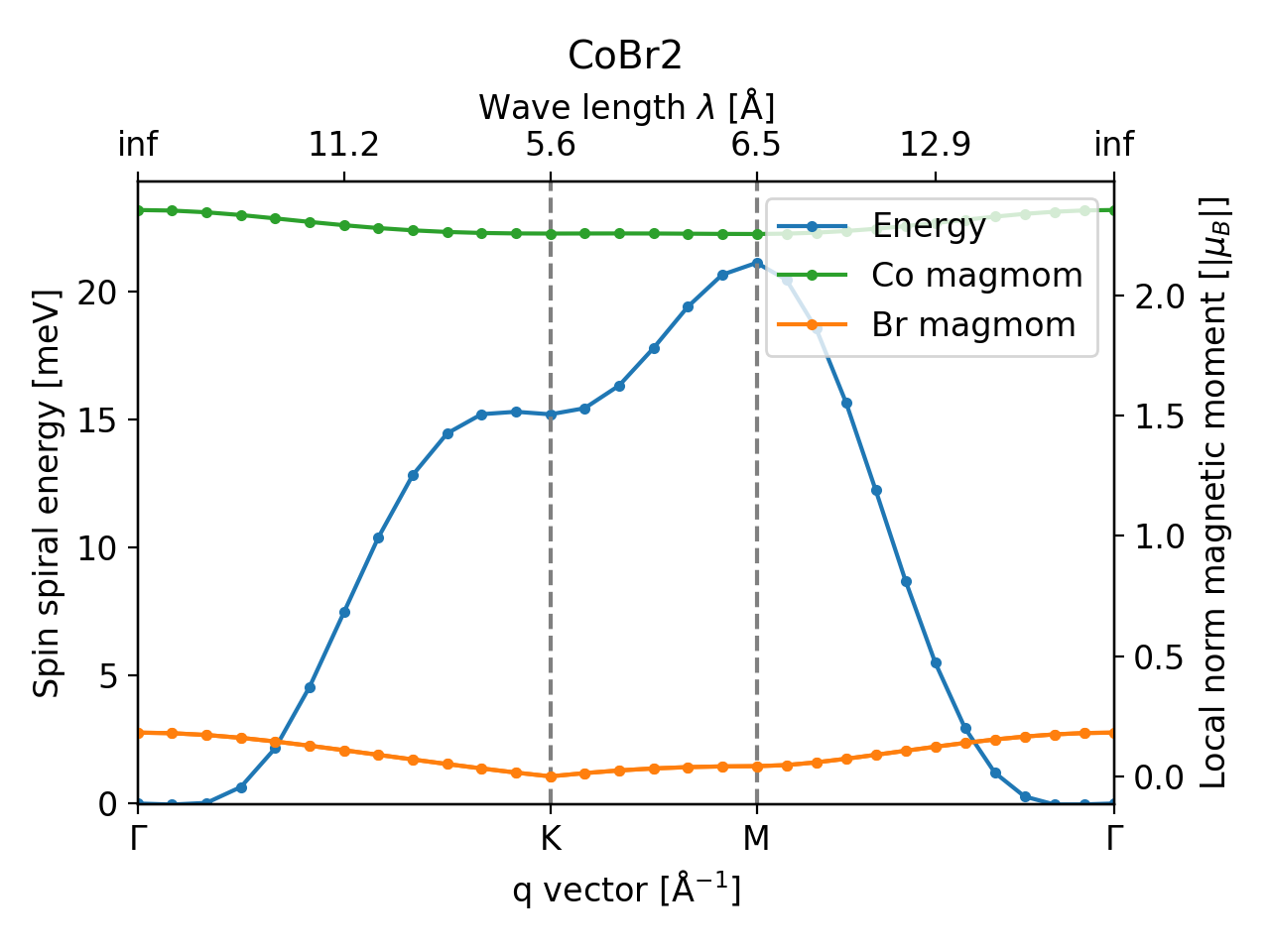}
    \includegraphics[width=0.32\textwidth]{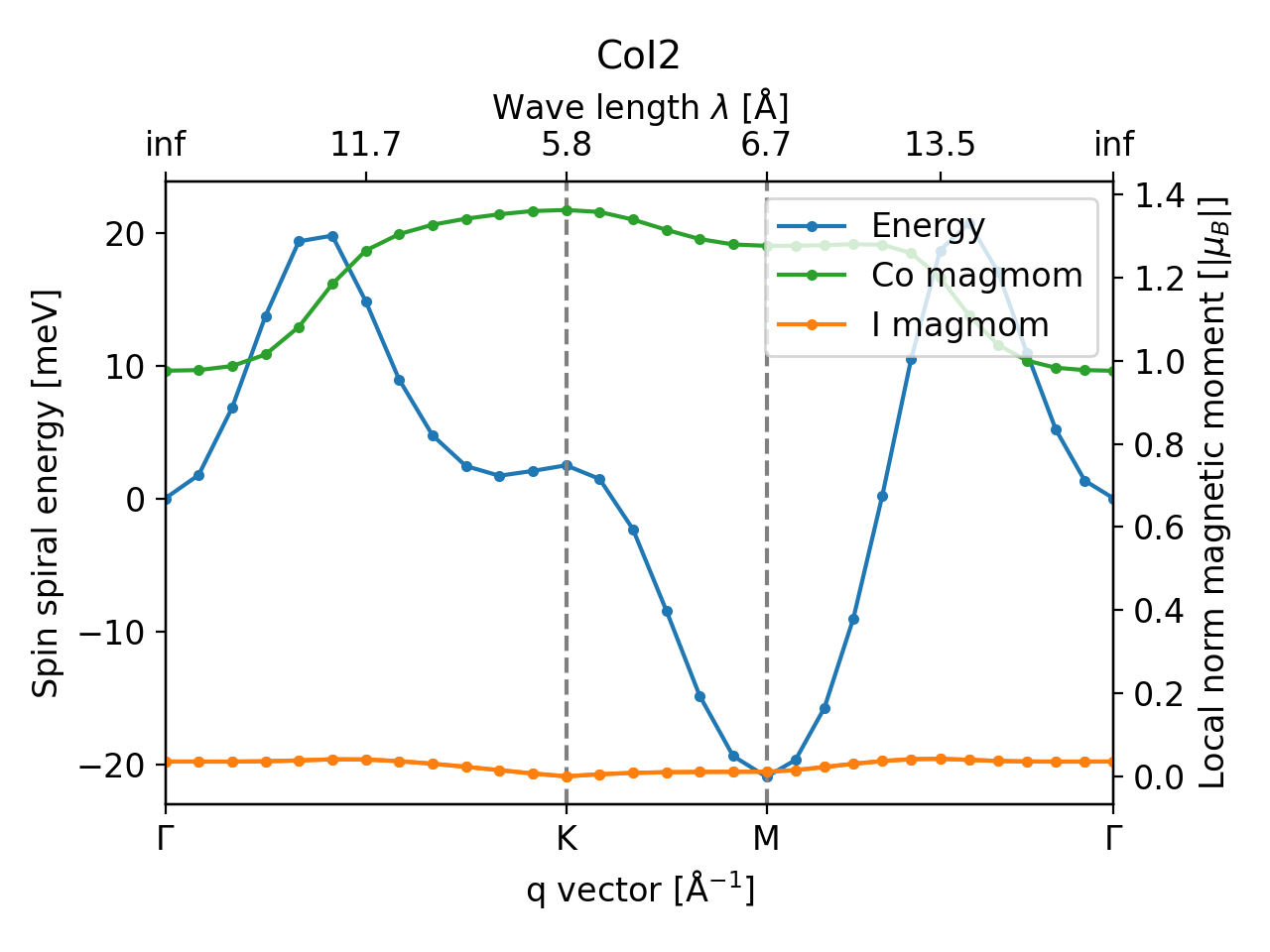}
    \includegraphics[width=0.32\textwidth]{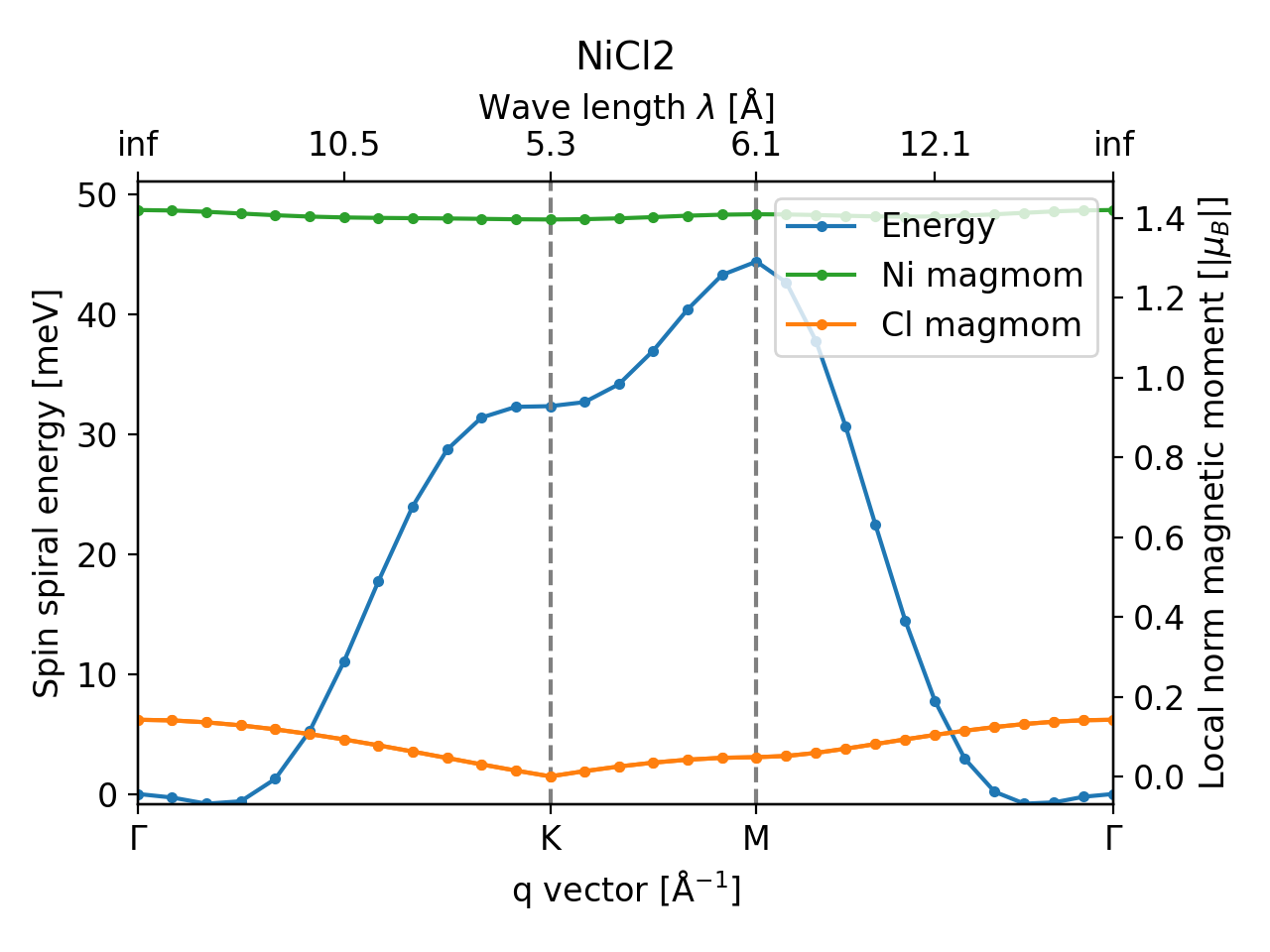}
    \includegraphics[width=0.32\textwidth]{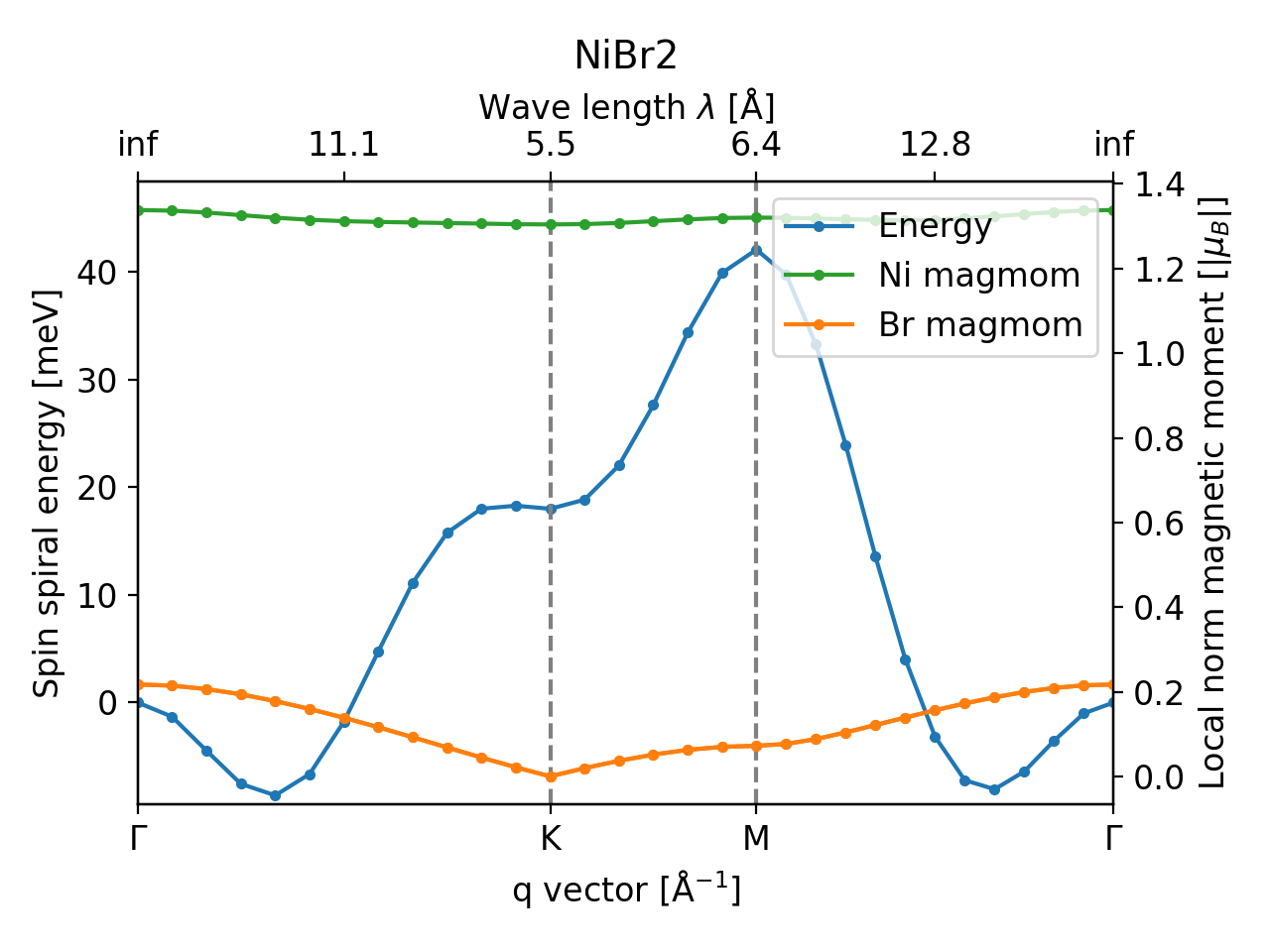}
    \includegraphics[width=0.32\textwidth]{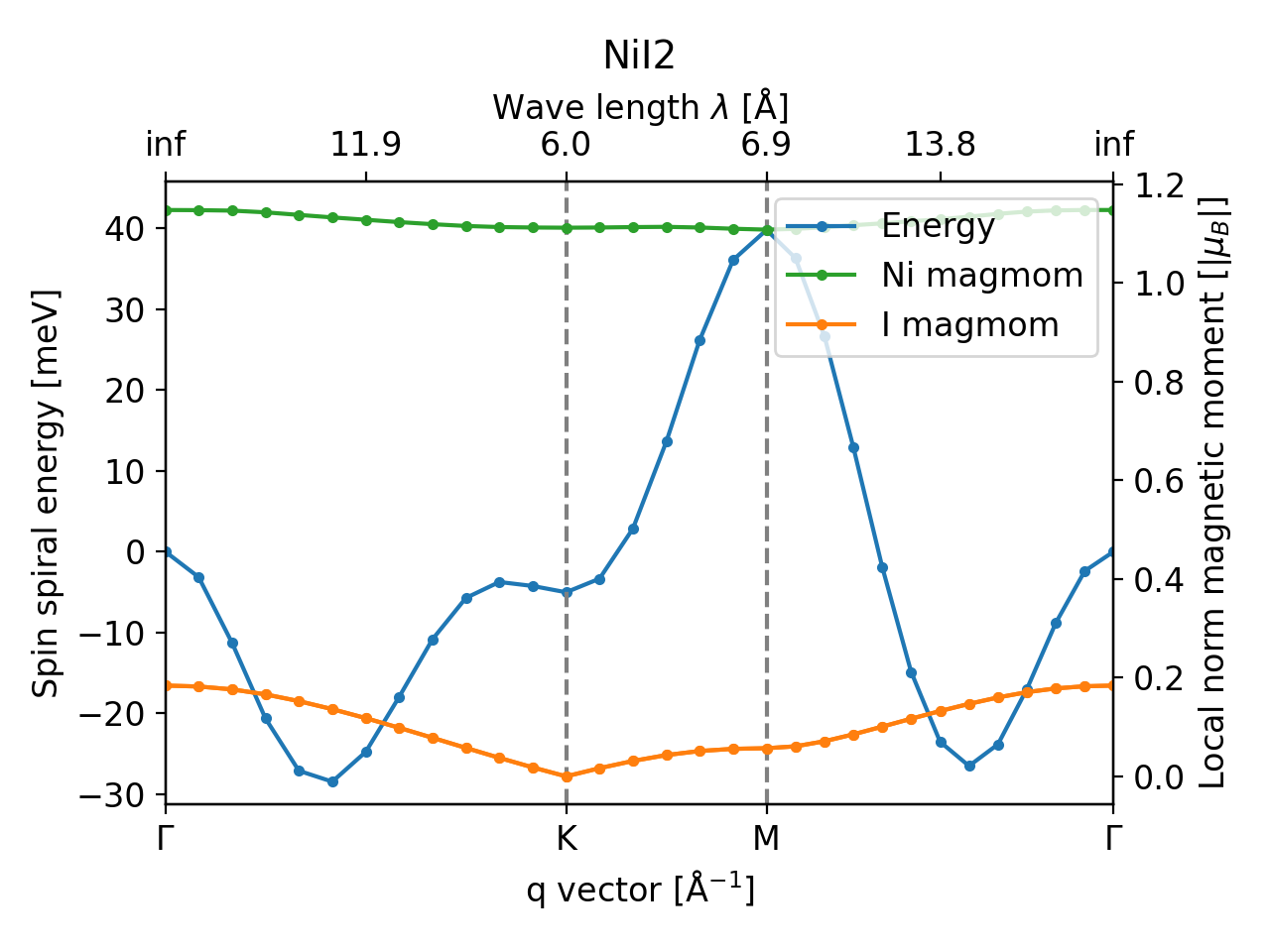}
    \caption{Spin spiral energies for AB$_2$ magnets and the local magnetic moments of the magnetic atoms and ligands. For ferromagnetic refer to Fig. \ref{fig:3}.}
    \label{fig:1}
\end{figure*}

\begin{figure*}[h!]
    \centering
    \includegraphics[width=0.32\textwidth]{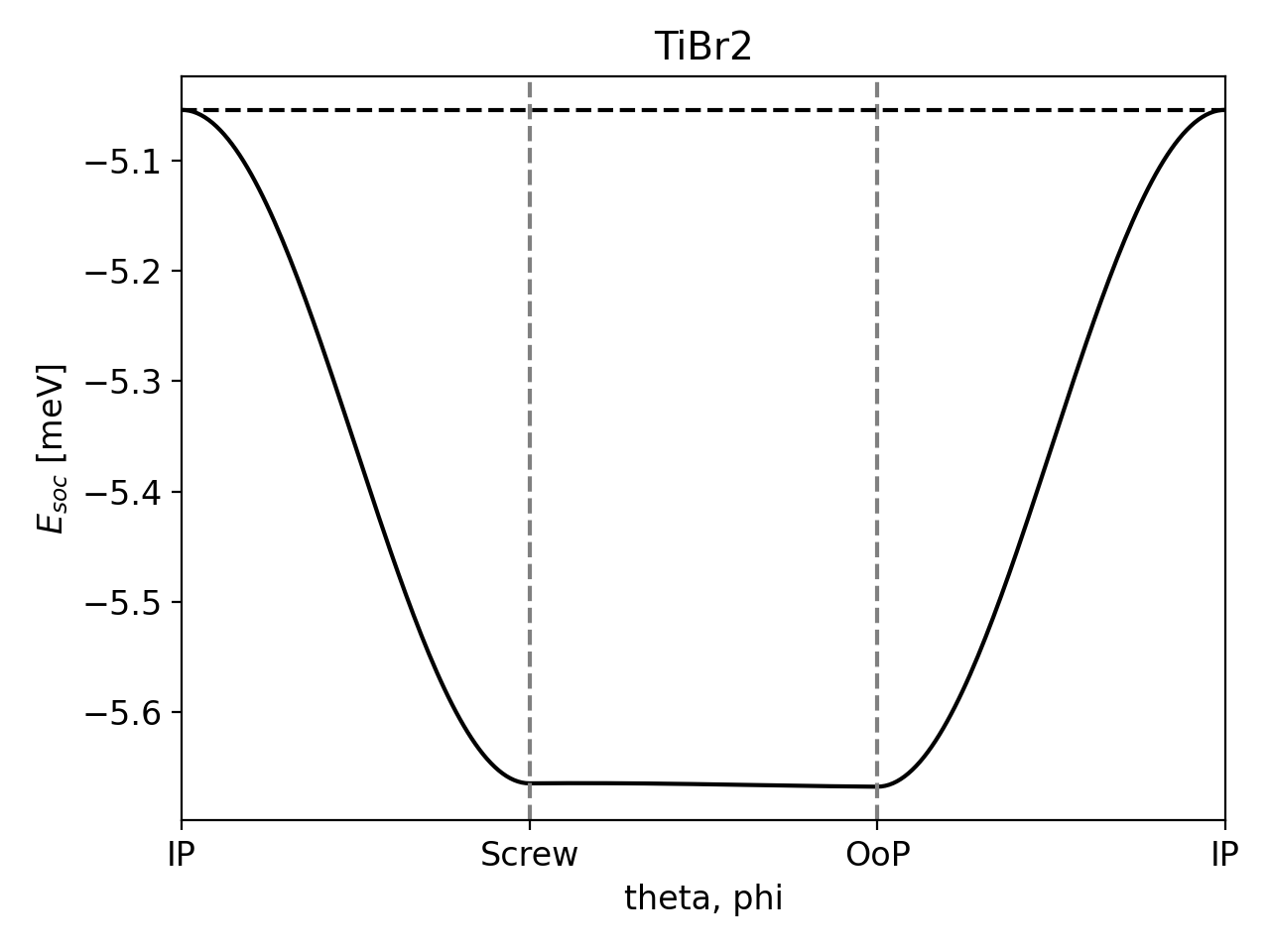}
    \includegraphics[width=0.32\textwidth]{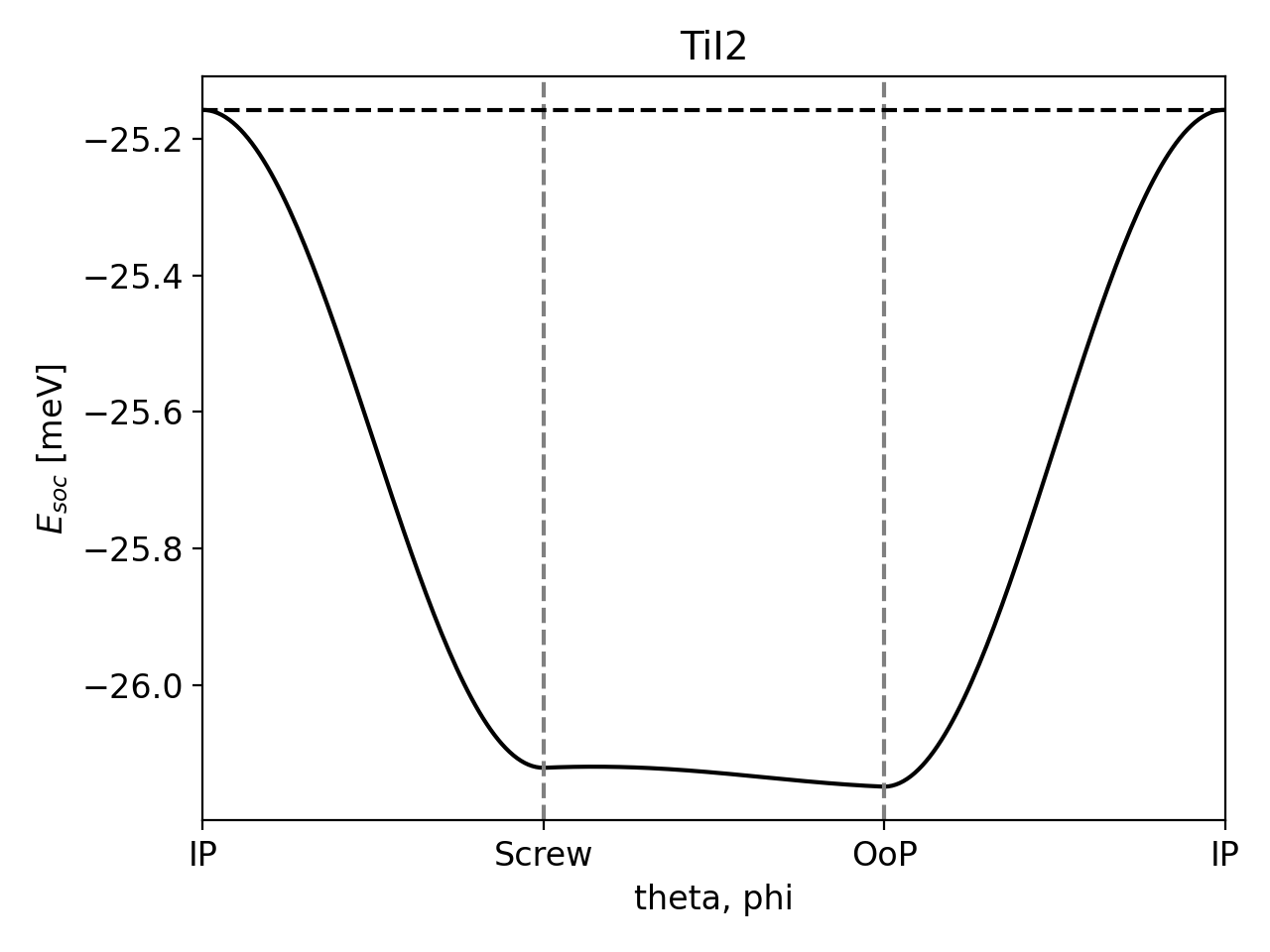}
    \includegraphics[width=0.32\textwidth]{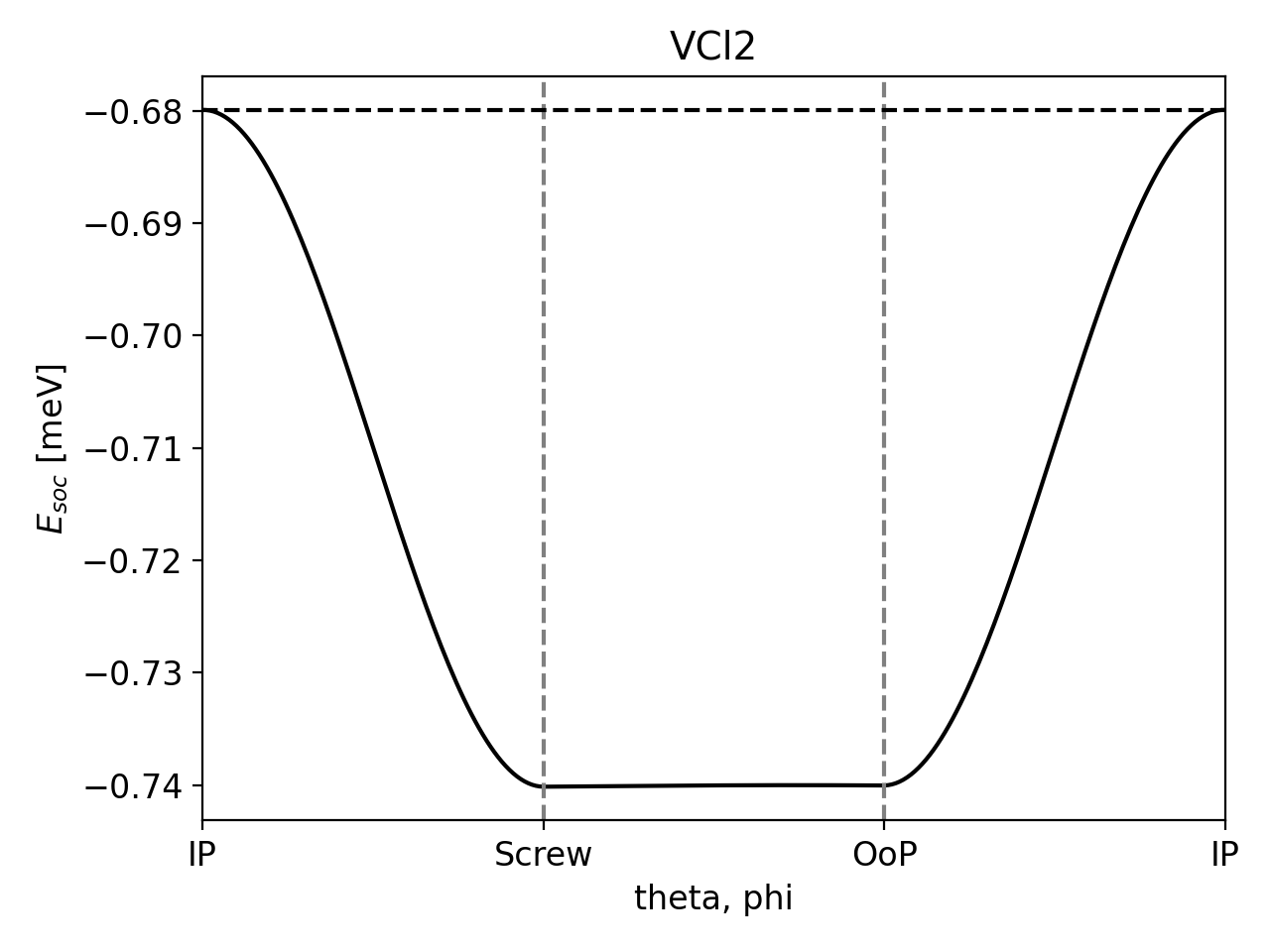}
    \includegraphics[width=0.32\textwidth]{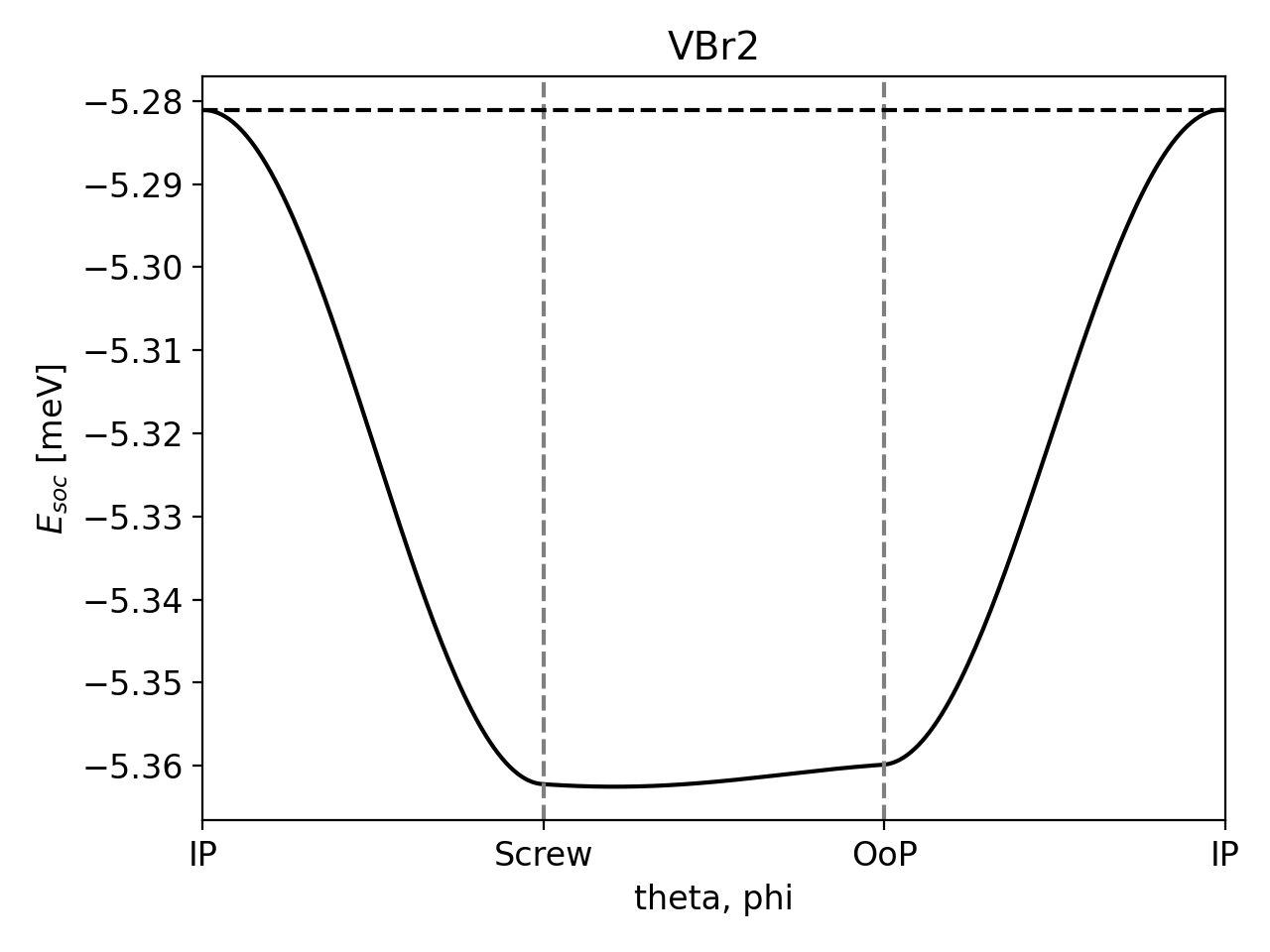}
    \includegraphics[width=0.32\textwidth]{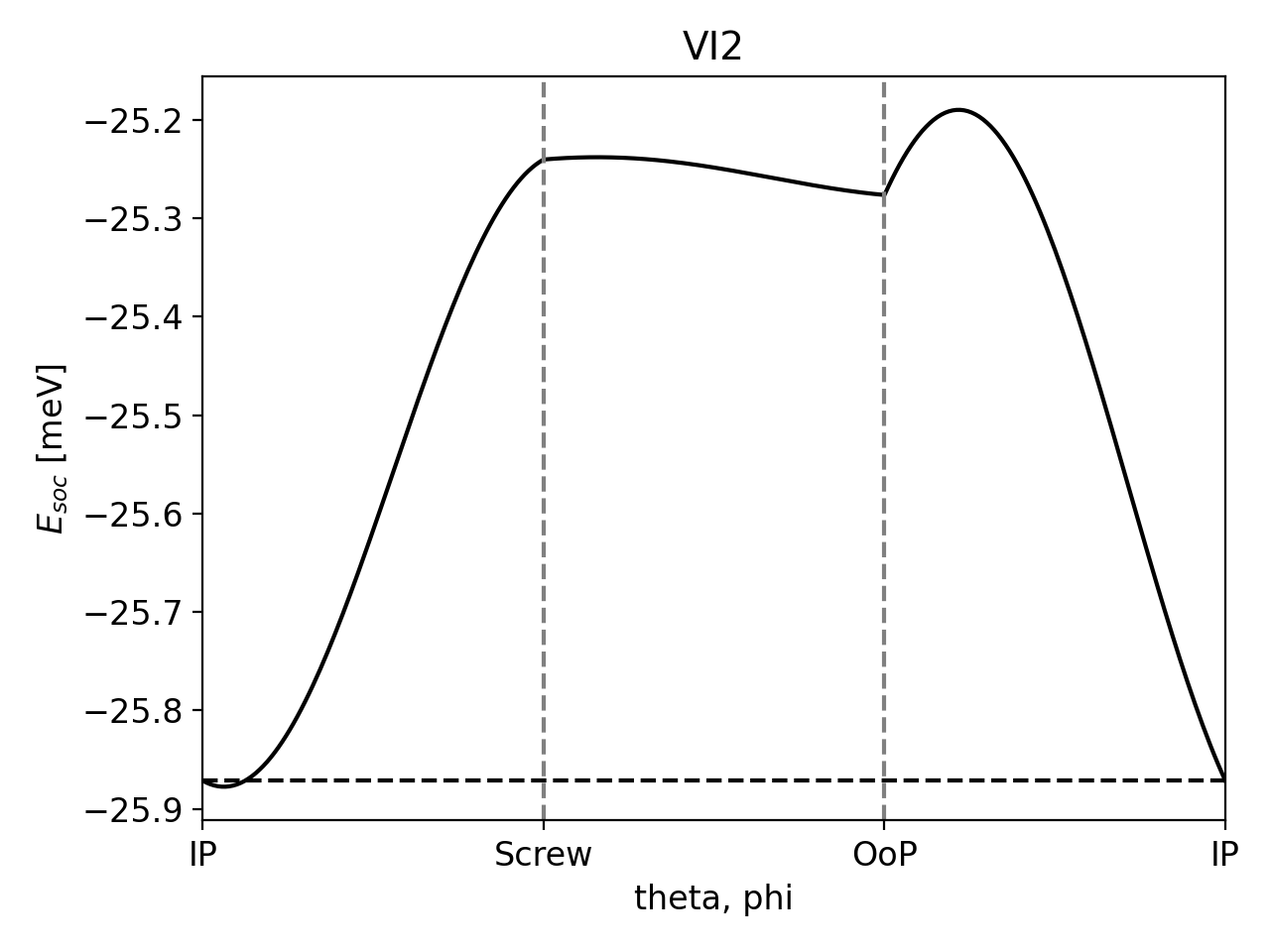}
    \includegraphics[width=0.32\textwidth]{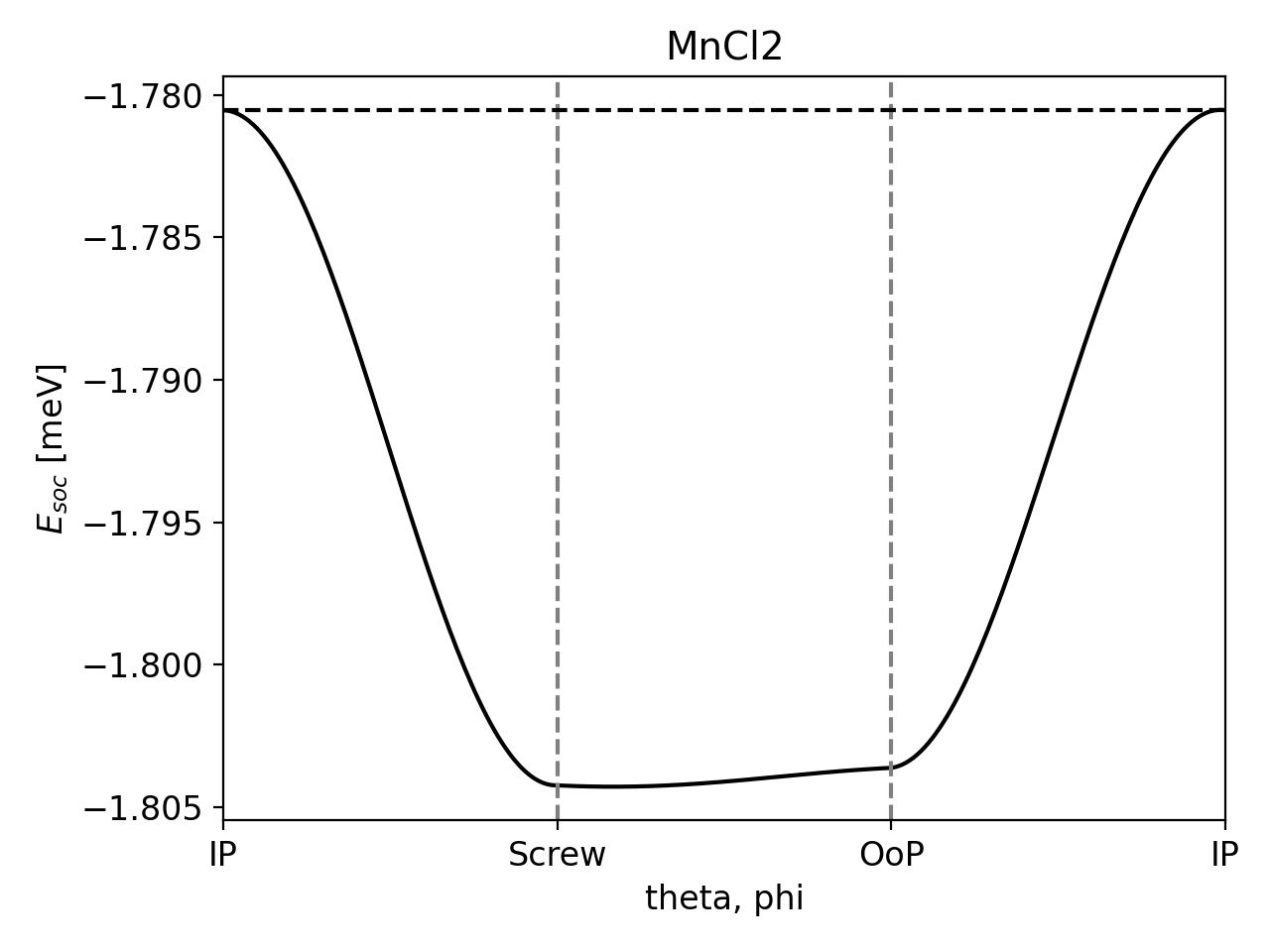}
    \includegraphics[width=0.32\textwidth]{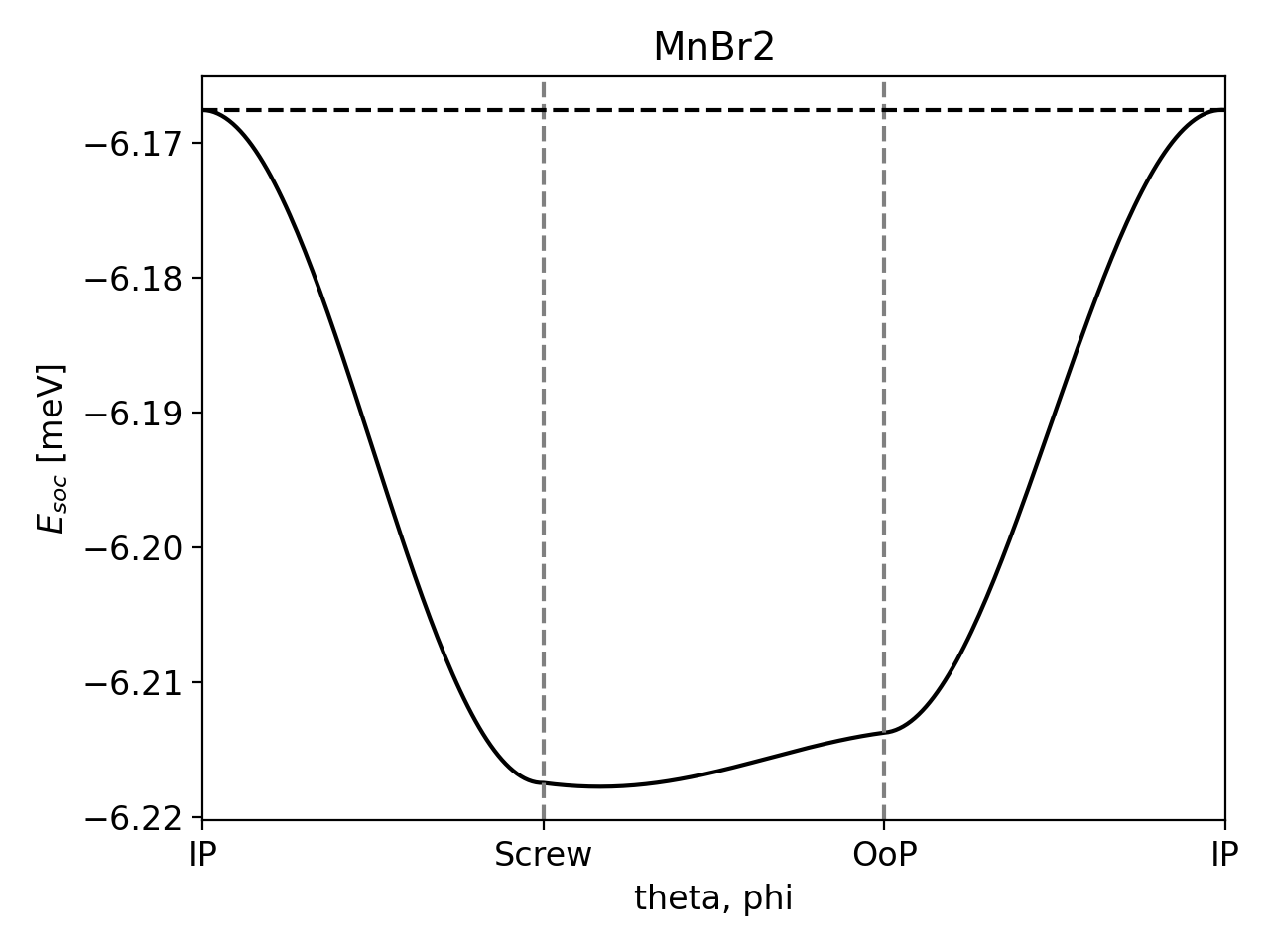}
    \includegraphics[width=0.32\textwidth]{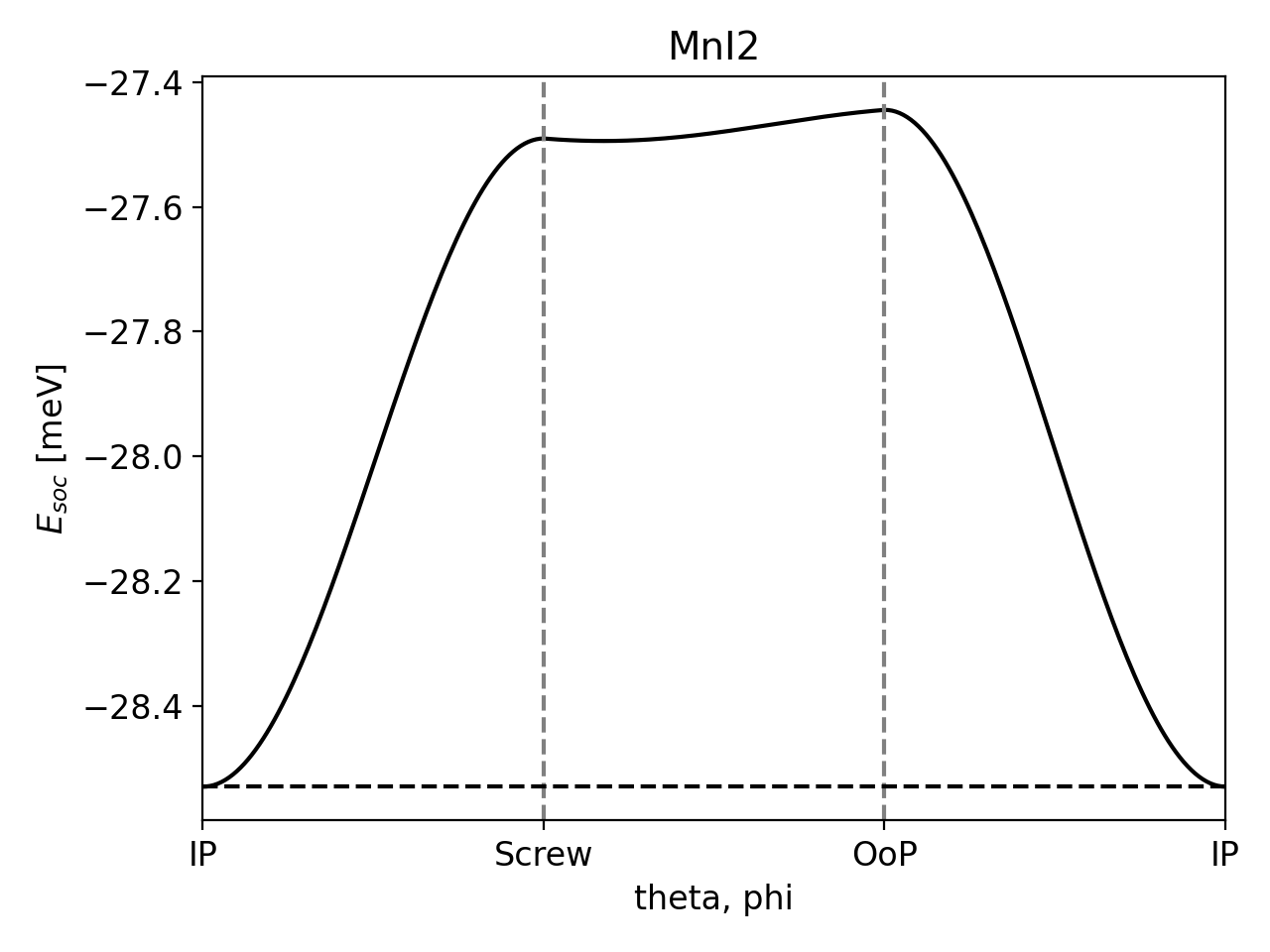}
    \includegraphics[width=0.32\textwidth]{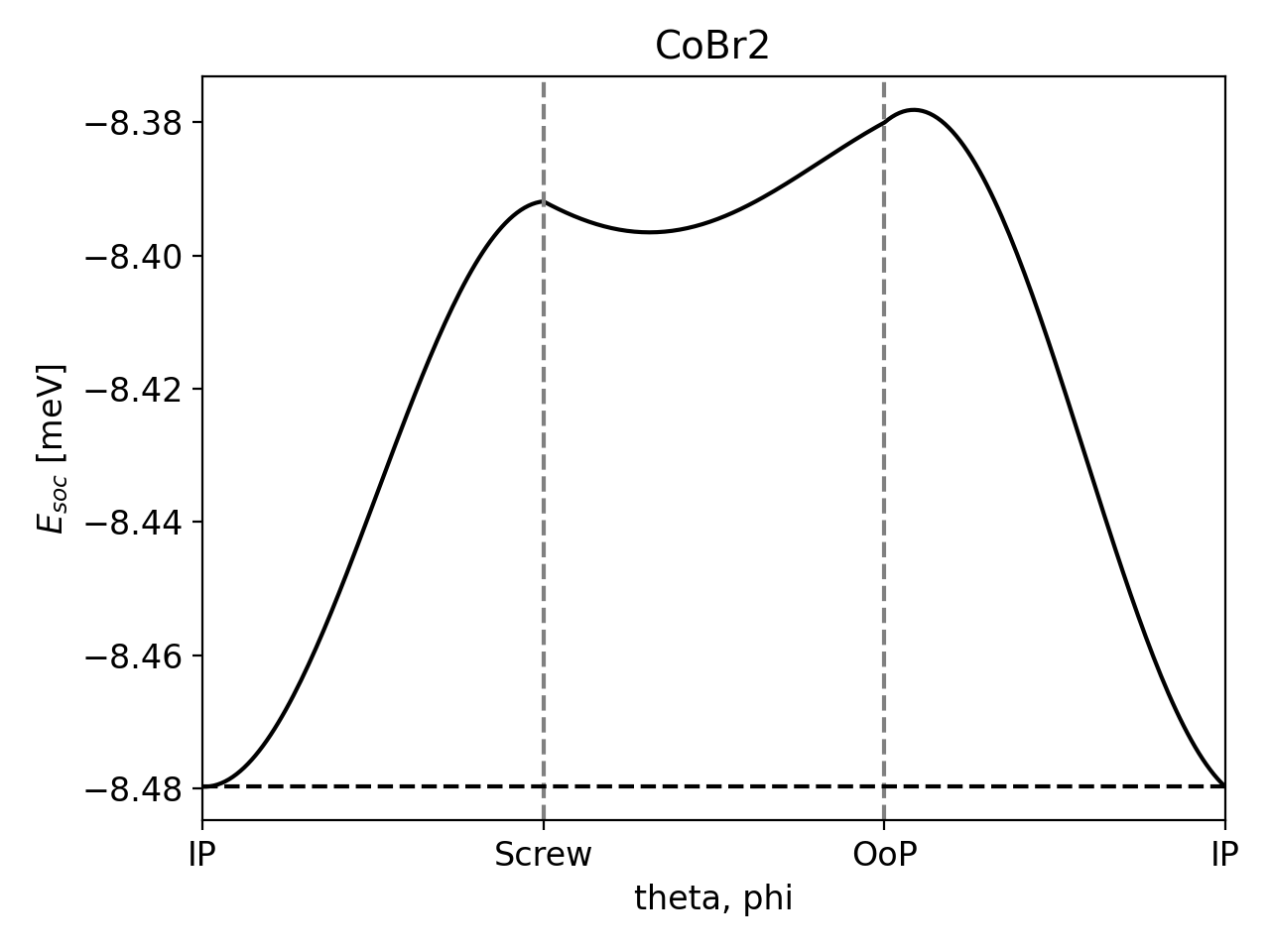}
    \includegraphics[width=0.32\textwidth]{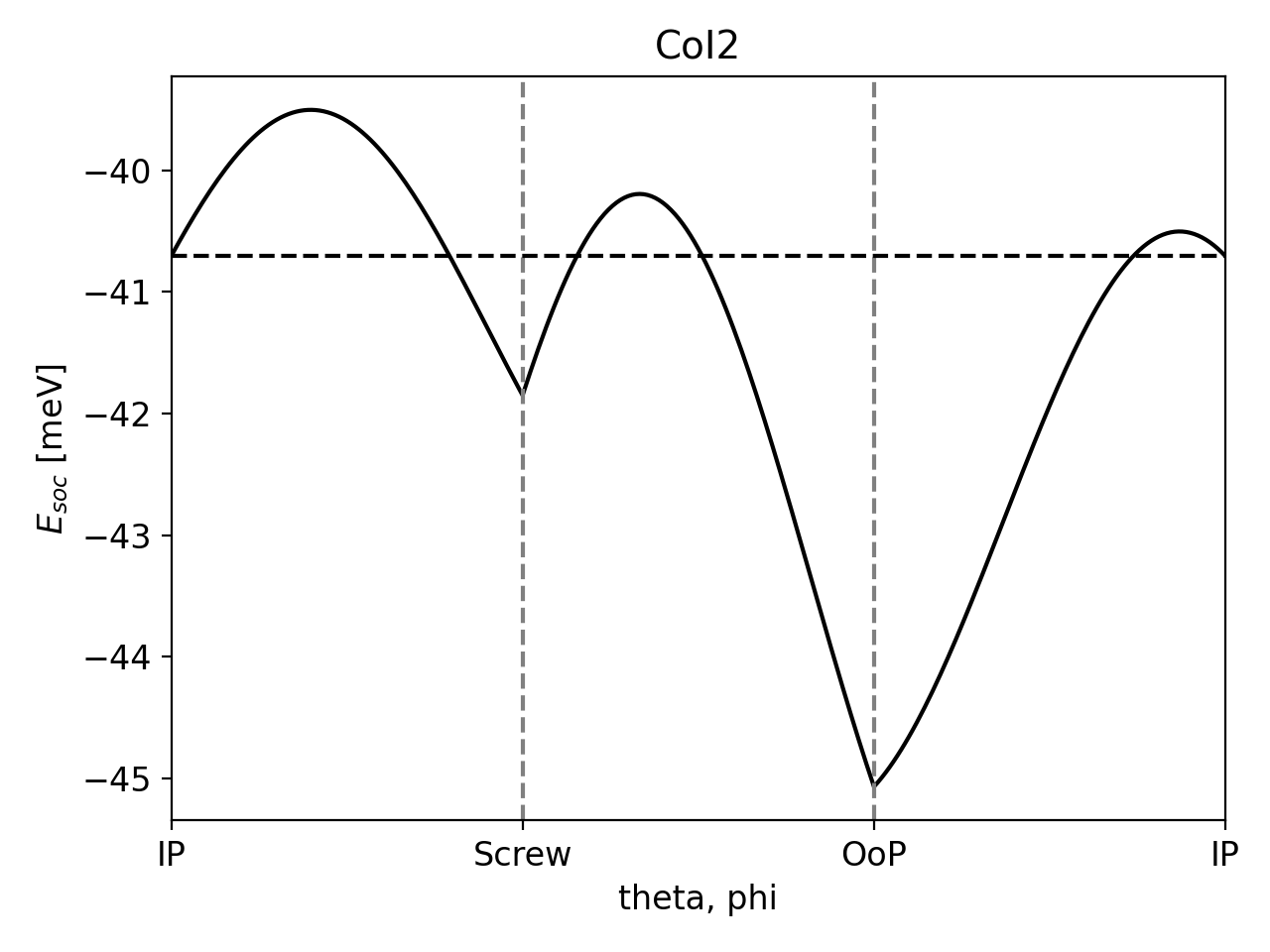}
    \includegraphics[width=0.32\textwidth]{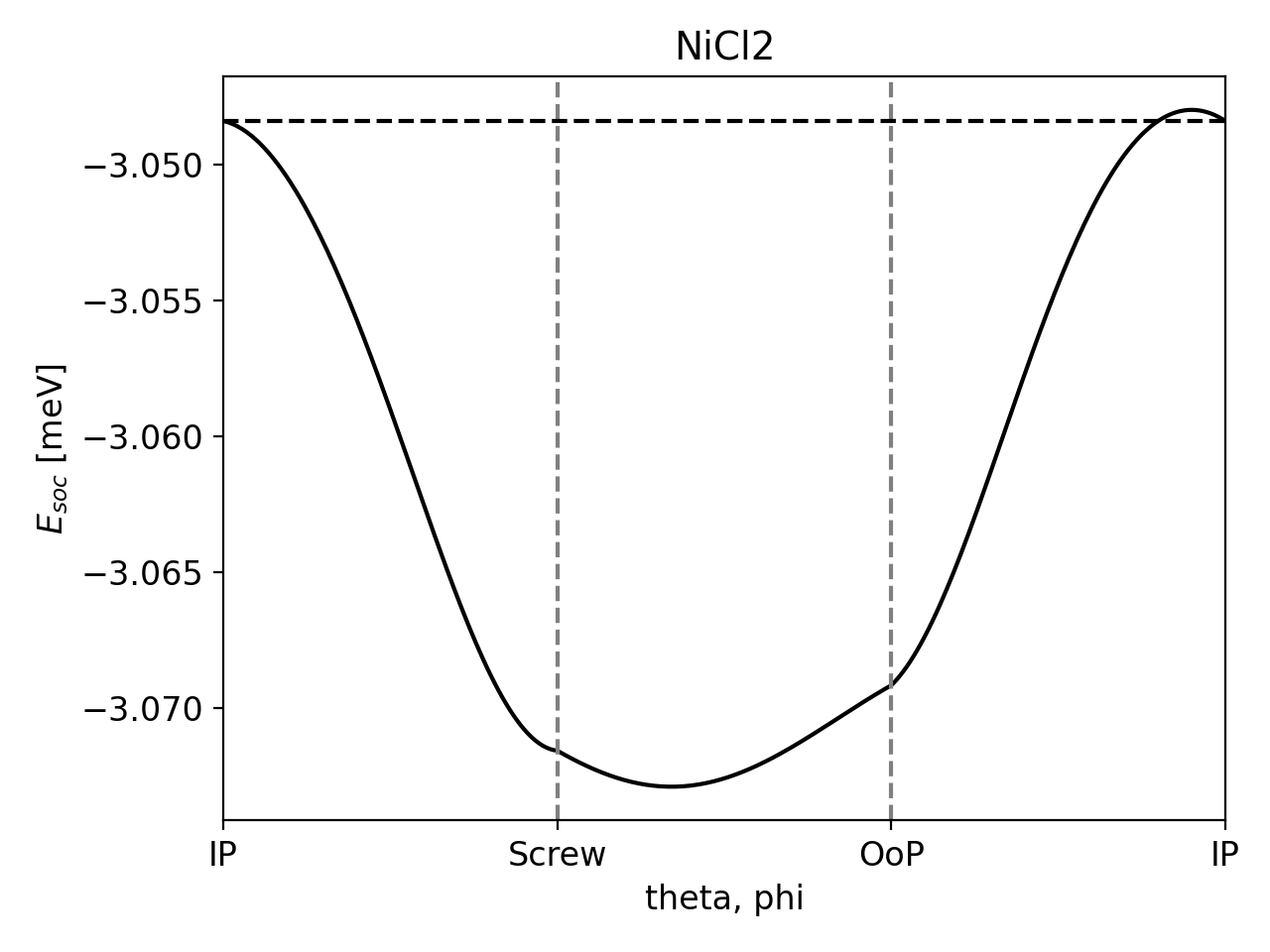}
    \includegraphics[width=0.32\textwidth]{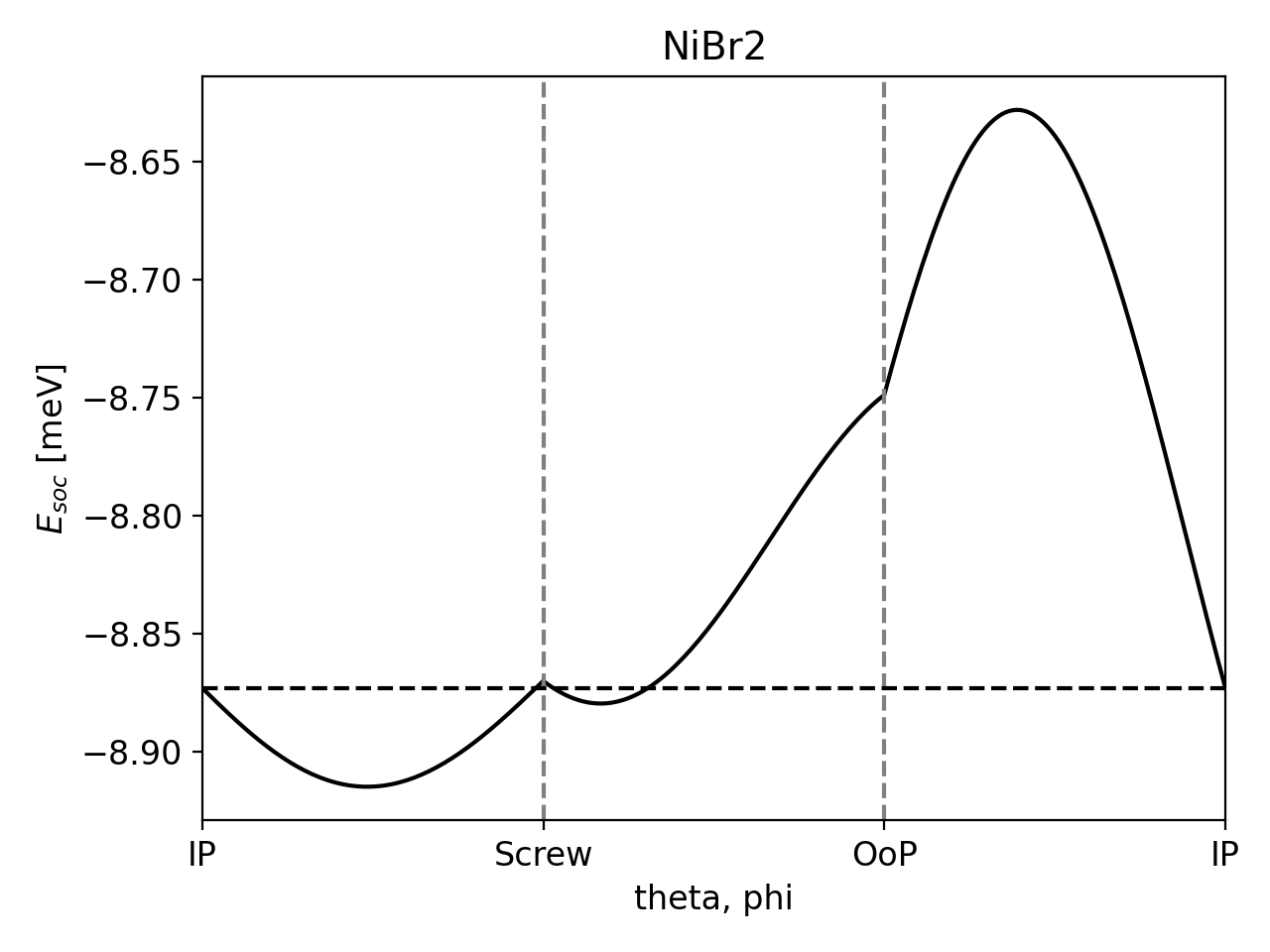}
    \includegraphics[width=0.32\textwidth]{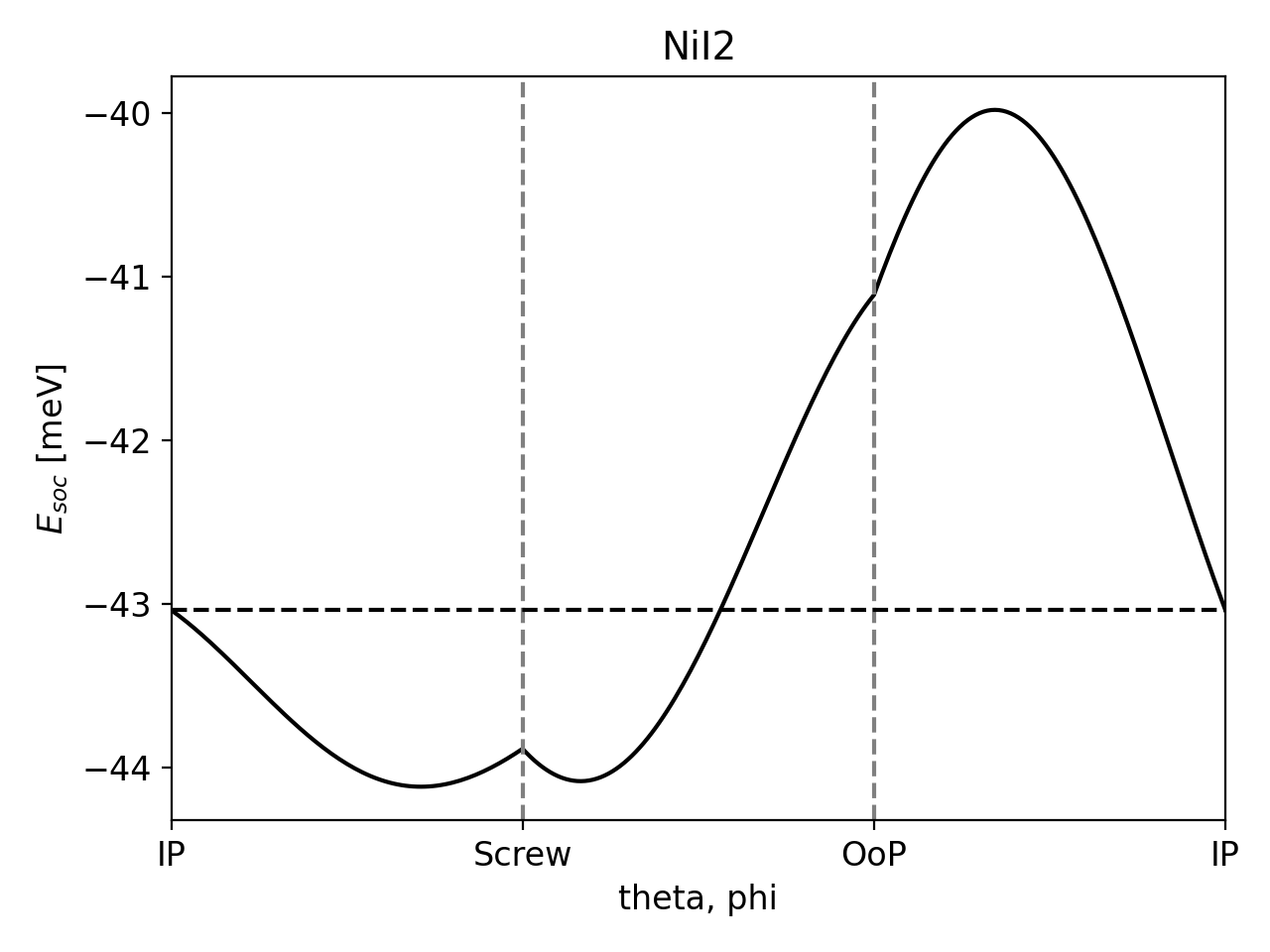}
    \caption{Projected spin orbit energies of the ground state found in Fig. \ref{fig:1}}
    \label{fig:2}
\end{figure*}

\begin{figure*}[h!]
    \centering
    \includegraphics[width=0.32\textwidth]{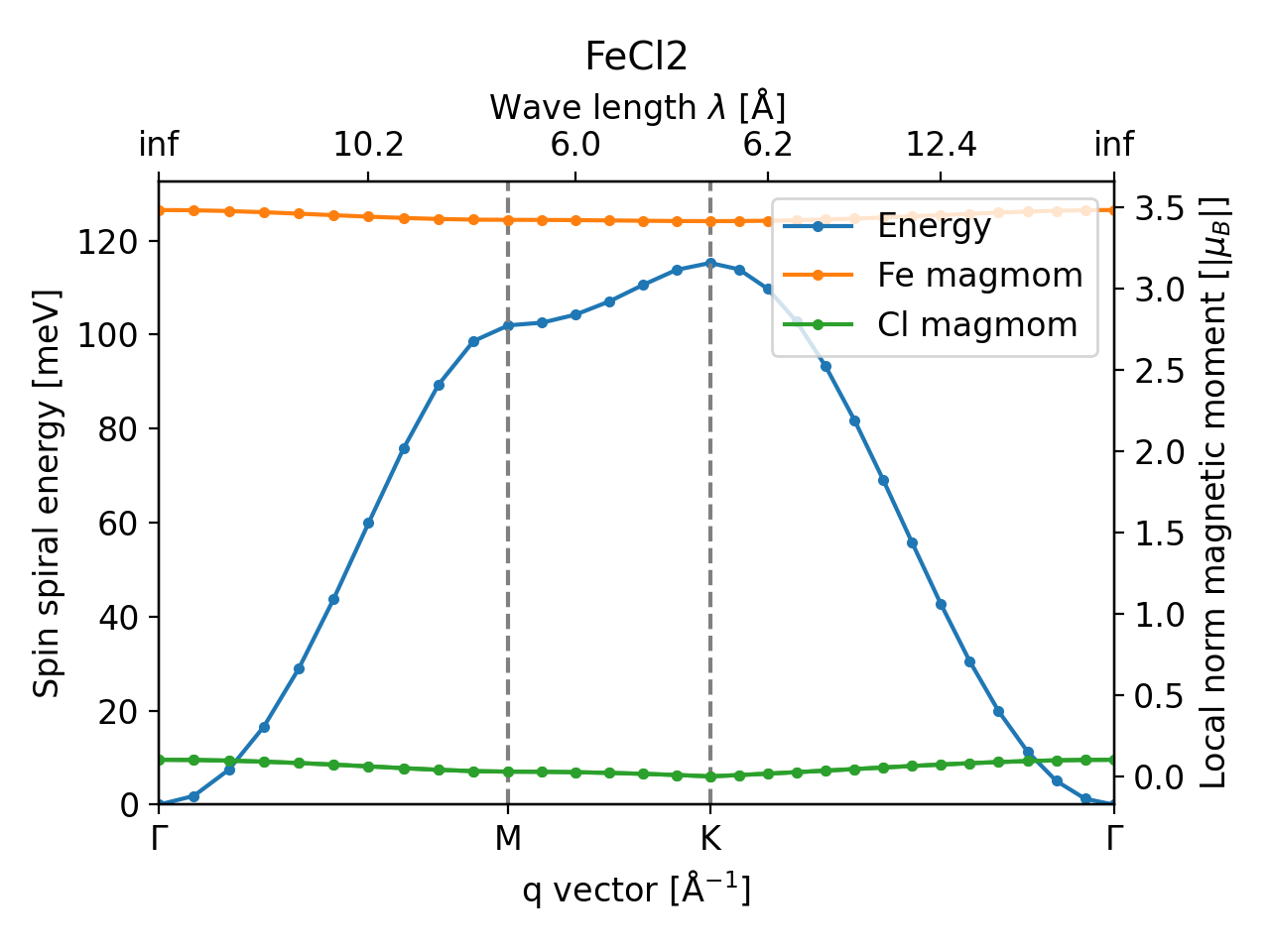}
    \includegraphics[width=0.32\textwidth]{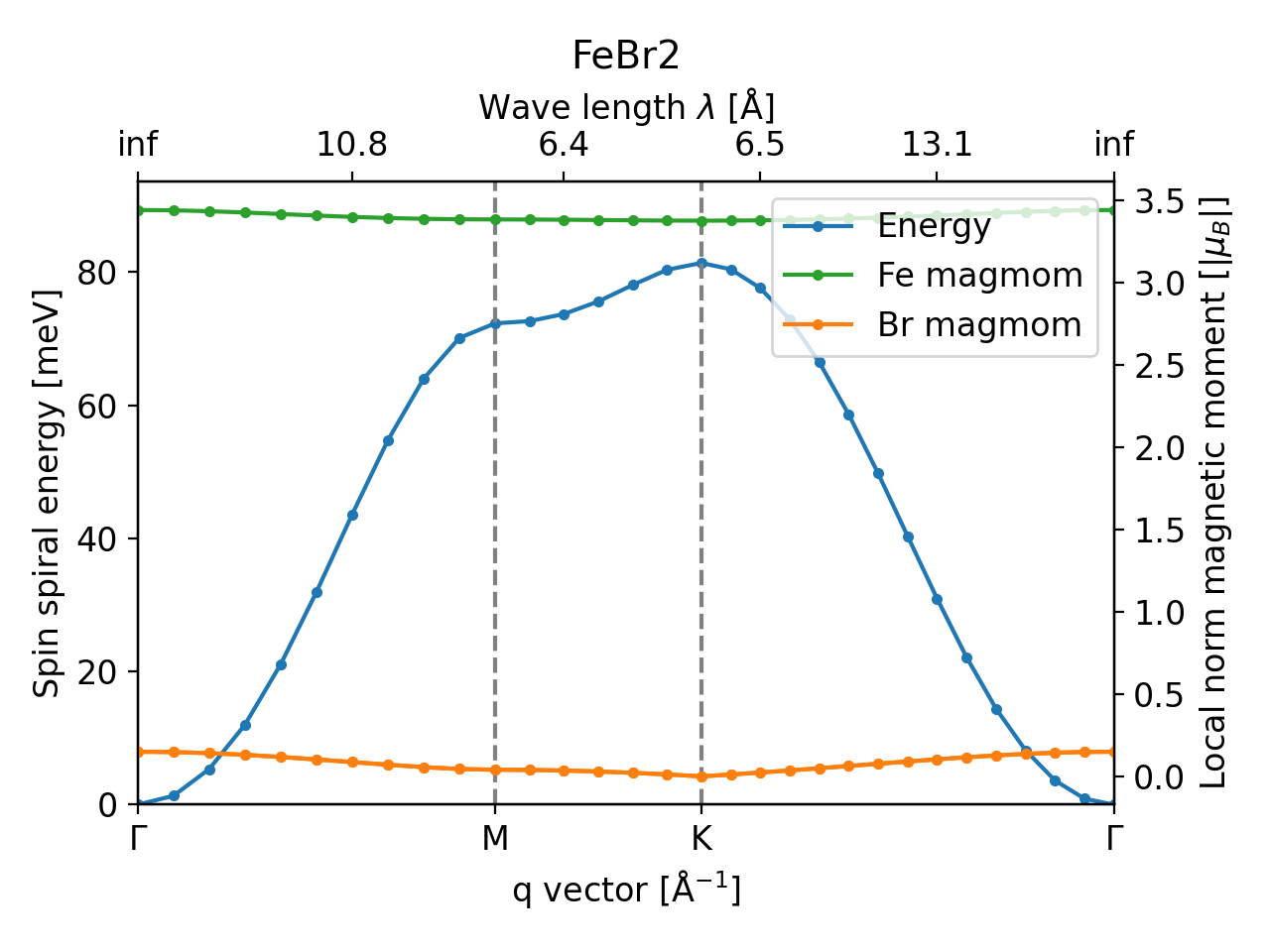}
    \includegraphics[width=0.32\textwidth]{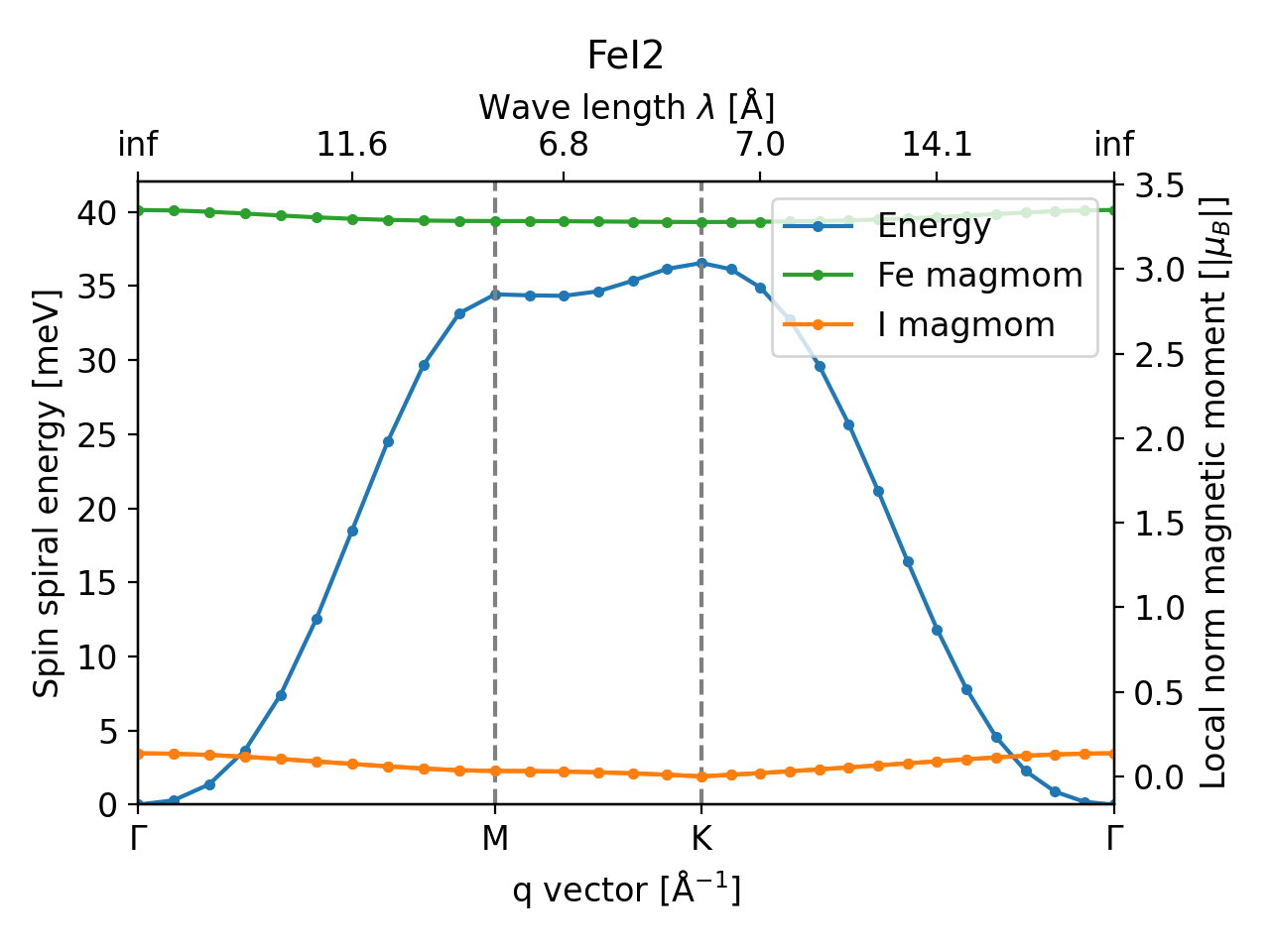}
    \includegraphics[width=0.32\textwidth]{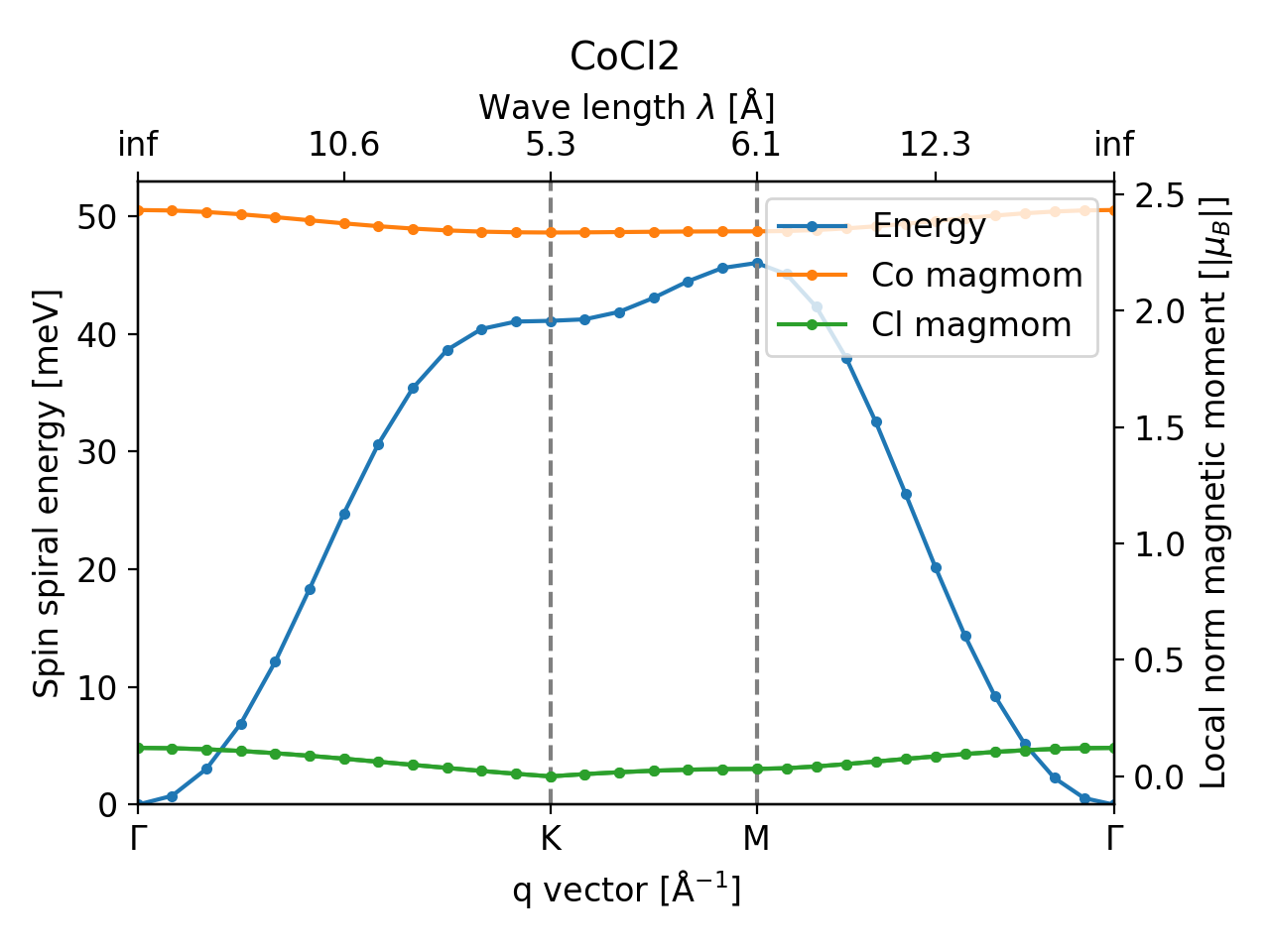}
    \caption{Spin spiral energies for ferromagnetic AB$_2$ magnets and the local magnetic moments on the atoms}
    \label{fig:3}
\end{figure*}

The entire spin spiral dispersion carry more information than just the energy minima was reported we reported in the main text, these are shown here for in figures \ref{fig:1} and \ref{fig:3}. One can find not only the stability with respect to the ferromagnetic configuration, but also compare to any other configuration in the energy landscape. Additionally, we can observe whether the remain magnetic moments are unchanged during self-consistent field cycle, and we find this is generally true except for CoI$_2$ and perhaps the titanium compounds. We note that the local magnetic moments found here are the integrated inside the PAW spheres of the respective atoms, thus these local moments does not integrate to the total moments reported in the main text since interstitial magnetization density is neglected.
We also provide the projected spin orbit energies in figure \ref{fig:2} for the lowest energy state, naturally the shape will depend very much on the specific spiral, in some cases such as the $\mathbf{q}=\mathbf{K}$ we find quite similar energies in the out-of-plane orientations, whereas incommensurate spirals tend to have more well defined minima. The in-plane orientations reported here are related by a 90$\degree$ phase shift, but the dashed line highlight that they are indeed degenerate as expected.
\section{Spin spiral convergence}
\begin{figure*}[h!]
    \centering
    \includegraphics[width=0.5\textwidth]{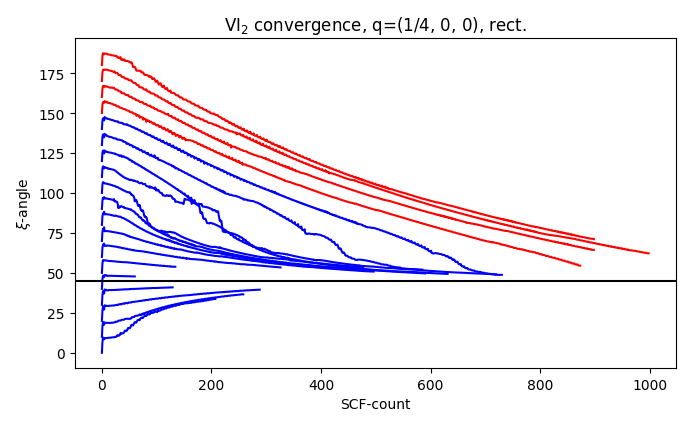}
    \caption{Convergence of the intracell angle $\xi$ in spin spiral ground state calculation of VI$_2$ at the spiral vector $\mathbf{q}= (1/4, 0, 0)$ at varying different initial conditions. The calculations shown in red, did not reach the convergence criteria on the density within the time-wall of the calculation, while the blue were considered converged. The black horizontal line is the expected angle for a smooth spin spiral as it if it was an equivalent spin spiral in the primitive unit cell.}
    \label{fig:VI2conv}
\end{figure*}

An example of convergence of a intracell angle in a rectangular supercell of the hexagonal VI$_2$ system is represented in \ref{fig:VI2conv}. We find that for all calculations which reach the convergence criteria on particularly the density, all converge the angle within some narrow region of the true angle.
We observe that the number of iterations required increase dramatically, when the initial guess is further away from the true angle, hence highlighting the importance of choosing initial conditions according to Eq. (8) in the main text.
\end{document}